\documentclass[fleqn,usenatbib]{mnras}

\usepackage{newtxtext,newtxmath}
\usepackage{graphicx}	
\usepackage{amsmath}	
\usepackage{caption} 
\usepackage{float}
\usepackage{newtxtext,newtxmath}
\usepackage{times}
\usepackage[normalem]{ulem} 
\usepackage{rotating}
\usepackage{epsfig}
\usepackage{fancyhdr}
\usepackage{epsfig}
\usepackage{graphicx}
\usepackage{amsmath}
\usepackage{theorem}
\usepackage{latexsym}
\usepackage{epic}
\usepackage{natbib} 
\usepackage[flushleft]{threeparttable}
\usepackage{tabulary}
\usepackage{tabularx}
\usepackage{pdflscape}
\usepackage{afterpage}
\usepackage{afterpage}
\usepackage{capt-of}
\usepackage{adjustbox,lipsum}
\usepackage{xcolor}
\usepackage{newtxmath}
\usepackage[normalem]{ulem} 
\usepackage{soul} 
\usepackage{enumitem}
\usepackage{rotating}

\newcommand{\yes}{\checkmark}
\newcommand{\no}{$\times$}

\usepackage[T1]{fontenc}

\DeclareRobustCommand{\VAN}[3]{#2}
\let\VANthebibliography\thebibliography
\def\thebibliography{\DeclareRobustCommand{\VAN}[3]{##3}\VANthebibliography}

\title[Properties of Low-Luminosity AGN]{Nuclear Activity and Host Galaxy Properties of Low-Luminosity AGN Identified from VLA Observations}

\author[]{
M. N. Rosli,$^{1}$
and A. Annuar$^{1}$\thanks{E-mail: adlyka@ukm.edu.my}
\\
$^{1}$Department of Applied Physics, Faculty of Science and Technology, Universiti Kebangsaan Malaysia, 43600, UKM Bangi, Selangor, Malaysia\\
}

\date{Accepted XXX. Received YYY; in original form ZZZ}

\pubyear{\the\year{}}

\begin{document}
\label{firstpage}
\pagerange{\pageref{firstpage}--\pageref{lastpage}}
\maketitle

\begin{abstract}
Low-luminosity active galactic nuclei (LLAGN; $L_{\rm bol} < 10^{42}$~erg~s$^{-1}$) may comprise a significant fraction of the local AGN population, yet their weak emission makes them difficult to detect. In this paper, we analyse 38 LLAGN identified from a 15~GHz sub-arcsecond Very Large Array survey and assess the effectiveness of X-ray, optical, and infrared wavelengths in identifying LLAGN. We found that optical emission-line diagnostics recovered $84.2^{+15.8}_{-22.9}$\% (32/38) of the sample, X-rays detected $63.2^{+25.7}_{-19.6}$\% (24/38), and infrared methods only identified $13.2^{+14.5}_{-8.0}$\% (5/38), reflecting limited X-ray sensitivity, weak or absent optical lines, and strong host galaxy contamination in the infrared. Compared to \textit{Swift}--BAT AGN, our LLAGN are $\sim$ 4.1~dex fainter in  bolometric luminosity  (log $L_{\mathrm{bol}} \approx$ 39.3 - 41.9 erg s$^{-1}$), host smaller black holes ($\sim$0.7~dex lower), and accrete at much lower rates (log $\lambda_{\text{Edd}} \approx$ -6.5 to -1.3, i.e., $\sim$ 4.2~dex lower). Host galaxies  span a broad range of morphologies, from disk- to bulge-dominated, with a subset exhibiting prominent bulges, potentially representing systems where nuclear activity has faded while the bulge remains dominant. LLAGN also  reside in galaxies with lower stellar masses ($\sim$0.3~dex) and suppressed star formation rates ($\sim$0.5~dex) relative to \textit{Swift}--BAT AGN. Overall, LLAGN in our sample systematically host smaller and weakly accreting black holes, residing in galaxies with diverse morphologies,  but lower stellar masses and reduced star formation activity, demonstrating the connection between low-level black hole accretion and host galaxy properties in the local Universe.

\end{abstract}


\section{Introduction}

Active Galactic Nuclei (AGN) are compact and energetic regions at the center of galaxies powered by accretion onto supermassive black holes (SMBHs).   
According to the AGN unified model first proposed by \citet{Antonucci1993}, the central engine consists of an SMBH, an accretion disk, a dusty torus, a hot corona and in some cases relativistic jets. 
This structure produces emission across the electromagnetic spectrum, from radio to $\gamma$-rays \citep{Ding2022, Murase2022}, with optical-UV radiation from the disk \citep{Jin2012, McHardy2018}, infrared emission from torus \citep{Kamraj2018, Balokovic2020, Fukazawa2020} and X-rays from the corona \citep{Kamraj2018, Balokovic2020, Fukazawa2020}. Jets, when present, radiate strongly in radio and high-energy $\gamma$-rays \citep{Araudo2010, Smith2016}.

AGN are generally characterized by high luminosities with bolometric luminosity ($L_{\text{bol}}$) exceeding $10^{42}\,\mathrm{erg\,s^{-1}}$ \citep{Mushotzky2004, Tozzi2006, Hickox2018, Annuar2020, Mohanadas2023}, variability across multiple timescales, and distinct emission-line features in their optical spectra.  
However, a subset of AGN known as \textit{low-luminosity AGN} (LLAGN) exhibit bolometric luminosities below $10^{42}\,\mathrm{erg\,s^{-1}}$ \citep{Ptak2001, Gu2009, Lambrides2020, Tomar2021}.  
These LLAGN are thought to accrete at extremely low Eddington ratios of $\lambda_{\mathrm{Edd}} < 10^{-3}$ \citep{Ho1995, Flohic2006, Ho2008}, often believed to accrete through radiatively inefficient accretion flows (RIAFs, \citealt{Narayan1994, Narayan1995, Ho2009, Jana2023, Yu-wei2025}).  
Such accretion flows lack the optically thick and geometrically thin disks typical of luminous AGN, leading to the absence or weakness of broad-line regions and dusty tori \citep{Elitzur2006, Gonzalez-Martin2017, Annuar2025}.  
Observations by \citet{Terashima2002} and subsequent theoretical work \citep{Mason2013, Yuan2014} suggest that LLAGN have truncated or missing accretion disks, marking a significant deviation from the classical unified model. However, several studies reported contradictory results where some LLAGN were found to retain significant torus and accretion disks, and in some cases even exhibit Compton-thick obscuration attributed to a geometrically thick torus \citep{Annuar2017, Brightman2018, daSilva2021, Boorman2025}. This shows that we still do not fully understand the structures of LLAGN.

From a physical perspective, LLAGN may provide key insight into the low-accretion regime of black hole growth and feedback.  
They represent the faint end of the AGN luminosity function and may account for a large fraction of supermassive black holes in the local Universe \citep{Ho2008}.  
Due to their weak emission, LLAGN are challenging to detect and can easily be confused with emission from other galactic sources such as X-ray binaries, supernova remnants, or star-forming regions \citep{Goulding2009, Baumgartner2013}.  
This makes the development of robust, multiwavelength diagnostics essential to identify and characterize LLAGN populations.

Each wavelength used to detect AGN offers unique advantages and limitations.  
Optical diagnostics, while powerful for luminous AGN, become unreliable for LLAGN because the host galaxy’s stellar emission can dominate over the faint accretion light, and dust obscuration can further attenuate the signal \citep{Rigby2006, Beckmann2009, Shao2017}.  
In the infrared, AGN signatures can be confused with intense star formation, as both produce strong dust emission at mid-IR wavelengths \citep{Mullaney2011, Assef2013, Satyapal2014, Kaviraj2019}.  
X-ray observations, though direct tracers of accretion power, are also limited by contamination from X-ray binaries and absorption by circumnuclear gas \citep{Brandt2005, Ichikawa2019, Masoura2020, Meulen2023}.  
Therefore, relying on a single wavelength can lead to incomplete or biased AGN samples.

Detecting AGN require a multiwavelength approach, as no single waveband provides a complete census of their population especially LLAGN, where their signature are much weaker.  
For instance, \citet{Goulding2009} identified AGN through infrared where they discovered more than half of their AGN were missed by optical detection.  
High-resolution mid-infrared study \citep{Asmus2011} detected compact nuclear emission in only 7 of 17 LLAGN, suggesting sensitivity limits and strong dependence on wavelength. X-ray surveys of nearby galaxies \citep{Zhang2009, Birchall2022} have reported nuclear detection fractions of 40--60\%, varying with host morphology. More recently, \citet{Yuk2025} demonstrated that among variability-selected LLAGN, only 3\% meet mid-infrared colour AGN criteria, while 18\% show X-ray signatures. This reflects the reduced detectability of key AGN tracers in LLAGN due to their weak nuclear emission and strong host galaxy dilution. 
Together, these findings show that a comprehensive understanding of LLAGN demographics can only be achieved through coordinated multiwavelength observations to overcome biases inherent to any single selection method. 
The differences in wavelength-dependent detection of LLAGN are closely linked to the properties of LLAGN themselves. Therefore, a multiwavelength approach is essential not only to recover the full LLAGN population, but also to reliably study how their nuclear activity relates to host structure.

AGN and their host galaxies are tightly connected through coevolution, with the properties of the host influencing black hole growth and vice versa \citep{Kormendy2013, Heckman2014}. Luminous AGN are generally found in massive, bulge-dominated galaxies with high stellar masses ($\log M_* / M_\odot \gtrsim 10.5$) and active star formation, whereas LLAGN are often hosted by disk-dominated, weak-bulge galaxies with lower stellar masses and subdued star formation \citep{Ho1997, Masegosa2011, Reines2015}. This structural difference implies that while luminous AGN are often fueled by large-scale processes such as mergers, LLAGN may be powered by secular processes or stellar mass loss \citep{Hopkins2006, Davis2011}. 

$L_{\text{bol}}$ provides a measure of the total radiative output of an AGN, serving as a key indicator of accretion power and efficiency. Luminous AGN typically exhibit $L_{\text{bol}} \gtrsim 10^{42}$~erg~s$^{-1}$, while LLAGN span $10^{39}$--$10^{41.5}$~erg~s$^{-1}$ and are often powered by RIAFs rather than standard thin disks \citep{Narayan1995, Yuan2014}.
The mass of the central black hole ($M_{\text{BH}}$) and the Eddington ratio ($\lambda_{\text{Edd}} = L_{\text{bol}}/L_{\text{Edd}}$) are fundamental parameters linking AGN activity to host properties. LLAGN typically host black holes of $10^5$--$10^8~M_\odot$, smaller than the $\gtrsim 10^8~M_\odot$ black holes in classical Seyferts and quasars \citep{Awaki2001, Shen2011}.  Their Eddington ratios are generally low ($\log \lambda_{\text{Edd}} \lesssim -2$), reflecting inefficient accretion through RIAFs \citep{Ho2009, Draper2010}. Correspondingly, LLAGN are found in galaxies with lower stellar masses ($\log M_*/M_\odot \sim 9$--$10$), indicating smaller bulge components and lower $M_{\text{BH}}/M_*$ ratios \citep{Reines2015, Lambrides2020}. These differences emphasize that the growth of black holes in LLAGN proceeds under different structural and mass constraints compared to luminous AGN.

Star formation activity and host morphology provide further insight into AGN--galaxy coevolution. While luminous AGN are often found in star-forming systems, LLAGN tend to reside in quiescent or gas-poor,  reflecting their lower accretion rates and structural immaturity \citep{Ho2008, Eracleous2010}. Morphologically, LLAGN hosts are primarily late-type galaxies (spirals and irregulars), suggesting that their nuclear activity is sustained by limited gas inflow rather than major mergers \citep{Schawinski2010, Reines2015}. Understanding these trends in stellar mass, star formation, and morphology is essential for tracing the full spectrum of black hole--galaxy coevolution at low luminosities.

Despite significant progress in understanding LLAGN, several fundamental challenges remain in constructing a complete and unbiased census \citep{Ho2008, Eracleous2010}. Their intrinsically low luminosities lead to strong host galaxy dilution, making it difficult to isolate nuclear emission from stellar and star-forming processes across all wavelengths \citep{Ho2008, Goulding2009, Trump2015}. In addition, the diversity of accretion structures introduces ambiguity in interpreting observational signatures \citep{Narayan1995, Yuan2014, Elitzur2006, Gonzalez-Martin2017}. Furthermore, the strong wavelength dependence of LLAGN detectability results in selection biases, as samples identified in different bands are often incomplete and not directly comparable \citep{Asmus2011, Ichikawa2019, Birchall2022, Yuk2025}. Together, these limitations hinder a unified view of LLAGN demographics and their role in low-accretion black hole growth \citep{Ho2009, Heckman2014}. Addressing this requires a systematic, homogeneous multiwavelength approach that accounts for selection biases while providing consistent measurements of both nuclear and host galaxy properties. 
Although previous studies have investigated LLAGN properties across multiple wavelengths, they are typically based on heterogeneous samples and non-uniform methodologies, limiting direct comparison with luminous AGN populations (e.g. \citealt{Ho2008, Goulding2009, Asmus2011, Ichikawa2019, Birchall2022}). In particular, a unified analysis combining multiwavelength diagnostics with consistently derived physical parameters for both LLAGN and luminous AGN remains lacking. In this work, we address this gap by performing a homogeneous analysis of LLAGN and directly comparing them with typical AGN samples, enabling a robust assessment of their accretion and host galaxy properties.

This study aims to investigate the efficiency of each wavelength in identifying LLAGN 
and to examine how their nuclear and host properties compare to those of their brighter counterparts.   
In particular, we evaluate the efficiency of standard diagnostic techniques spanning X-ray, optical and infrared wavelength in identifying LLAGN that may be missed when relying on a single band.  
We further characterize the key physical properties of these nuclei, including their bolometric luminosities, black hole masses, and Eddington ratios, in order to assess their accretion states and compare them with standard AGN populations.  
In parallel, we analyze the host galaxy characteristics such as morphology, stellar mass, and star formation rate to explore how LLAGN activity is connected to the structural and evolutionary properties of their galaxies.  
Through this comprehensive multiwavelength approach, we aim to establish where LLAGN lie within the broader AGN–host connection and to provide a more complete understanding of low-level black hole activities in nearby galaxies and its role in galaxy evolution.

\section{Initial AGN Sample}

Our initial sample was drawn from the sub-arcsecond 15~GHz Very Large Array (VLA) survey of nearby LLAGN by \citet{Saikia2018}. In this study, we adopted the 45 detected sources from Table~A.1 of \citet{Saikia2018} as our initial AGN sample. Originally, their study targeted 76 nuclei from the optically selected \textit{Palomar spectroscopic survey} \citep{Ho1995, Ho1997}, focusing on galaxies that had not been previously detected at high radio frequencies. \citet{Saikia2018} achieved a typical rms sensitivity of $\sim$11.5~$\mu$Jy~beam$^{-1}$ at 15~GHz and detected compact nuclear radio emission in 45 of the 76 sources ($\approx$60\%). These detections exhibit parsec-scale radio cores largely uncontaminated by star formation, providing one of the most complete and homogeneous high-resolution radio samples of nearby LLAGN to date. \citet{Saikia2018} sample was chosen as the starting point of this work because its high spatial resolution and sensitivity at 15~GHz isolating emission from the central parsec-scale region, ensuring that the radio flux traces genuine AGN activity rather than circumnuclear star formation. This parent sample represents the active subset of the Palomar galaxies, encompassing Seyferts, LINERs, and transition objects with confirmed compact nuclear radio emission. 

The defining criterion in \citet{Saikia2018} was radio detection and optical classification, rather than a strict limit in bolometric luminosity. Their goal was to establish a census of actively accreting nuclei in nearby galaxies by identifying compact, high-brightness radio cores which are signatures of jets or outflows powered by accretion onto a central black hole. At 15~GHz, extended star-forming emission is heavily suppressed, making such compact detections a robust indicator of AGN activity. The optical classifications (Seyfert, LINER, or transition) were inherited from the \textit{Palomar spectroscopic survey}, which employed emission-line ratio diagnostics (BPT diagrams; \citealt{Ho1997}) to distinguish AGN-photoionized nuclei from H~II regions.

In this context, the ``LLAGN'' labelled in \citet{Saikia2018} primarily denotes low-power, radiatively inefficient accretors inferred from their optical line ratios and compact radio morphologies, rather than strictly luminosity-defined systems. Consequently, their sample spans a relatively broad range of luminosities from genuinely weak to higher luminosity.

\section{Final Low-Luminosity AGN Sample}

Radio emission at 15~GHz is largely unaffected by dust obscuration and free–free absorption, making it an excellent tracer of compact nuclear activity even in galaxies where optical or X-ray signatures are weak. However, a radio-detected sample can include both relatively luminous and weak AGN, including the jet-dominated AGN. Therefore, a bolometric luminosity cut was required to isolate the faint nuclei from the more luminous AGN. This ensures our final sample consists of only LLAGN.
From the 45 radio-detected AGN of \citet{Saikia2018}, we identified the truly low-luminosity members by estimating their bolometric luminosities using available X-ray and optical measurements (see Section \ref{sec: Multiwavelength Analysis} and \ref{sec: AGN properties}). 

We adopted a bolometric luminosity threshold of $L_{\mathrm{bol}} < 10^{42}~\mathrm{erg~s^{-1}}$, a value that separates the regime of radiatively efficient, Seyfert-like accretion from the RIAFs characteristic of LLAGN \citep{Ptak2001, Ho2008, Saikia2018, Lambrides2020}.  
Sources exceeding this luminosity limit either in X-ray or optical were excluded from the subsequent low-luminosity analysis, since their energy output is more consistent with typical AGN activity than with the faint nuclei targeted in this study.

Applying this luminosity criterion to the 45 radio-detected galaxies yields a final sample of 38 LLAGN, corresponding to approximately 84\% of the original set. 
The remaining 7 galaxies have luminosities $L_{\mathrm{bol}}  \, \geq \, 10^{42}~\mathrm{erg~s^{-1}}$, and are therefore excluded from our sample. Those are NGC 410, NGC 2832, NGC 3735, NGC 3982, NGC 4051, NGC 5273 and NGC 6482 (refer to Table \ref{Tab: Luminosity for Excluded 7 LLAGN}).
These excluded sources do not exhibit any systematic differences from our final LLAGN sample (i.e. compact radio cores and optical classifications). They may represent the higher-luminosity tail of the same population rather than a distinct class of objects.

Within our final LLAGN sample, only three galaxies (NGC~1055, NGC~4395, and NGC~6951) show extended radio morphologies suggestive of jet emission. This corresponds to only $7.9^{12.5}_{-5.7}\%$ (3/38) of the LLAGN, while the majority of sources appear compact at the resolution of the 15\,GHz observations.\footnote{Uncertainties were derived using the small-number Poisson approximations of \citet{Gehrels1986} for a 90\% double-sided (95\% single-sided) confidence interval, with the upper limit capped at 100\%.}
In addition, as the parent sample is selected at 15 GHz, it preferentially includes LLAGN hosting compact radio cores associated with weak jets or outflows. This is consistent with the predominance of compact sources in our sample, where only a small fraction exhibit extended radio morphology. As a result, our sample may miss intrinsically radio-quiet or radio-faint nuclei. Therefore, both bolometric luminosities and Eddington ratios of our sample may be systematically higher compared to an unbiased LLAGN population, due to the selection bias toward radio-loud sources.

Our sample of 38 LLAGN represent nearby, energetically faint nuclei that provide ideal laboratories for studying the properties of LLAGN.
The full list of LLAGN used in this work, including their names, position, distances, AGN and host galaxy properties, especially the bolometric luminosities, is presented in Table \ref{tab:basic properties 38 LLAGN}.

\begin{table*}\textcolor{white}{[]}
\centering
\begin{adjustbox}{angle=90, width=\textwidth, totalheight=0.9\textheight, keepaspectratio}
\begin{minipage}{\textheight}
\caption{Basic properties for the 38 LLAGN, including coordinates, optical emission line ratios, WISE photometry, and optical/near-infrared magnitudes.}
\begin{tabular}{lccccccccccccccc}
\hline
Name & z & R.A. & Dec. & log \(([\mathrm{O\,III}]/\mathrm{H}\beta\)) & log \(([\mathrm{O\,I}]/\mathrm{H}\alpha\))& log \(([\mathrm{N\,II}]/\mathrm{H}\alpha\))& log \(([\mathrm{S\,II}]/\mathrm{H}\alpha\))& W1 & W2 & W3 & B & V & R & K \\
& & (h:m:s) & (d:m:s) & & & & & (mag) & (mag) & (mag) & (mag) & (mag) & (mag) & (mag) \\
(1) & (2) & (3) & (4) & (5) & (6) & (7) & (8) & (9) & (10) & (11) & (12) & (13) & (14) & (15) \\
\hline
NGC488 & 0.007579 & 01:21:46.80 & 05:15:24.6 & -0.022 & -0.886 & 0.176 & -0.328 & 9.185 & 9.197 & 8.509 & 15.00 & -- & -- & 6.96 \\
NGC521 & 0.016738 & 01:24:33.77 & 01:43:52.9 & 0.114 & -1.066 & 0.076 & -0.481 & 10.257 & 10.297 & 8.956 & 12.50 & -- & 12.30 & 8.58 \\
NGC660 & 0.002829 & 01:43:02.32 & 13:38:44.9 & 0.403 & -1.328 & -0.071 & -0.367 & 8.000 & 6.847 & 3.514 & 12.02 & 11.16 & -- & 7.34 \\
NGC777 & 0.016728 & 02:00:14.91 & 31:25:45.9 & 0.566 & -0.409 & 0.253 & -0.004 & 9.442 & 9.498 & 8.888 & 12.49 & 11.45 & -- & 8.37 \\
NGC841 & 0.01516 & 02:11:17.37 & 37:29:49.7 & 0.314 & -0.237 & 0.155 & 0.124 & 9.997 & 9.971 & 8.020 & 12.80 & -- & -- & 9.38 \\
NGC1055 & 0.003322 & 02:41:45.20 &  00:26:38.1 & -- & -1.194 & -0.180 & -0.387 & 9.277 & 8.961 & 5.001 & 11.40 & 10.59 & 10.50 & 7.15 \\
NGC1169 & 0.007966 &  03:03:34.74 & 46:23:11.1 & 0.225 & -0.495 & 0.450 & 0.241 & 9.791 & 9.781 & 9.109 & 13.20 & -- & -- & 8.06 \\
NGC1961 & 0.013122 & 05:42:04.65 & 69:22:42.4 & 0.072 & -0.678 & 0.292 & 0.029 & 9.752 & 9.618 & 6.724 & 11.73 & 10.99 & -- & 7.73 \\
NGC2681 & 0.002308 &  08:53:32.72 &  51:18:49.2 & 0.238 & -0.721 & 0.373 & -0.125 & 8.502 & 8.422 & 5.809 & 11.09 & 10.29 & -- & 7.45 \\
NGC2683 & 0.001371 & 08:52:41.30 & 33:25:18.7 & 0.467 & -0.481 & 0.173 & 0.143 & 8.779 & 8.688 & 6.564 & 10.68 & 9.79 & 9.38 & 6.33 \\
NGC2859 & 0.005631 &  09:24:18.53 & 34:30:48.6 & 0.431 & -1.036 & 0.267 & 0.025 & 9.035 & 9.056 & 8.328 & 11.80 & -- & -- & 8.04 \\
NGC2985 & 0.004416 & 09:50:22.18 & 72:16:44.2 & -0.013 & -0.921 & -0.066 & -0.215 & 9.061 & 9.019 & 6.991 & 11.37 & 10.61 & -- & 7.36 \\
NGC3642 & 0.005287 & 11:22:17.90 & 59:04:28.3 & 0.121 & -0.745 & -0.149 & 0.017 & 10.370 & 10.235 & 7.733 & 14.86 & 14.04 & -- & 8.97 \\
NGC3898 & 0.003859 & 11:49:15.24 & 56:05:04.3 & 0.322 & -1.041 & 0.170 & 0.079 & 8.914 & 8.927 & 8.130 & 11.70 & -- & -- & 7.66 \\
NGC3992 & 0.003496 & 11:57:35.96 & 53:22:29.0 & 0.299 & -0.886 & 0.238 & -0.009 & 9.639 & 9.677 & 8.763 & 10.94 & -- & 9.57 & 6.94 \\
NGC4036 & 0.00462 & 12:01:26.75 & 61:53:44.6 & 0.267 & -0.337 & 0.352 & 0.258 & 8.826 & 8.813 & 7.720 & 12.18 & 11.20 & -- & 7.56 \\
NGC4111 & 0.002628 &  12:07:03.13 & 43:03:56.3 & -0.076 & -0.721 & 0.117 & 0.013 & 8.225 & 8.221 & 7.009 & 11.63 & 10.74 & -- & 7.55 \\
NGC4145 & 0.003366 & 12:10:01.56 &  39:53:00.9 & 0.037 & -0.886 & -0.215 & -0.097 & 12.037 & 12.062 & 9.515 & -- & -- & -- & 8.48 \\
NGC4346 & 0.002775 &  12:23:27.96 &  46:59:37.6 & 0.107 & -0.102 & 0.571 & 0.210 & 9.095 & 9.126 & 8.561 & 12.30 & -- & -- & 8.18 \\
NGC4395 & 0.001064 & 12:25:48.88 & 33:32:48.7 & 0.794 & -0.444 & -0.357 & -0.018 & 12.624 & 11.839 & 8.605 & 10.54 & 10.11 & 9.98 & 9.98 \\
NGC4750 & 0.005404 &  12:50:07.32 & 72:52:28.6 & 0.241 & -0.222 & 0.456 & 0.301 & 9.791 & 9.636 & 7.365 & 11.80 & -- & -- & 8.12 \\
NGC5194 & 0.001534 &  13:29:52.71 & 47:11:42.8 & 0.952 & -0.796 & 0.462 & -0.066 & 8.840 & 8.664 & 6.190 & 9.26 & 8.36 & 8.40 & 5.50 \\
NGC5195 & 0.001518 & 13:29:59.53 &  47:15:58.3 & 0.086 & -0.260 & 0.735 & 0.301 & 7.543 & 7.671 & 4.531 & 10.45 & 9.55 & 9.31 & 6.25 \\
NGC5395 & 0.01158 &  13:58:37.96 & 37:25:28.2 & 0.481 & -0.770 & 0.090 & -0.119 & 10.632 & 10.653 & 8.780 & 13.26 & 12.48 & -- & 8.57 \\
NGC5448 & 0.006725 &  14:02:50.05 & 49:10:21.2 & -0.032 & -0.699 & 0.013 & 0.061 & 10.147 & 10.114 & 7.452 & 12.70 & -- & -- & 8.79 \\
NGC5485 & 0.006352 & 14:07:11.35 & 55:00:06.0 & 0.407 & -0.678 & 0.262 & -0.081 & 9.725 & 9.753 & 8.645 & 13.82 & 12.22 & -- & 8.40 \\
NGC5566 & 0.005027 & 14:20:19.89 &  03:56:01.4 & 0.267 & -0.699 & 0.243 & 0.137 & 8.635 & 8.610 & 7.385 & 11.46 & 10.55 & -- & 7.39 \\
NGC5631 & 0.006484 &  14:26:33.29 & 56:34:57.4 & 0.538 & -0.357 & 0.417 & 0.420 & 9.486 & 9.503 & 8.607 & 13.51 & 12.55 & -- & 8.47 \\
NGC5746 & 0.005764 & 14:44:55.98 & 01:57:18.1 & 0.344 & -1.000 & 0.318 & -0.060 & 8.852 & 8.800 & 7.383 & 12.30 & -- & -- & 6.88 \\
NGC5850 & 0.008489 & 15:07:07.68 & 01:32:39.3 & 0.377 & -0.658 & 0.228 & 0.155 & 9.697 & 9.732 & 8.310 & 11.50 & -- & 10.79 & 8.10 \\
NGC5921 & 0.004937 &  15:21:56.49 & 05:04:14.3 & 0.004 & -0.959 & -0.046 & -0.215 & 9.877 & 9.815 & 6.964 & 12.70 & -- & -- & 8.10 \\
NGC5985 & 0.008412 & 15:39:37.06 & 59:19:55.2 & 0.423 & -0.523 & 0.489 & 0.179 & 10.813 & 10.901 & 10.000 & 15.24 & 14.22 & -- & 8.15 \\
NGC6340 & 0.004026 & 17:10:24.84 & 72:18:15.9 & 0.185 & -0.367 & 0.033 & 0.083 & 9.706 & 9.700 & 8.781 & 11.87 & 11.01 & -- & 8.39 \\
NGC6703 & 0.007565 & 18:47:18.82 & 45:33:02.3 & 0.146 & -0.444 & 0.477 & 0.167 & 9.228 & 9.240 & 8.699 & 12.29 & 11.34 & -- & 8.25 \\
NGC6951 & 0.00475 & 20:37:14.12 &  66:06:20.0 & 0.821 & -0.638 & 0.394 & -0.041 & 8.980 & 8.783 & 5.170 & 11.64 & 10.65 & -- & 7.22 \\
NGC7814 & 0.003506 & 00:03:14.90 & 16:08:43.2 & 0.000 & -0.222 & -0.119 & 0.292 & 8.438 & 8.380 & 6.850 & 12.00 & -- & -- & 7.08 \\
IC356 & 0.002985 & 04:07:46.89 & 69:48:44.7 & 0.389 & -0.959 & 0.225 & -0.027 & 8.731 & 8.708 & 7.577 & 13.30 & -- & -- & 6.04 \\
IC520 & 0.011628 &  08:53:42.25 & 73:29:27.4 & 0.393 & -1.013 & 0.220 & -0.125 & 10.048 & 10.048 & 7.714 & 11.90 & -- & -- & 8.72 \\
\hline
\end{tabular}
\begin{tablenotes}
    \footnotesize
        \item[] \textit{Notes}. (1) Galaxy name; (2) Redshift; (3)-(4) radio position of LLAGN as detected by VLA \citep{Saikia2018} in equatorial coordinates (J2000), where right ascension is given in hours:minutes:seconds and declination in degrees:arcminutes:arcseconds; (5)–(8) Logarithmic values of emission line ratios from optical spectroscopy based on \citet{Ho1997}; (9)–(11) WISE magnitudes for W1 (3.4 $\mu$m), W2 (4.6 $\mu$m), and W3 (12.0 $\mu$m) bands; (12)-(15) Optical and near-infrared magnitudes (B, V, R, K bands) for the galaxy sample compiled from SIMBAD.
    \end{tablenotes}
\label{tab:basic properties 38 LLAGN}
\end{minipage}
\end{adjustbox}
\end{table*}

\section{Multiwavelength Analysis}
\label{sec: Multiwavelength Analysis}

AGN often require a multiwavelength approach for reliable identification, as emission from accretion processes manifests differently across the electromagnetic spectrum. In this study, we combine X-ray, optical, and infrared diagnostics to assess how effectively standard AGN identification techniques recover the 38 LLAGN confirmed by compact radio nuclei. 

\subsection{X-ray}

X-ray observations are among the most direct probes of accretion activity in AGN, tracing high-energy processes that occur near the supermassive black hole. X-rays can penetrate large columns of gas and dust, allowing the detection of obscured or weakly accreting nuclei that may be undetectable at optical or infrared wavelengths. In this work, we utilized archival observations from three major telescopes which are \textit{Chandra}, \textit{XMM-Newton}, and \textit{Swift}--BAT to provide comprehensive X-ray coverage of the sample. Each instrument contributes uniquely to the characterization of the LLAGN population. For soft X-rays, we prioritize \textit{Chandra} data when available since it have finer angular resolution ($\sim$0.5$^{\prime\prime}$, \citealt{Weisskopf2002}) than \textit{XMM-Newton} ($\sim$6$^{\prime\prime}$, \citealt{Jansen2001}) which allows clearer separation of faint nuclear emission from surrounding host galaxy contamination. For sources lacking \textit{Chandra} observations, we used \textit{XMM-Newton} data.

\begin{figure}
    \centering
    \includegraphics[width=1 \linewidth]{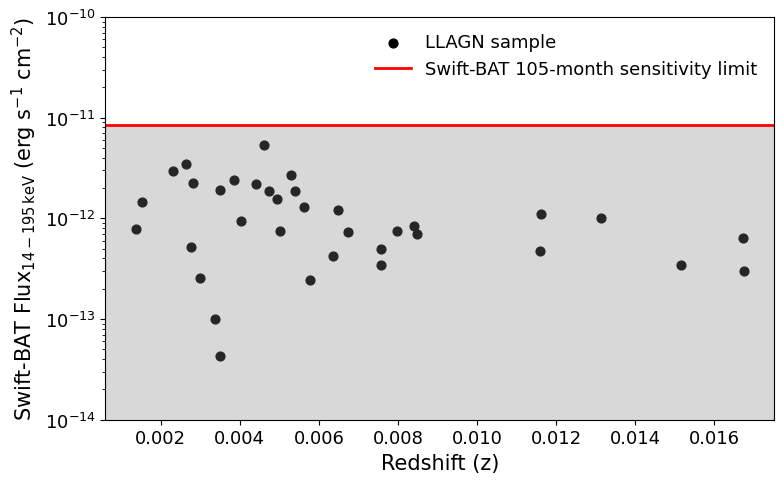}
\caption{\textit{Swift}--BAT 14--195 keV flux versus redshift for the LLAGN sample. The red horizontal line represents the approximate $5\sigma$ sensitivity limit of the 105-month \textit{Swift}--BAT survey.}
    \label{fig:Swift-BAT sensitivity}
\end{figure}

\subsubsection{Chandra}
\label{subsubsec: Chandra}

The \textit{Chandra X-ray Observatory} provides sub-arcsecond angular resolution, making it particularly well suited for isolating compact nuclear emission from surrounding diffuse gas or X-ray binaries. Out of 38 galaxies, only 24 have \textit{Chandra} data. These data 
were obtained from the High Energy Astrophysics Science Archive Research Center (HEASARC)\footnote{HEASARC data are publicly available at \url{https://heasarc.gsfc.nasa.gov/cgi-bin/W3Browse/w3browse.pl}}. All data were reprocessed using the \texttt{CIAO} software package \citep{Fruscione2007} with the latest calibration files via the \texttt{chandra\_repro} command.

For sources with single observations, we extracted events in the 2.0-8.0~keV range using \texttt{dmcopy} and produced corresponding exposure maps with \texttt{fluximage}. This energy range is used to minimize contamination at soft X-rays such as diffuse gas or star formation activity. 
The point spread function (PSF) map was calculated using \texttt{mksfmap} for 90\% encircled energy. For galaxies with multiple observations, \texttt{merge\_obs} was used to combine the datasets, filter to the same energy range, and generate merged exposure and PSF maps.

\begin{figure*}
    \centering
    \includegraphics[width=1\linewidth]{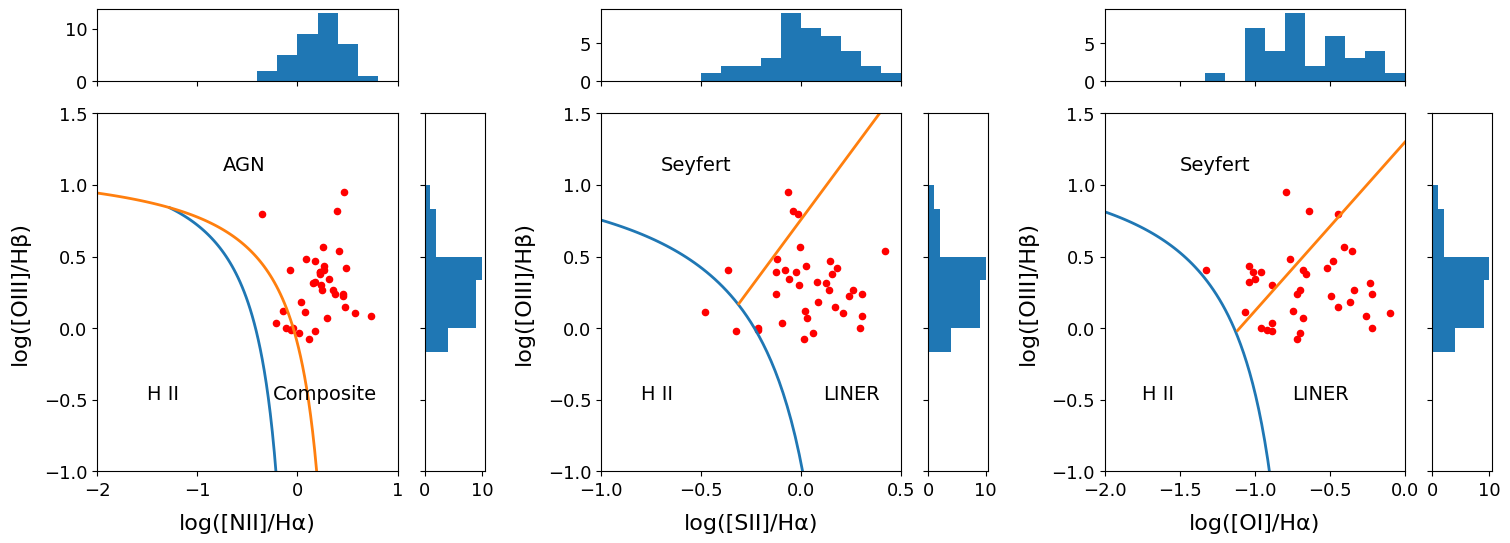}
        \caption{BPT originally from \citet{Baldwin1981}; \citet{Kewley2006} provided classification scheme. \textit{Left}: The log ([\ion{O}{iii}] / H$\beta$) vs log ([\ion{N}{ii}] / H$\alpha$) diagram, \textit{Centre}: log ([\ion{O}{iii}] / H$\beta$) vs log ([\ion{S}{ii}] / H$\alpha$) diagram and \textit{Right}: log ([\ion{O}{iii}] / H$\beta$) vs log ([\ion{O}{i}] / H$\alpha$) diagram. The marginal histograms on the top and right of each panel show the number distribution of galaxies along each emission-line ratio axis.}
    \label{fig:BPT DIAGRAM}
\end{figure*}

Source detection was performed using the \texttt{wavdetect} algorithm with wavelet scales of 1, 2, 4, and 8 pixels and a significance threshold of \texttt{sigthresh} = $10^{-6}$ which are appropriate for minimizing spurious detections in ACIS imaging. To ensure that detected X-ray sources correspond to the galaxy nucleus, we compared the X-ray positions with optical images from the Sloan Digital Sky Survey (SDSS) using the \texttt{DS9} visualization software \citep{Joye2003}. Sources that are positionally coincident with the optical nucleus were retained as AGN detections.

Out of the 24 LLAGN with available \textit{Chandra} data, 22 show clear nuclear X-ray detections within the central few arcseconds of their optical nuclei, corresponding to a detection rate of $91.7^{+8.3}_{-29.6}$\%.  The remaining two sources exhibit no significant point-like emission at the nucleus above the local background level and are treated as X-ray non-detections in subsequent analysis.  This high detection fraction demonstrates the effectiveness of \textit{Chandra}'s angular resolution and sensitivity in identifying faint nuclear activity even in low-luminosity systems.
However, the high detection rate may also be influenced by the fact that these galaxies were targeted by Chandra because they had been previously identified as active.

After the galaxies are detected, we calculate the X-ray flux for each source, which is then used to estimate the bolometric luminosity.
The X-ray fluxes for each galaxy were estimated by combining the measured count rates with the known redshift, Galactic column density ($N_{\mathrm{H,Gal}}$), and an assumed photon index of $\Gamma = 1.8$, which is typical for AGN. The $N_{\mathrm{H,Gal}}$ were obtained using the \texttt{XSPEC} tool \citep{Arnaud1996}, ensuring consistency across the sample.
We then used \texttt{WebPIMMS} to convert the instrumental count rates into unabsorbed fluxes in the energy band of 2-10 keV. Once the fluxes were determined, the X-ray luminosities were calculated and subsequently used to estimate the bolometric luminosity as in Section \ref{sec:Lbol}. 
Although previous studies suggested that LLAGN can exhibit harder X-ray spectra (e.g., \citealt{Emmanoulopoulos2012, Ford2025}), we calculated the hardness ratio using the 2--10 keV and 0.5--2 keV bands. We found that the average hardness ratio of our sample, $\mathrm{HR} = -0.22 \pm 0.45$, which is consistent with typical AGN spectral slopes \citep{Nandra1997, Pons2016} within the uncertainties and showed no evidence for extreme spectral hardness or softness.

\subsubsection{XMM-Newton}

The \textit{XMM-Newton} telescope provides high sensitivity due to its large effective collecting area, complementing \textit{Chandra}'s superior spatial resolution. Only two  of the remaining 14 galaxies have \textit{XMM-Newton} observations. 
These galaxies were analyzed using the \texttt{HEASoft} (v6.32) and \texttt{SAS} (v21.0) software packages. Event files for the MOS1, MOS2, and PN cameras were generated using the \texttt{emchain} and \texttt{epchain} tasks, with the parameter \texttt{with\_outoftime = true} enabled for the PN detector to correct for out-of-time events. The combined data from all cameras were merged using the \texttt{merge} command, and images were created with \texttt{evselect}. Point-source detection was performed using the \texttt{edetect\_chain} algorithm, which applies a sliding-cell method to identify statistically significant X-ray sources in the 2.0-10 keV energy range and provides their positions, counts, and detection likelihoods. Detected X-ray sources were again cross-matched with SDSS optical nuclei using \texttt{DS9} to identify AGN. 
Both LLAGN observed with \textit{XMM-Newton} were successfully detected. 
These detections allow us to identify their nuclei as hosting candidate LLAGN. The X-ray luminosities for the \textit{XMM-Newton} sources were then determined using the same method discussed for \textit{Chandra} in Section \ref{subsubsec: Chandra}. 

\subsubsection{\textit{Swift}--BAT}

To complement the soft X-ray coverage from \textit{Chandra} and \textit{XMM-Newton}, we included hard X-ray data provided by the \textit{Swift} Burst Alert Telescope (BAT) catalog \citep{Oh2018}.
Our sample was compared with the \textit{Swift}--BAT 105-month catalog to identify possible hard X-ray counterparts. 
Only two ($5.3^{+11.3}_{-4.3}$\%) of the LLAGN in our sample (NGC~4395 and NGC~5194) were detected in the hard X-ray band by \textit{Swift}--BAT.  
These detections are consistent with previous studies identifying both systems as bona fide AGN with persistent hard X-ray emission \citep{Xu2016, Kammoun2019}.  
The remaining LLAGN were not detected by BAT, which shows the challenges of identifying such faint nuclei in the hard X-ray band and the need for more sensitive or deeper hard X-ray observations to fully complement the available soft X-ray data.

The low detection rate of LLAGN in the hard X-ray band may arise either from intrinsically low luminosities or from heavy obscuration that suppresses the observed emission. To investigate this, we estimated the expected hard X-ray fluxes of our LLAGN using the extinction-corrected [O~\textsc{iii}] emission line as a proxy for intrinsic AGN power. The [O~\textsc{iii}] line originates from the narrow-line region (NLR), which lies outside the obscuring torus and is therefore relatively unaffected by nuclear obscuration, making it a reliable tracer of intrinsic luminosity. We adopted [O~\textsc{iii}] fluxes from \citet{Ho1997} and converted them to \textit{Swift}--BAT 14--195~keV fluxes using the empirical scaling factor from \citet{Berney2015}. Based on this conversion, we compared the inferred hard X-ray fluxes with the $5\sigma$ sensitivity limit of the 105-month \textit{Swift}--BAT survey. As shown in Fig.\ref{fig:Swift-BAT sensitivity}, the majority of our LLAGN are predicted to lie below the \textit{Swift}--BAT detection threshold. This suggests that their non-detection is primarily driven by intrinsically low luminosities, although heavy obscuration may also contribute in some cases. Therefore, as these sources fall below the \textit{Swift}--BAT sensitivity limit, they remain undetected in current hard X-ray surveys.

\subsection{Optical}

AGN detection through optical wavelengths can be achieved using spectroscopic diagnostics such as the Baldwin, Phillips, and Terlevich (BPT) diagram \citep{Baldwin1981}.  
The BPT diagram distinguishes between different ionizing sources in galaxies by comparing the strengths of key emission-line ratios, particularly the high-ionization lines relative to the hydrogen Balmer lines.  
In this parameter space, distinct regions correspond to ionization dominated by star formation (H~\textsc{ii} region), composite systems (mixed AGN and star formation), Seyfert-type AGN, and low-ionization nuclear emission-line regions (LINERs).

\begin{figure*}
    \centering
    \includegraphics[width=0.9\linewidth]{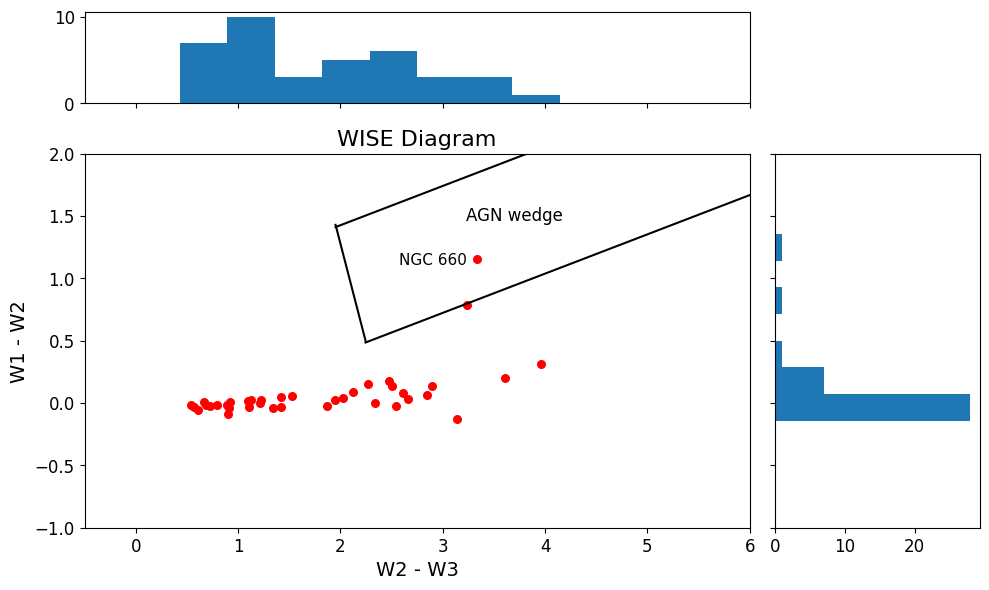}
    \caption{Infrared colour–colour diagram constructed using W1, W2, and W3 WISE bands (3.4, 4.6, and 12~$\mu$m, respectively). The AGN selection wedge defined by \citet{Mateos2012} is shown by the black solid line where sources within this region are classified as AGN, while sources outside the wedge are expected to not host an AGN, suggesting mid-IR emission that is dominated by star-formation activity. The top and right panels show the marginal histograms, illustrating the number distribution of galaxies along the $W2-W3$ and $W1-W2$ color axes, respectively.}
    \label{fig:WISE Diagram}
\end{figure*}

For this work, we constructed BPT diagrams using the emission-line ratios [O~\textsc{iii}]/H$\beta$, [O~\textsc{i}]/H$\alpha$, [N~\textsc{ii}]/H$\alpha$, and [S~\textsc{ii}]/H$\alpha$ (see Table \ref{tab:basic properties 38 LLAGN}), with line fluxes taken from the optical spectroscopic catalog of \citet{Ho1997}.  
We adopted the classification scheme of \citet{Kewley2006}, which refines earlier empirical boundaries from \citet{Veilleux1987} as used by \citet{Ho1997} by incorporating theoretical photoionization models and empirical demarcation curves from \citet{Kewley2001} and \citet{Kauffmann2003}.  
The \citet{Kewley2006} framework provides a more accurate separation of composite galaxies from Seyfert and LINER classes, and also introduces the ``Seyfert–LINER'' division line, which was not present in earlier diagrams.

Each galaxy was plotted on the three standard BPT diagnostic planes which are [O~\textsc{iii}]/H$\beta$ vs. [N~\textsc{ii}]/H$\alpha$, [O~\textsc{iii}]/H$\beta$ vs. [S~\textsc{ii}]/H$\alpha$, and [O~\textsc{iii}]/H$\beta$ vs. [O~\textsc{i}]/H$\alpha$. All three BPT diagrams were used because each probes different ionization conditions and help to distinguish AGN from star formation and composite (see Fig \ref{fig:BPT DIAGRAM}). Sources lacking sufficient emission-line measurements were flagged as ``no data'' cases. Among the 38 LLAGN in our sample, one galaxy (NGC~1055) lacked sufficient optical emission-line data, leaving 37 galaxies available for BPT classification.  In the [O~\textsc{iii}]/H$\beta$ vs. [N~\textsc{ii}]/H$\alpha$ diagram, 32 of the 37 ($86.5^{13.5}_{-23.5}$\%) galaxies were identified as AGN, while five ($13.5^{+0.1}_{-0.2}$\%) were categorized as composite objects.  No sources were located in the H~\textsc{ii} region of this diagram.  In the [O~\textsc{iii}]/H$\beta$ vs. [S~\textsc{ii}]/H$\alpha$ plane, four ($10.8^{+13.9}_{-7.1}$\%) galaxies were classified as Seyfert, 31 ($83.8^{+16.2}_{-23.1}$\%) as LINERs, and two ($5.4^{+11.6}_{-4.4}$\%) as H~\textsc{ii} galaxies.  Meanwhile, in the [O~\textsc{iii}]/H$\beta$ vs. [O~\textsc{i}]/H$\alpha$ diagram, 12 ($32.4^{+20.1}_{-13.7}$\%) galaxies were Seyfert and 25 ($67.6^{+26.8}_{-20.6}$\%) were LINER, with none found in the H~\textsc{ii} region. The detailed classification of each LLAGN in all three BPT diagrams is listed in Table \ref{tab:galaxy_classification}.

Because LLAGN can occupy ambiguous regions in individual diagnostic planes, we explored the impact of different classification thresholds by varying the number of BPT diagrams required for AGN identification. Adopting the 1/3 criterion (i.e. AGN/\ion{H}{ii}/LINER  or AGN/LINER/LINER) yields an optical AGN fraction of 32/38 ($84^{+16}_{-23}$\%), while requiring consistency in at least two diagrams (i.e. AGN/\ion{H}{ii}/Seyfert or AGN/Seyfert/LINER or AGN/LINER/Seyfert) reduces this to 12/38 ($32^{+19.6}_{-13.4}$\%), and a strict 3/3 (AGN/Seyfert/Seyfert) agreement further reduces it to 4/38 ($10.5^{+13.6}_{-6.9}$\%). The stricter criteria preferentially select higher-ionization Seyfert systems but exclude a large fraction of LINER-like nuclei. This is particularly relevant as many of our sources fall within the LINER or composite regions of the BPT diagrams. While LINERs may arise from alternative ionization mechanisms in some cases (e.g., \citealt{Lopez2024, Yuk2025}), previous studies have shown that they are frequently associated with low-luminosity AGN (e.g., \citealt{Ho1995, Young2012, Nemer2025}). In addition, our final LLAGN sample is selected based on compact, parsec-scale 15 GHz radio core detections, which are difficult to reproduce through stellar photoionization alone, further supporting an accretion-driven origin for the nuclear activity.

Therefore, we adopt a conservative approach by classifying a source as an AGN if it appears as either an AGN or a Seyfert in any of the diagnostic diagrams.
We adopt this 1/3-diagram criterion because LLAGN often exhibit weak emission lines and can occupy different regions across diagnostic planes due to low ionization and radiatively inefficient accretion (e.g. \citealt{Ho2008,Ho2009}). Requiring agreement across multiple diagrams would therefore bias against genuine low-luminosity nuclei. Furthermore, the resulting AGN fraction is broadly consistent, within uncertainties, with the Palomar based on \citet{Ho1997} if we consider both Seyfert and LINER nuclei in their sample as AGN-like systems (25/38, 66$^{+26}_{-20}$\%), indicating that our conclusions are not driven by the adopted criterion.

\begin{table*}
\centering
\caption{Optical nuclear classification based on the BPT diagram}
\label{tab:galaxy_classification}
\begin{tabular}{ccccc}
\hline
Name & log ([\ion{O}{iii}] / H$\beta$) vs & log ([\ion{O}{iii}] / H$\beta$) vs  & log ([\ion{O}{iii}] / H$\beta$) vs  & Final Classification  \\
 &  log ([\ion{N}{ii}] / H$\alpha$) & log ([\ion{S}{ii}] / H$\alpha$) &  log ([\ion{O}{i}] / H$\alpha$) &  \\
(1) & (2) & (3) & (4) & (5)\\
\hline
NGC488 & AGN & \ion{H}{ii} & LINER & AGN-LINER \\
NGC521 & AGN & \ion{H}{ii} & Seyfert & AGN-Seyfert \\
NGC660 & AGN & Seyfert & Seyfert & AGN-Seyfert \\
NGC777 & AGN & LINER & LINER & AGN-LINER \\
NGC841 & AGN & LINER & LINER & AGN-LINER \\
NGC1055 & - & - & - & - \\
NGC1169 & AGN & LINER & LINER & AGN-LINER \\
NGC1961 & AGN & LINER & LINER & AGN-LINER \\
NGC2681 & AGN & LINER & LINER & AGN-LINER \\
NGC2683 & AGN & LINER & LINER & AGN-LINER \\
NGC2859 & AGN & LINER & Seyfert & AGN-Seyfert \\
NGC2985 & Composite & LINER & LINER & LINER \\
NGC3642 & Composite & LINER & LINER & LINER \\
NGC3898 & AGN & LINER & Seyfert & AGN-Seyfert \\
NGC3992 & AGN & LINER & Seyfert & AGN-Seyfert \\
NGC4036 & AGN & LINER & LINER & AGN-LINER \\
NGC4111 & AGN & LINER & LINER & AGN-LINER \\
NGC4145 & Composite & LINER & LINER & LINER \\
NGC4346 & AGN & LINER & LINER & AGN-LINER \\
NGC4395 & AGN & Seyfert & Seyfert & AGN-Seyfert \\
NGC4750 & AGN & LINER & LINER & AGN-LINER \\
NGC5194 & AGN & Seyfert & Seyfert & AGN-Seyfert \\
NGC5195 & AGN & LINER & LINER & AGN-LINER \\
NGC5395 & AGN & LINER & Seyfert & AGN-Seyfert \\
NGC5448 & AGN & LINER & LINER & AGN-LINER \\
NGC5485 & AGN & LINER & LINER & AGN-LINER \\
NGC5566 & AGN & LINER & LINER & AGN-LINER \\
NGC5631 & AGN & LINER & LINER & AGN-LINER \\
NGC5746 & AGN & LINER & Seyfert & AGN-Seyfert \\
NGC5850 & AGN & LINER & LINER & AGN-LINER \\
NGC5921 & Composite & LINER & LINER & LINER \\
NGC5985 & AGN & LINER & LINER & AGN-LINER \\
NGC6340 & AGN & LINER & LINER & AGN-LINER \\
NGC6703 & AGN & LINER & LINER & AGN-LINER \\
NGC6951 & AGN & Seyfert & Seyfert & AGN-Seyfert \\
NGC7814 & Composite & LINER & LINER & LINER \\
IC356 & AGN & LINER & Seyfert & AGN-Seyfert \\
IC520 & AGN & LINER & Seyfert & AGN-Seyfert \\
\hline
\end{tabular}
\begin{itemize} 
\item[] \textit{Notes}. (1) Galaxy name; (2) classification based on log ([\ion{O}{iii}] / H$\beta$) vs log ([\ion{N}{ii}] / H$\alpha$) diagram - AGN, Composite, \ion{H}{ii}; (3) classification based on log ([\ion{O}{iii}] / H$\beta$) vs log ([\ion{S}{ii}] / H$\alpha$) diagram - Seyfert, LINER, \ion{H}{ii} ; (4) classification based on log ([\ion{O}{iii}] / H$\beta$) vs log ([\ion{O}{i}] / H$\alpha$) diagram - Seyfert, LINER, \ion{H}{ii} ; and (5) final classification of the source in optical based on the three BPT diagrams. We considered them as AGN if they are classified as AGN or Seyfert in any of the BPT diagram.
\end{itemize}
\end{table*}

By combining the classifications from all three diagnostic planes, we conclude that the overall optical breakdown of our LLAGN sample comprises 32 AGN ($84.2^{+15.8}_{-22.9}$\%), five LINERs ($13.2^{+14.5}_{-8.0}$\%), with one galaxy lacking sufficient data ($2.6^{+9.9}_{-2.5}$\%) for classification. Refer to Table \ref{tab:galaxy_classification} for final classification determined based on the three BPT diagrams.No galaxy falls within the pure H~\textsc{ii}-region domain, indicating that all systems exhibit at least some level of nuclear activity. Although our LLAGN sample is classified as AGN based on the three BPT diagrams, many sources show LINER-like line ratios in at least one diagram. This differs from typical AGN samples, such as the \textit{Swift}--BAT survey, where the majority of sources are consistently classified as Seyfert across diagnostic diagrams. The predominance of LINER-like features in LLAGN reflects their lower ionizing luminosity compared to the more luminous Seyfert-dominated \textit{Swift}--BAT AGN.

\subsection{Infrared}

IR diagnostics provide an important complementary view of AGN activity, particularly in systems where optical and X-ray signatures may be obscured by dust.  
Two independent IR-based approaches were used in this study: a photometric color–color selection using data from the \textit{Wide-field Infrared Survey Explorer} (WISE) and an emission-line search for [Ne~\textsc{v}]$\uplambda14.32\,\mu\mathrm{m}$ features using \textit{Spitzer} spectroscopy.  
These methods probe different physical mechanisms, including thermal dust emission in the mid-IR and high-ionization lines produced by the AGN radiation field, which allows a more complete assessment of nuclear activity in low-luminosity systems.

\subsubsection{WISE Diagram}

The first method employs mid-IR photometry from the WISE telescope, which provides all-sky coverage in four bands: W1 (3.4~$\mu$m), W2 (4.6~$\mu$m), W3 (12.0~$\mu$m), and W4 (22.0~$\mu$m).  
Photometric data for our 38 LLAGN were retrieved from the NASA/IPAC Infrared Science Archive (IRSA)\footnote{WISE data can be accessed through \url{https://irsa.ipac.caltech.edu/cgi-bin/Gator/nph-dd}.}.
AGN candidates were identified using the color–color selection method developed by \citet{Mateos2012}, which defines an ``AGN wedge'' region in the W1–W2 versus W2–W3 plane where mid-IR emission is dominated by hot dust heated by the AGN torus.  

As shown in Fig. \ref{fig:WISE Diagram}, only one galaxy (NGC~660, $2.6^{+9.9}_{-2.5}$\%) lies within the AGN wedge, while the remaining 37 galaxies fall outside the AGN region.  
One additional source (NGC~4395) lies close to the AGN boundary, suggesting weak but non-negligible hot-dust emission.  
The predominance of sources outside the wedge indicates that most LLAGN in our sample lack the strong mid-IR colors typical of luminous AGN, likely because their mid-IR output is dominated by stellar or star-forming emission rather than AGN-heated dust.
This result is expected as \textit{WISE} color selection is primarily calibrated for more luminous AGN with strong hot dust emission, and is therefore not well suited for detecting low-luminosity systems \citep{Hviding2022}. Although recent SED modelling can improve the separation of weak AGN from host galaxy emission in LLAGN, the AGN infrared contribution can remain very small (e.g. $\sim$0.1--30\%; \citealt{Lopez2024}). Since the host galaxy often dominates the mid-infrared emission, even advanced SED techniques may only modestly improve the identification of faint nuclei.

\subsubsection{[Ne~\textsc{v}]$\uplambda14.32\,\mu\mathrm{m}$ Emission}

\begin{figure*}
    \centering
    \includegraphics[width=1 \linewidth]{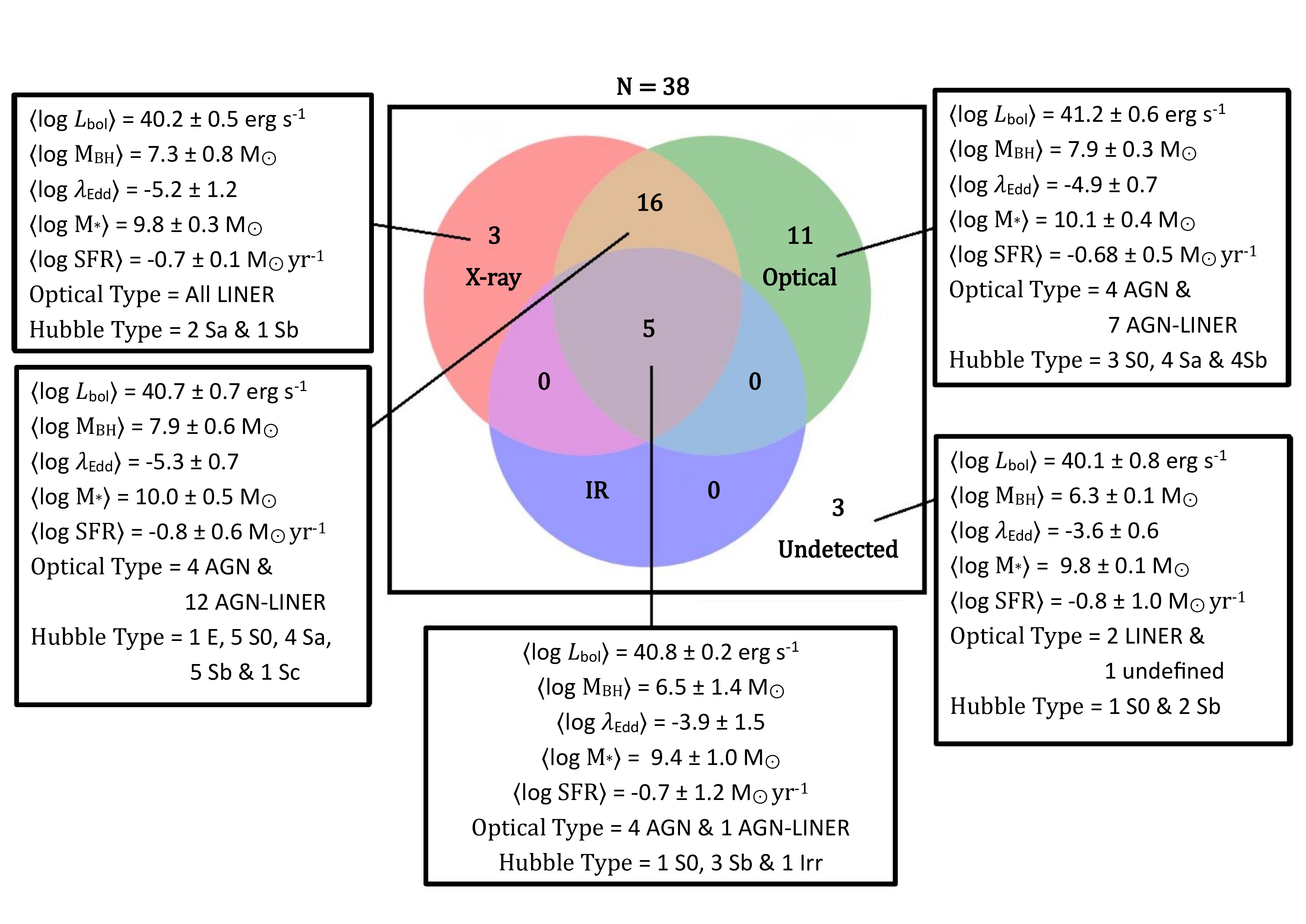}
\caption{Venn diagram showing the distribution of our LLAGN detected across X-ray, optical and IR wavelengths. The numbers indicate the amount of source detected in individual wavebands and their overlaps, with five sources detected in all bands and three sources remain undetected at any of those wavelengths. Each subset is annotated with its average AGN and host galaxy properties (e.g., bolometric luminosity, Eddington ratio, stellar mass, and star formation rate), enabling a direct comparison of their physical characteristics across different detection groups.}
    \label{fig:Venn Diagram}
\end{figure*}

The second method relies on the detection of the [Ne~\textsc{v}]$\uplambda14.32\,\mu\mathrm{m}$ emission line. The fine-structure line at 14.32~$\mu$m originates mainly in the NLR of AGN where [Ne~\textsc{v}] ions are created by ionization from the primary radiation of the central nuclear source. As the NLR is located outside the obscuring torus, its emission provides a relatively direct probe of nuclear activity. 
The [Ne~\textsc{v}] line is a robust tracer of $\text{AGN}$ photoionization because the ionization potential of $\text{Ne}^{4+}$ ($97.1 \text{ eV}$) exceeds that attainable in H~\textsc{ii} regions, making its presence a clear indicator of $\text{AGN}$ activity \citep{Weedman2005, Goulding2009, Bernard-Salas2009, Iwasawa2011, Izotov2012,Negus2023}.
While NLR emission lines can also be detected at optical wavelengths, they can suffer obscuration by dust from host galaxy and interstellar medium. In contrast, mid-infrared NLR lines suffer significantly less extinction, allowing AGN signatures to be detected even in dusty or highly inclined systems.

Among infrared diagnostics, lines such as [O~\textsc{iv}]~$\lambda25.89 \mu$m are commonly used as AGN tracers. However, [O~\textsc{iv}] can receive non-negligible contributions from intense star formation and shocks, particularly in low-luminosity systems. The [Ne~\textsc{v}] line, owing to its much higher ionization potential is effectively free from contamination by stellar processes and therefore provides a more secure and conservative indicator of AGN photoionization. Therefore, this study will identify AGN  based on the presence of the [Ne~\textsc{v}]$\uplambda14.32\,\mu\mathrm{m}$ emission line.

In order to identify AGN, we compared 38 LLAGN with the IR spectroscopic catalogs of \citet{Spoon2022} and \citet{Goulding2009}, where both were based on \textit{Spitzer} observations. IR emission-line data were available for 10 galaxies from \citet{Spoon2022} of which two galaxies from it show [Ne~\textsc{v}] detections. While, in \citet{Goulding2009}, five galaxies had  available data and three of these galaxies were detected. Overall, five galaxies out of a total sample of 38 galaxies, equivalent to $13.2^{+14.5}_{-8.0}$\% (5/38) detected in at least one of the [Ne~\textsc{v}]$\uplambda14.32\,\mu\mathrm{m}$ emission lines catalogs. However, if the detection rate is calculated based only on the number of galaxies with data, the detection rate would be $33.3^{+36.8}_{-20.2}$\% (5/15).
The relatively low [Ne V] detection fraction does not necessarily imply the absence of AGN activity in the remaining galaxies. In LLAGN, the [Ne~\textsc{v}] emission can be intrinsically weak due to low accretion rates and reduced ionizing photon output, and may fall below the sensitivity limits of the available Spitzer spectroscopy. As a result, a significant fraction of AGN identified through X-ray and optical diagnostics may remain undetected in [Ne~\textsc{v}] which shows the limitations of this method when applied to low-luminosity systems.
We also investigated the [O~\textsc{iv}] line as AGN diagnostic. However, all galaxies with detectable [O~\textsc{iv}] also exhibit [Ne~\textsc{v}] emission, meaning that [O~\textsc{iv}] reveals no additional AGN.

\begin{table*}
    \centering
\caption{AGN detection at each wavelength.}
    \label{tab:Multiwavelength AGN}
    \begin{tabular}{cccccc}
        \hline
        Galaxy & Soft X-ray  & Hard X-ray & WISE diagram & [\ion{Ne}{V}] emission & BPT diagram\\
        & (2-10 keV) & (14-195 keV) & & & \\
        (1) & (2) & (3) & (4) & (5) & (6)\\
        \hline
        NGC488   & -  & \no  & \no  & -  & \yes \\
        NGC521   & -  & \no  & \no  & -  & \yes \\
        NGC660   & \yes  & \no  & \yes  & \yes  & \yes \\
        NGC777   & \yes  & \no  & \no  & \no  & \yes \\
        NGC841   & -  & \no  & \no  & -  & \yes \\
        NGC1055  & \no  & \no  & \no  & \no  & - \\
        NGC1169  & -  & \no  & \no  & -  & \yes \\
        NGC1961  & \yes  & \no  & \no  & \no  & \yes \\
        NGC2681  & \yes  & \no & \no  & \no  & \yes \\
        NGC2683  & \yes  & \no  & \no  & -  & \yes \\
        NGC2859  & -  & \no  & \no  & -  & \yes \\
        NGC2985  & \yes  & \no  & \no  & -  & \no \\
        NGC3642  & \yes  & \no  & \no  & \no  & \no \\
        NGC3898  & \yes  & \no  & \no  & \no  & \yes \\
        NGC3992  & \yes  & \no  & \no  & -  & \yes \\
        NGC4036  & \yes  & \no  & \no  & \no  & \yes \\
        NGC4111  & \yes  & \no  & \no  & -  & \yes \\
        NGC4145  & -  & \no  & \no  & -  & \no \\
        NGC4346  & \yes  & \no  & \no  & -  & \yes \\
        NGC4395  & \yes  & \yes  & \no  & \yes  & \yes \\
        NGC4750  & \yes  & \no  & \no  & -  & \yes \\
        NGC5194  & \yes  & \yes  & \no  & \yes  & \yes \\
        NGC5195  & \yes  & \no  & \no  & \yes  & \yes \\
        NGC5395  & \yes  & \no  & \no  & -  & \yes \\
        NGC5448  & \yes  & \no  & \no  & -  & \yes \\
        NGC5485  & \no  & \no  & \no  & -  & \yes \\
        NGC5566  & -  & \no  & \no  & -  & \yes \\
        NGC5631  & \yes  & \no & \no  & \no  & \yes \\
        NGC5746  & \yes  & \no  & \no  & -  & \yes \\
        NGC5850  & -  & \no  & \no  & -  & \yes \\
        NGC5921  & -  & \no  & \no  & -  & \no \\
        NGC5985  & \yes  & \no  & \no  & -  & \yes \\
        NGC6340  & -  & \no  & \no  & -  & \yes \\
        NGC6703  & \yes  & \no  & \no  & \no  & \yes \\
        NGC6951  & \yes  & \no  & \no  & \yes  & \yes \\
        NGC7814  & \yes  & \no  & \no  & -  & \no \\
        IC356    & -  & \no  & \no  & -  & \yes \\
        IC520    & -  & \no  & \no  & -  & \yes \\

        \hline
    \end{tabular}
    \begin{itemize} 
\item[] \textit{Notes}. (1) Galaxy name; (2) Soft X-ray AGN detection using \textit{Chandra} or \textit{XMM-Newton}; (3) Hard X-ray AGN detection using \textit{Swift}--BAT; (4) IR AGN detection using WISE colour-colour diagram ; (5) IR AGN detection using the high ionisation [\ion{Ne}{V}] emission line at 14.32 $\umu$m; and (6) Optical AGN detection using the BPT diagram. The symbols in the table have the following meanings: "\yes" = detected; "\no" = undetected; and "-" = insufficient data. 
\end{itemize}
\end{table*}

\subsection{Detection Effectiveness Across Wavelengths}

In this section, we evaluated the performance of each wavelength by referring directly to Table \ref{tab:Multiwavelength AGN}, which lists the detection outcome for every galaxy in the sample, and Figure \ref{fig:Venn Diagram}, which visualizes the overlap of AGN detections in a Venn diagram.

Based on the X-ray detection, a total of 38 galaxies in the sample have available X-ray observations from  \textit{Chandra}, \textit{XMM-Newton}, or \textit{Swift}--BAT, as summarized in Table \ref{tab:Multiwavelength AGN}. Among these, 24 demonstrate clear nuclear X-ray detections, indicating a detection fraction of $63.2^{+25.7}_{-19.6}$\% for the observed sample which is comparable to normal AGN detection rates from past paper (20\%-60\%, \citealt{Desroches2009, Zhang2009, Miller2015, She2017}). When considering only the soft X-ray data availablity, this will correspond to a high detection rate fraction of $92.3^{+7.7}_{-28.7}$\% (24/26). 
Although the total number of X-ray detections is not the highest among the three wavelengths, the X-ray band remains a highly reliable indicator of genuine AGN activity due to its lower contamination, even though it lacks of soft X-ray data and hard X-ray sensitivity. This reliability is clearly reflected in Figure \ref{fig:Venn Diagram}, where X-ray detections overlap strongly with optical (16 galaxies, $42.1^{+21.8}_{-15.7}$\%) and form part of the central region of galaxies detected in all three wavelengths (5 galaxies, $13.2^{+14.5}_{-8.0}$\%). These overlaps indicate that X-ray detections are tightly associated with other AGN indicators, reinforcing that X-rays probe the accretion flow most directly. Only two X-ray non-detections are reported in Table \ref{tab:Multiwavelength AGN} (NGC 5921 and NGC 5850). These may represent either undetected or heavily obscured systems potentially hosting Compton-thick AGN. These cases show the limitations of current X-ray sensitivity when dealing with extremely weak or deeply buried LLAGN.

Optical emission-line diagnostics provide the largest number of AGN identifications in our sample.  
Table \ref{tab:Multiwavelength AGN} shows that 32 out of 38 galaxies display emission-line ratios consistent with AGN-like, yielding an overall optical detection fraction of $84.2^{+15.8}_{-22.9}$\%, the highest among the three wavebands. This detection fraction is significantly higher than that reported in previous optical surveys of normal AGN, where typical optical identification rates are only $\approx$ 10\%–42\% \citep{Ho1997, Zaw2019, Mickaelian2021, Mickaelian2024, Mezcua2024, Mazzolari2025}.  
However, if we restrict the sample to sources optically classified as AGN–Seyfert (Table~\ref{tab:galaxy_classification}) and consistently identified as Seyferts across all diagnostic diagrams (AGN/Seyfert/Seyfert), the resulting fraction is $10.5^{+13.6}_{-6.9}\%$ (4/38). Furthermore, if we include the sources that fall into the LINER region in at least one diagram (i.e. AGN/LINER/Seyfert or AGN/Seyfert/LINER) which are commonly considered part of the LLAGN population (e.g., \citealt{Ho1995, Young2012, Nemer2025}), the fraction will increase to $28.9^{+19.0}_{-12.7}\%$ (11/38). Both fractions are consistent with the results reported in previous studies, indicating that our results are not significantly biased by the initial optical selection of the parent sample.
In the paper, we adopt the overall optical detection fraction of $84.2^{+15.8}_{-22.9}\%$ (32/38) as the primary reference for AGN identification in this work since the optical signature might not be very dominant in the BPT diagrams.
The dominance by optical identification is clearly illustrated in Figure \ref{fig:Venn Diagram}, where the optical-only region contains the largest number of uniquely detected sources (11 galaxies, $28.9^{+19.0}_{-12.7}$\%). These galaxies show AGN-like line ratios but no corroborating X-ray or IR signatures. This shows both the strengths and limitations of optical diagnostics, i.e., they excel at detecting low-level ionization around galactic nuclei, but may be influenced by old stellar populations or shock excitation. Thus, optical methods are highly sensitive but not always uniquely diagnostic of accretion. Hence, while optical detections dominate numerically, they must be interpreted with caution, especially in low-luminosity systems.

While for IR diagnostics, including WISE color selection and the high-ionization [Ne V] 14.32 µm line, are the least effective in identifying AGN signatures in our LLAGN sample. Only one galaxy (NGC 660) meets the WISE color AGN criteria, and five galaxies show detectable [Ne V] emission. Combined, this results in only five ($13.2^{+14.5}_{-8.0}$\%) IR-detected AGN, consistent with Figure \ref{fig:Venn Diagram}, which shows no IR-only detections. The absence of unique IR detections indicates that IR-based AGN identification in LLAGN is severely hindered by host galaxy contamination and the intrinsically weak dust emission expected from RIAF-dominated accretion flows. 
In such faint nuclei, the mid-IR emission from circumnuclear star formation often dominates over any AGN contribution. As a result, the IR band tends to confirm rather than independently identify LLAGN in the local Universe.

A particularly important subset of the sample is the group of five galaxies ($13.2^{+14.5}_{-8.0}$\%) detected in all three wavelengths (NGC 660, NGC 4395, NGC 5194, NGC 5195, and NGC 6951). These galaxies occupy the central overlap region in Figure \ref{fig:Venn Diagram}, representing the most secure LLAGN in the entire sample. The presence of X-ray, optical, and IR AGN signatures in all five cases strongly supports the existence of active accretion despite their low luminosities. These multiwavelength detections demonstrate that, when the AGN is sufficiently luminous or unobscured, the signatures of accretion appear consistently across X-ray, optical, and infrared regimes. As such, this subset provides a benchmark for understanding the full range of LLAGN behavior and serves as a reference against which weaker or ambiguous detections can be compared.

At the opposite extreme, three galaxies ($7.9^{+12.5}_{-5.7}$\%) in the sample which are NGC 1055, NGC 4145, and NGC 5921 show no AGN signatures in any of the three wavelengths. Neither X-ray, optical, nor IR diagnostics identify any indication of nuclear activity. 
Nevertheless, the 15\,GHz VLA detections reported by \citet{Saikia2018} are statistically significant, with signal-to-noise ratios well above the $\gtrsim 5\sigma$ threshold typically adopted for reliable radio detections. Therefore, we infer that these three sources are likely to host AGN, despite the lack of detection at any wavelengths.
Their non-detections, indicated by the empty set in Figure \ref{fig:Venn Diagram}, suggest two possibilities. 
Firstly, any existing AGN may be far below current detection limits, producing emission so weak that it remains undetectable across all wavelengths. Secondly, host galaxy contamination may overwhelm faint AGN signatures, especially in optical and IR regimes. 
Their properties provide important context for the lower boundary of accretion activity in nearby galaxies.

\begin{figure*}
\centering
\includegraphics[width=1 \linewidth]{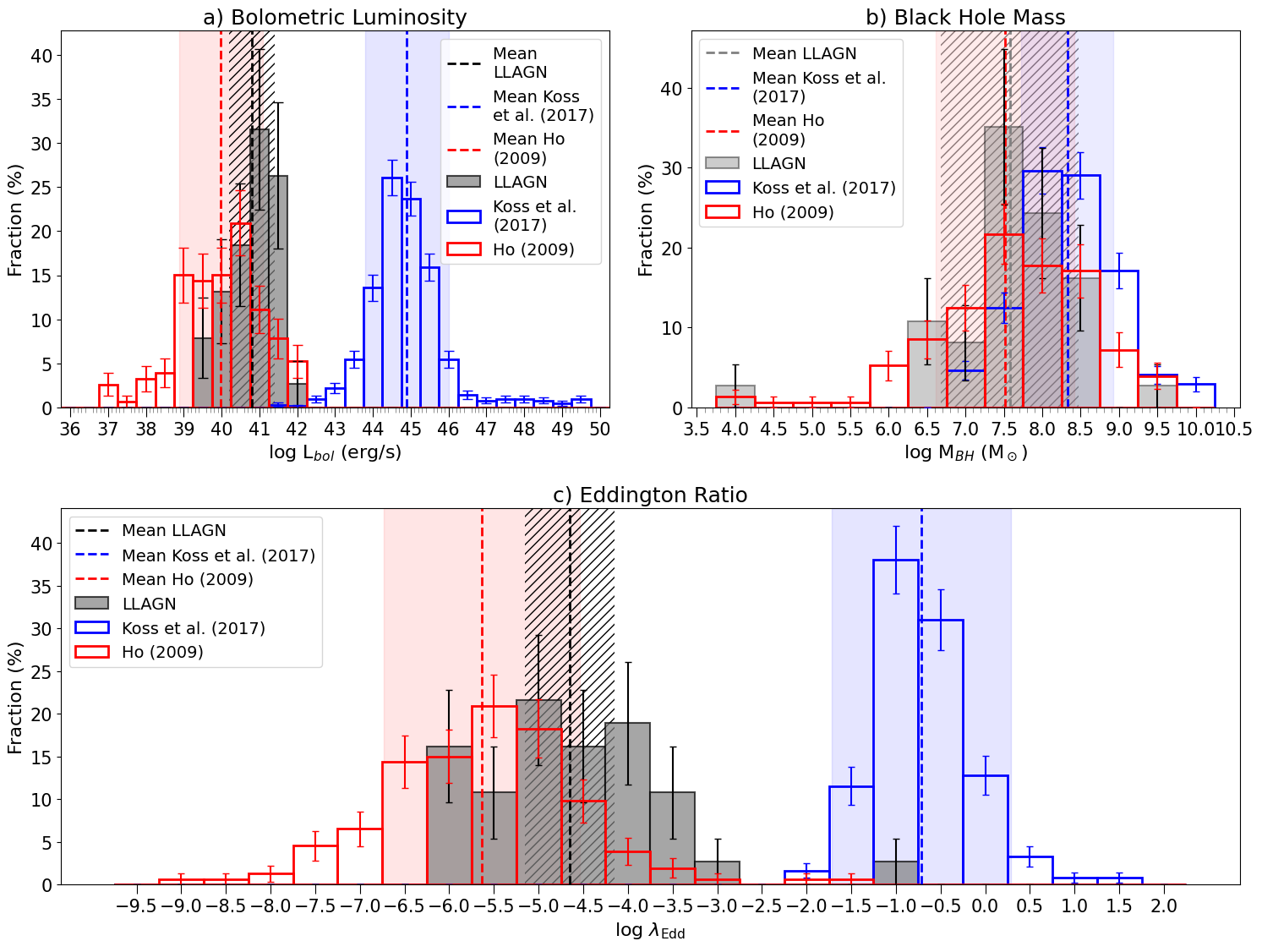}
\caption{Distributions of AGN properties for our LLAGN sample (grey) compared with the \textit{Swift}--BAT AGN (blue; \citealt{Koss2017}) and an additional LLAGN sample from \citealt{Ho2009} (red). Panels show bolometric luminosity (a), black hole mass (b), and Eddington ratio (c).}
\label{fig:AGN_properties}
\end{figure*}

\begin{table*}
\centering
\caption{Physical properties of the LLAGN and their host galaxies}.
\label{tab:Physical And host LLAGN properties}
\begin{tabular}{l c c c c c c c}
\hline
\textbf{Name} & \textbf{log $L_{\mathrm{bol}}$} & \textbf{log $M_{\mathrm{BH}}$} & log \textbf{$\lambda_{\mathrm{Edd}}$} & \textbf{log $M_\ast$} & \textbf{log SFR} & \textbf{Optical Type} & \textbf{Hubble Type} \\
& (erg s$^{-1}$) & ($M_\odot$) & & ($M_\odot$) & ($M_\odot yr^{-1}$) & & \\
(1) & (2) & (3) & (4) & (5) & (6) & (7) & (8) \\
\hline
NGC 488  &  41.02$^{+0.18}_{-0.30}$$^{a}$  &  $8.17^{+0.19}_{-0.10}$  & $-5.25^{+0.08}_{-0.20}$  &  -  &  $-0.74 \pm {0.09}$  &  AGN-LINER  &  Sb \\
NGC 521  &  41.66$^{+0.30}_{-0.30}$$^{a}$  &  $8.42^{+0.18}_{-0.09}$  &  $-4.86^{+0.21}_{-3.91}$  &  $10.54 \pm 0.10$  &  $-0.20 \pm {0.09}$  &  AGN  &  Sb \\
NGC 660  &   $40.87^{+0.01}_{-0.01}$$^{b}$  &  $7.30^{+0.21}_{-0.11}$  &   $-4.53^{-0.09}_{-0.09}$  &  $9.62 \pm 0.10$  &  $0.43 \pm {0.11}$  &  AGN  &  Sa \\
NGC 777  &   $41.45^{+0.10}_{-0.12}$$^{b}$  &  $9.36^{+0.15}_{-0.07}$  &  $-6.02^{+0.03}_{-0.05}$  &  $10.79 \pm 0.10$  &  $-0.18 \pm {0.09}$  &  AGN-LINER  &  E \\
NGC 841  &   41.63$^{+0.30}_{-0.30}$$^{a}$  &  $7.78^{+0.23}_{-0.12}$  &   $-4.25^{+0.19}_{-3.88}$  &  -  &  $0.09 \pm {0.10}$  &  AGN-LINER  &  Sa \\
NGC 1055  &   $40.30^{+0.08}_{-0.11}$$^{b}$  &  $6.25^{+0.84}_{-0.46}$  &   $-4.05^{-0.30}_{-0.35}$  &  $9.83 \pm 0.10$  &  $-0.03 \pm{0.10} $  &  -  &  Sb \\
NGC 1169  &   41.40$^{+0.11}_{-0.16}$$^{a}$  &  $8.07^{+0.17}_{-0.09}$  &   $-4.77^{+0.03}_{-0.07}$  &  -  &  $-0.94 \pm {0.08}$  &  AGN-LINER  &  Sb \\
NGC 1961  &   $41.20^{+0.03}_{-0.03}$$^{b}$  &  $8.71^{+0.17}_{-0.09}$  &   $-5.61^{-0.05}_{-0.05}$  &  $10.79 \pm 0.10$  &  $0.49 \pm {+0.10}$  &  AGN-LINER  &  Sc \\
NGC 2681  &   $40.69^{+0.01}_{-0.01}$$^{b}$  &  $6.94^{+0.27}_{-0.14}$  &   $-4.35^{-0.12}_{-0.13}$  &  $9.39 \pm 0.10$  &  $-0.69^{+0.09}_{-0.10}$  &  AGN-LINER  &  Sa \\
NGC 2683  &   $40.06^{+0.02}_{-0.02}$$^{b}$  &  $7.34^{+0.24}_{-0.12}$  &   $-5.37^{-0.10}_{-0.10}$  &  $9.41 \pm 0.10$  &  $-1.46 \pm{0.09}$  &  AGN-LINER  &  Sb \\
NGC 2859  &   41.34$^{+0.11}_{-0.16}$$^{a}$  &  $8.15^{+0.19}_{-0.10}$  &   $-4.92^{+0.02}_{-0.06}$  &  -  &  $-0.93 \pm {0.09} $  &  AGN  &  S0 \\
NGC 2985  &   $40.40^{+0.07}_{-0.08}$$^{b}$  &  $7.51^{+0.15}_{-0.08}$  &   $-5.21^{-0.00}_{-0.01}$  &  $9.99 \pm 0.10$  &  $-0.60 \pm {0.10}$  &  LINER  &  Sa \\
NGC 3642  &   $40.53^{+0.08}_{-0.10}$$^{b}$  &  $6.39^{+0.40}_{-0.21}$  &   $-3.96^{-0.11}_{-0.11}$  &  $9.51 \pm 0.10$  &  $-0.74 \pm {0.09}$  &  LINER  &  Sb \\
NGC 3898  &   $39.95^{+0.05}_{-0.06}$$^{b}$  &  $8.36^{+0.16}_{-0.08}$  &   $-6.51^{-0.03}_{-0.02}$  &  -  &  $-1.19 \pm {0.09}$  &  AGN  &  Sa \\
NGC 3992  &   $39.30^{+0.20}_{-0.37}$$^{b}$  &  $7.63^{+0.21}_{-0.11}$  &   $-6.43^{+0.09}_{-0.27}$  &  $9.99 \pm 0.10$  &  $-1.54 \pm {0.08}$  &  AGN  &  Sb \\
NGC 4036  &   $40.87^{+0.04}_{-0.05}$$^{b}$  &  $8.45^{+0.17}_{-0.09}$  &  $-5.69^{-0.04}_{-0.04}$  &  $9.98 \pm 0.10$  &  $-0.86 \pm {0.09}$  &  AGN-LINER  &  S0 \\
NGC 4111  &   $41.28^{+0.02}_{-0.02}$$^{b}$  &  $7.62^{+0.12}_{-0.06}$  &  $-4.44^{-0.04}_{-0.04}$  &  $9.48 \pm 0.10$  &  $-1.07 \pm {0.09}$  &  AGN-LINER  &  S0 \\
NGC 4145  &   39.78$^{+0.11}_{-0.16}$$^{a}$  &  -  &  -  &  -  &  $-1.88\pm{0.07}$  &  LINER  &  S0 \\
NGC 4346  &   $39.62^{+0.12}_{-0.17}$$^{b}$  &  $7.60^{+0.22}_{-0.11}$  &   $-6.08^{+0.01}_{-0.05}$  &  -  &  $-1.66 \pm {0.08}$  &  AGN-LINER  &  S0 \\
NGC 4395  &   $40.83^{+0.01}_{-0.01}$$^{b}$  & $4.07^{+0.34}_{-0.34}$  &   $-1.34^{+0.34}_{-0.34}$  &  $7.66 \pm 0.10$  &  $-2.54 \pm {0.07}$  &  AGN  &  S0 \\
NGC 4750  &   $41.05^{+0.07}_{-0.08}$$^{b}$  &  $7.43^{+0.23}_{-0.12}$  &   $-4.49^{-0.05}_{-0.04}$  &  -  &  $-0.57 \pm {0.09}$  &  AGN-LINER  &  Sa \\
NGC 5194  &   $40.85^{+0.01}_{-0.01}$$^{b}$  &  $6.66^{+0.40}_{-0.21}$  &   $-3.91^{-0.19}_{-0.21}$  &  $9.84 \pm 0.10$  &  $-1.21 \pm {0.09}$  &  AGN  &  Sb \\
NGC 5195  &   $40.46^{+0.44}_{-0.46}$$^{b}$  &  $7.24^{+0.29}_{-0.15}$  &   $-4.88^{-0.58}_{-0.31}$  &  $9.52 \pm 0.10$  &  $-0.54 \pm {0.10}$  &  AGN-LINER  &  Irr \\
NGC 5395  &  $41.13^{+0.07}_{-0.09}$$^{b}$  &  $7.58^{+0.25}_{-0.13}$  &   $-4.55^{-0.05}_{-0.04}$  &  $10.35 \pm 0.10$  &  $-0.47 \pm {0.09}$  &  AGN  &  Sb \\
NGC 5448  &   $40.34^{+0.12}_{-0.18}$$^{b}$  &  $7.24^{+0.29}_{-0.15}$  &   $-4.99^{-0.02}_{-0.03}$  &  -  &  $-0.41 \pm {0.10}$  &  AGN-LINER  &  Sa \\
NGC 5485  &   $40.38^{+0.11}_{-0.16}$$^{b}$  &  $8.37^{+0.18}_{-0.09}$  &   $-6.09^{+0.03}_{-0.06}$  &  $10.00 \pm 0.10$  &  $-0.95 \pm {0.09}$  &  AGN-LINER  &  S0 \\
NGC 5566  &   41.00$^{+0.11}_{-0.16}$$^{a}$  &  $7.79^{+0.21}_{-0.11}$  &   $-4.89^{+0.01}_{-0.05}$  &  $10.11 \pm 0.10$  &  $-0.64^{+0.09}_{-0.10}$  &  AGN-LINER  &  Sa \\
NGC 5631  &  $40.20^{+0.14}_{-0.21}$$^{b}$  &  $7.90^{+0.22}_{-0.11}$  &   $-5.80^{+0.03}_{-0.09}$  &  $9.91 \pm 0.10$  &  $-0.92 \pm{0.09}$  &  AGN-LINER  &  S0 \\
NGC 5746  &   $41.26^{+0.02}_{-0.02}$$^{b}$  &  $8.29^{+0.18}_{-0.09}$  &   $-5.13^{-0.07}_{-0.07}$  &  -  &  $-0.52^{+0.09}_{-0.10}$  &  AGN  &  Sb \\
NGC 5850  &   41.43$^{+0.11}_{-0.16}$$^{a}$  &  $7.50^{+0.22}_{-0.11}$  &   $-4.17^{+0.01}_{-0.04}$  &  $10.20 \pm 0.10$  &  $-0.55 \pm {0.09} $  &  AGN-LINER  &  Sb \\
NGC 5921  &  41.30$^{+0.18}_{-0.30}$$^{a}$  &  $6.36^{+0.49}_{-0.26}$  &   $-3.16^{-0.06}_{-0.04}$  &  -  &  $-0.49 \pm {0.10}$  &  LINER  &  Sb \\
NGC 5985  &  $41.36^{+0.11}_{-0.00}$$^{b}$  &  $7.76^{+0.21}_{-0.11}$  &   $-4.50^{+0.01}_{-0.11}$  &  $10.27 \pm 0.10$  &  $-1.26 \pm {0.07}$  &  AGN-LINER  &  Sb \\
NGC 6340  &   40.91$^{+0.11}_{-0.16}$$^{a}$  &  $7.56^{+0.17}_{-0.09}$  &   $-4.75^{+0.03}_{-0.07}$  &  $9.51 \pm 0.10$  &  $-1.41 \pm {0.08}$  &  AGN-LINER  &  S0 \\
NGC 6703  &  $41.14^{+0.02}_{-0.03}$$^{b}$  &  $8.05^{+0.09}_{-0.05}$  &   $-5.02^{-0.02}_{-0.02}$  &  $10.13 \pm 0.10$  &  $-0.82 \pm {0.09}$  &  AGN-LINER  &  S0 \\
NGC 6951  &   $40.76^{+0.02}_{-0.03}$$^{b}$  &  $7.29^{+0.32}_{-0.16}$  &   $-4.63^{-0.13}_{-0.14}$  &  $10.14 \pm 0.10$  &  $0.22 \pm {0.10}$  &  AGN  &  Sb \\
NGC 7814  &   $39.63^{+0.07}_{-0.08}$$^{b}$  &  $7.96^{+0.20}_{-0.10}$  &   $-6.43^{-0.03}_{-0.02}$  &  -  &  $-0.75 \pm {0.09}$  &  LINER  &  Sa \\
IC 356  &   40.08$^{+0.11}_{-0.16}$$^{a}$  &  $7.75^{+0.23}_{-0.12}$  &   $-5.77^{+0.00}_{-0.04}$  &  -  &  $-1.19 \pm {0.09}$  &  AGN  &  Sa \\
IC 520  &   41.90$^{+0.18}_{-0.30}$$^{a}$  &  $7.47^{+0.23}_{-0.12}$  &   $-3.67^{+0.06}_{-0.18}$  &  -  &  $-0.02 \pm {0.10}$  &  AGN  &  Sa \\
\hline
\end{tabular}
\\[3pt]
\begin{itemize}
\item[] \textit{Notes}. (1) Galaxy name; (2) Logarithm of bolometric luminosity in erg s$^{-1}$; (3) Logarithm of the black hole mass relative to the mass of the Sun ($M_\odot$) measured from velocity dispersion; (4) Logarithm of Eddington ratio expressed as $L_{\mathrm{bol}}/L_{\mathrm{Edd}}$; (5) Logarithm of the stellar mass relative to the mass of the Sun ($M_\odot$); (6) Logarithm of the star formation rate in $M_\odot yr^{-1}$; (7) final optical spectroscopic classification based on the BPT diagram; (8) Hubble classification of the host galaxy.
\item[] $^{a}$ Optical-derived bolometric luminosity
\item[] $^{b}$ X-ray-derived bolometric luminosity
\end{itemize}
\end{table*}

\section{AGN Properties}
\label{sec: AGN properties}

In this section, we examine the physical properties as listed in Table \ref{tab:Physical And host LLAGN properties} for the final LLAGN sample.
We focus on three key parameters that describe the nuclear energy output and accretion state which are  
the bolometric luminosity ($L_{\mathrm{bol}}$), the black hole mass ($M_{\mathrm{BH}}$), and the Eddington ratio ($\lambda_{\mathrm{Edd}} = L_{\mathrm{bol}} / L_{\mathrm{Edd}}$).  
These quantities are compared with those of the \textit{Swift}--BAT AGN sample, which represents a population of nearby luminous AGN. This will allow us to see how our LLAGN differ from a more luminous AGN population.

\subsection{Bolometric Luminosity}
\label{sec:Lbol}

\begin{table*}
\centering
\caption{Comparison of Bolometric Luminosities Derived from Different Methods}
\label{tab:lbol_comparison}
\begin{tabular}{l c c c}
\hline 
Galaxy & 
$\log L_{\mathrm{bol}}$ (X-ray; \citealt{Lopez2024}) & 
$\log L_{\mathrm{bol}}$ (X-ray; \citealt{Vasudevan2010}) & 
$\log L_{\mathrm{bol}}$ (Optical; \citealt{Heckman2014}) \\
 & (erg s$^{-1}$) & (erg s$^{-1}$) & (erg s$^{-1}$) \\
\hline
NGC488  & – & – & $41.02^{+0.18}_{-0.30}$ \\
NGC521  & – & – & $41.66^{+0.30}_{-0.30}$ \\
NGC660  & $40.87^{+0.01}_{-0.01}$ & $41.19^{+0.01}_{-0.01}$ & $40.98^{+0.11}_{-0.16}$ \\
NGC777  & $41.45^{+0.10}_{-0.12}$ & $41.77^{+0.10}_{-0.12}$ & $41.99^{+0.11}_{-0.16}$ \\
NGC841  & – & – & $41.63^{+0.30}_{-0.30}$ \\
NGC1055 & $40.30^{+0.08}_{-0.11}$ & $40.62^{+0.08}_{-0.11}$ & – \\
NGC1169 & – & – & $41.40^{+0.11}_{-0.16}$ \\
NGC1961 & $41.20^{+0.03}_{-0.03}$ & $41.51^{+0.03}_{-0.03}$ & $41.97^{+0.11}_{-0.16}$ \\
NGC2681 & $40.69^{+0.01}_{-0.01}$ & $41.01^{+0.01}_{-0.01}$ & $40.92^{+0.18}_{-0.30}$ \\
NGC2683 & $40.06^{+0.02}_{-0.02}$ & $40.38^{+0.02}_{-0.02}$ & $39.90^{+0.11}_{-0.16}$ \\
NGC2859 & – & – & $41.34^{+0.11}_{-0.16}$ \\
NGC2985 & $40.40^{+0.07}_{-0.08}$ & $40.72^{+0.07}_{-0.08}$ & $41.35^{+0.11}_{-0.16}$ \\
NGC3642 & $40.53^{+0.08}_{-0.10}$ & $40.85^{+0.08}_{-0.10}$ & $41.60^{+0.11}_{-0.16}$ \\
NGC3898 & $39.95^{+0.05}_{-0.06}$ & $40.27^{+0.05}_{-0.06}$ & $41.28^{+0.11}_{-0.16}$ \\
NGC3992 & $39.30^{+0.20}_{-0.37}$ & $39.62^{+0.20}_{-0.37}$ & $41.09^{+0.11}_{-0.16}$ \\
NGC4036 & $40.87^{+0.04}_{-0.05}$ & $41.19^{+0.04}_{-0.05}$ & $41.78^{+0.11}_{-0.16}$ \\
NGC4111 & $41.28^{+0.02}_{-0.02}$ & $41.60^{+0.02}_{-0.02}$ & $41.10^{+0.11}_{-0.16}$ \\
NGC4145 & – & – & $39.78^{+0.11}_{-0.16}$ \\
NGC4346 & $39.62^{+0.12}_{-0.17}$ & $39.94^{+0.12}_{-0.17}$ & $40.33^{+0.30}_{-0.30}$ \\
NGC4395 & $40.83^{+0.01}_{-0.01}$ & $41.15^{+0.01}_{-0.01}$ & $41.36^{+0.11}_{-0.16}$ \\
NGC4750 & $41.05^{+0.07}_{-0.08}$ & $41.37^{+0.07}_{-0.08}$ & $41.46^{+0.11}_{-0.16}$ \\
NGC5194 & $40.85^{+0.01}_{-0.01}$ & $41.17^{+0.01}_{-0.01}$ & $41.58^{+0.11}_{-0.16}$ \\
NGC5195 & $40.46^{+0.44}_{-0.46}$ & $40.78^{+0.44}_{-0.46}$ & $40.24^{+0.11}_{-0.16}$ \\
NGC5395 & $41.13^{+0.07}_{-0.09}$ & $41.45^{+0.07}_{-0.09}$ & $41.53^{+0.11}_{-0.16}$ \\
NGC5448 & $40.34^{+0.12}_{-0.18}$ & $40.66^{+0.12}_{-0.18}$ & $41.25^{+0.11}_{-0.16}$ \\
NGC5485 & $40.38^{+0.11}_{-0.16}$ & $40.70^{+0.11}_{-0.16}$ & $40.96^{+0.18}_{-0.30}$ \\
NGC5566 & – & – & $41.00^{+0.11}_{-0.16}$ \\
NGC5631 & $40.20^{+0.14}_{-0.21}$ & $40.52^{+0.14}_{-0.21}$ & $41.43^{+0.11}_{-0.16}$ \\
NGC5746 & $41.26^{+0.02}_{-0.02}$ & $41.58^{+0.02}_{-0.02}$ & $40.64^{+0.18}_{-0.30}$ \\
NGC5850 & – & – & $41.43^{+0.11}_{-0.16}$ \\
NGC5921 & – & – & $41.30^{+0.18}_{-0.30}$ \\
NGC5985 & $41.36^{+0.11}_{-0.00}$ & $41.68^{+0.11}_{-0.00}$ & $41.50^{+0.11}_{-0.16}$ \\
NGC6340 & – & – & $40.91^{+0.11}_{-0.16}$ \\
NGC6703 & $41.14^{+0.02}_{-0.03}$ & $41.46^{+0.02}_{-0.03}$ & $41.18^{+0.11}_{-0.16}$ \\
NGC6951 & $40.76^{+0.02}_{-0.03}$ & $41.08^{+0.02}_{-0.03}$ & $41.35^{+0.11}_{-0.16}$ \\
NGC7814 & $39.63^{+0.07}_{-0.08}$ & $39.95^{+0.07}_{-0.08}$ & $39.44^{+0.11}_{-0.16}$ \\
IC356   & – & – & $40.08^{+0.11}_{-0.16}$ \\
IC520   & – & – & $41.90^{+0.18}_{-0.30}$ \\
\hline
\end{tabular}
\end{table*}

Bolometric luminosities for our LLAGN were estimated using both X-ray and optical indicators, depending on data availability. X-ray luminosities were prioritized because X-rays provide the most direct and least contaminated tracer of AGN power. They are also less affected by dust extinction and host galaxy emission which make them more reliable for weak nuclei such as LLAGN. For X-ray--detected sources, intrinsic 2--10~keV luminosities were derived as described in Section \ref{sec: Multiwavelength Analysis} and converted to bolometric luminosities using a bolometric correction factor, $k_{\mathrm{bol}}$, which is calculated as $L_{\mathrm{bol}} = k_{\mathrm{bol}} L_{\mathrm{2-10\,keV}}$. We primarily adopted the LLAGN-specific correction of \citet{Lopez2024}  where $k_{\mathrm{bol}} = 9.59$, which yields systematically lower values in the low-accretion regime. For comparison, we also computed $L_{\mathrm{bol}}$ using the widely adopted correction of \citet{Vasudevan2010} with $k_{\mathrm{bol}} = 20$.
For sources lacking X-ray data but possess optical measurements, $L_{\mathrm{bol}}$ was estimated from the extinction-corrected [O\,\textsc{iii}]~$\lambda5007$ line luminosity using the \citet{Heckman2014} where they adopted a bolometric correction factor of $k_{\mathrm{bol}} = 600$, with an intrinsic scatter of $\approx 0.3$--$0.4~\mathrm{dex}$. However, this calibration is primarily based on Seyfert~2 galaxies. Since most of our AGN are classified as LINERs, applying this calibration may introduce systematic uncertainties. This is because, in LINERs, the lower ionization state implies that [O\,\textsc{iii}] traces a smaller fraction of the total ionizing luminosity, which can lead to an overestimation of $L_{\mathrm{bol}}$. This further reinforces our preference for X-ray-derived values when available.

Our LLAGN classification is intentionally strict rather than conservative, as we exclude any galaxy exceeding $L_{\mathrm{bol}} = 10^{42}\,\mathrm{erg\,s^{-1}}$ in any of the three bolometric estimators (see Table~\ref{tab:lbol_comparison} and \ref{Tab: Luminosity for Excluded 7 LLAGN} for a comparison of the derived $L_{\mathrm{bol}}$ values). Although the LLAGN-specific correction places some additional sources below this limit, they are excluded for surpassing the threshold under the other methods. Our final sample thus demands robustness across all three estimators rather than relying on the most permissive correction.

Our 38 LLAGN\footnote{Before defining our final LLAGN sample, we first examined the bolometric luminosities of all 45 \citet{Saikia2018} galaxies with available multiwavelength data. As listed in Table~\ref{Tab: Luminosity for Excluded 7 LLAGN}, seven galaxies in the parent sample exhibit $\log L_{\rm bol} > 42$~erg~s$^{-1}$, which are significantly more luminous than the low-power nuclei targeted in this study. These galaxies were excluded from further analysis (refer Appendix~\ref{Appendix A}).} galaxies exhibit bolometric luminosities ranging from $\log L_{\mathrm{bol}} =$ $39.3$ to $41.9$~erg~s$^{-1}$, with a mean value of $\langle \log L_{\mathrm{bol}} \rangle =$ 40.8 $\pm 0.6$. For comparison, the \textit{Swift}--BAT AGN sample \citep{Koss2017} spans $\log L_{\mathrm{bol}} = 41.3$--$49.5$~erg~s$^{-1}$ with an average of $\langle \log L_{\mathrm{bol}} \rangle = 44.9 \pm 1.1$. Thus, the mean bolometric luminosity of our LLAGN is lower by over 4$\times$ relative to the \textit{Swift}--BAT AGN, consistent with the expected low-power nature of LLAGN. As shown in Fig. \ref{fig:AGN_properties}(a), the two samples display clearly separated luminosity distributions, with little overlap between the LLAGN and the luminous \textit{Swift}--BAT AGN.  A two-sample Kolmogorov--Smirnov (KS) test between the two distributions yields a statistic of $D = 0.995$ and a p-value of $7.8 \times 10^{-59}$, confirming that the two samples occupy different luminosity regimes.

We further compared our LLAGN sample with the local LLAGN population by \citet{Ho2009}, which has an average of $\langle \log L_{\mathrm{bol}} \rangle = 40.0 \pm 1.1$, consistent with our sample. While both samples occupy the low-luminosity regime, our LLAGN are systematically more luminous on average by $\sim$0.8 dex. A KS test between the two LLAGN samples yields $D = 0.363$ and a p-value of $4.5 \times 10^{-4}$, indicating that the two distributions are statistically different.
However, the mean value remain consistent between these two populations.

\subsection{{Black Hole Mass}} 
\label{subsec: black hole mass}

We estimated black hole masses for our LLAGN by adopting the stellar velocity dispersions from the \citet{Ho2009a} optical catalog and applying the $M_{\mathrm{BH}}$–$\sigma_{\ast}$ relation established by \citet{McConnell2011}, which have intrinsic scatter of 0.34 dex. As shown in Fig. \ref{fig:AGN_properties}(b), the resulting $M_{\mathrm{BH}}$ values span $\log(M_{\mathrm{BH}}/M_{\odot})$ {$\approx$} $4.1$ - $9.0$, with a mean of $\langle \log(M_{\mathrm{BH}}/M_{\odot}) \rangle = $ $7.6 \pm 0.9$.
This range indicates that most LLAGN in our sample host intermediate-mass to low-mass black holes. In contrast, the \textit{Swift}--BAT AGN sample from \citet{Koss2017} has a mean $\langle \log(M_{\mathrm{BH}}/M_{\odot}) \rangle =$ $ 8.3 \pm 0.6$, reflecting a population dominated by massive systems.
The offset between the \textit{Swift}--BAT AGN and our LLAGN sample is approximately $\sim 0.7$ dex. Accounting for the intrinsic scatter in the $M_{\rm BH}$--$\sigma_{\ast}$ relation, the typical uncertainty in $\log(M_{\rm BH}/M_{\odot})$ is $\sim 0.35$--$0.40$ dex. This intrinsic scatter is significantly smaller than the observed offset, and therefore does not affect our results.

A two-sample KS test yields $D = 0.396$ and $p = 3.1 \times 10^{-5}$, confirming a statistically significant difference between the two distributions. The LLAGN mass distribution is skewed toward lower $M_{\mathrm{BH}}$ values, with several sources extending below $10^{7}\,M_{\odot}$, especially NGC 4395, a dwarf galaxy with $M_{\mathrm{BH}} = 10^{4}\,M_{\odot}$  \citep{Yang2022, Nandi2023}, whereas the BAT sample is concentrated above $10^{8}\,M_{\odot}$. To check whether this result is driven by the extremely low-mass outlier NGC 4395, we repeated the KS test after removing this source. The resulting statistic remains strongly significant ($D = 0.9730$, $p = 6.6 \times 10^{-40}$), indicating that the discrepancy between the two mass distributions is not dominated by this outlier.

To place our results in the broader LLAGN context, we further compared our sample with the LLAGN population by \citet{Ho2009}. The \citet{Ho2009} sample has a mean of $\langle \log(M_{\mathrm{BH}}/M_{\odot}) \rangle = 7.6 \pm 0.9$, consistent with our value. A two-sample KS test between the two LLAGN populations yields $D = 0.0994$ and $p = 0.8958$, indicating no statistically significant difference between the distributions. This suggests that our LLAGN sample is representative of the general low-luminosity AGN population rather than a biased subset. This contrast highlights the lower-mass nature of black holes in nearby LLAGN compared to more luminous AGN.

\begin{figure*}
\centering
\includegraphics[width=1 \linewidth]{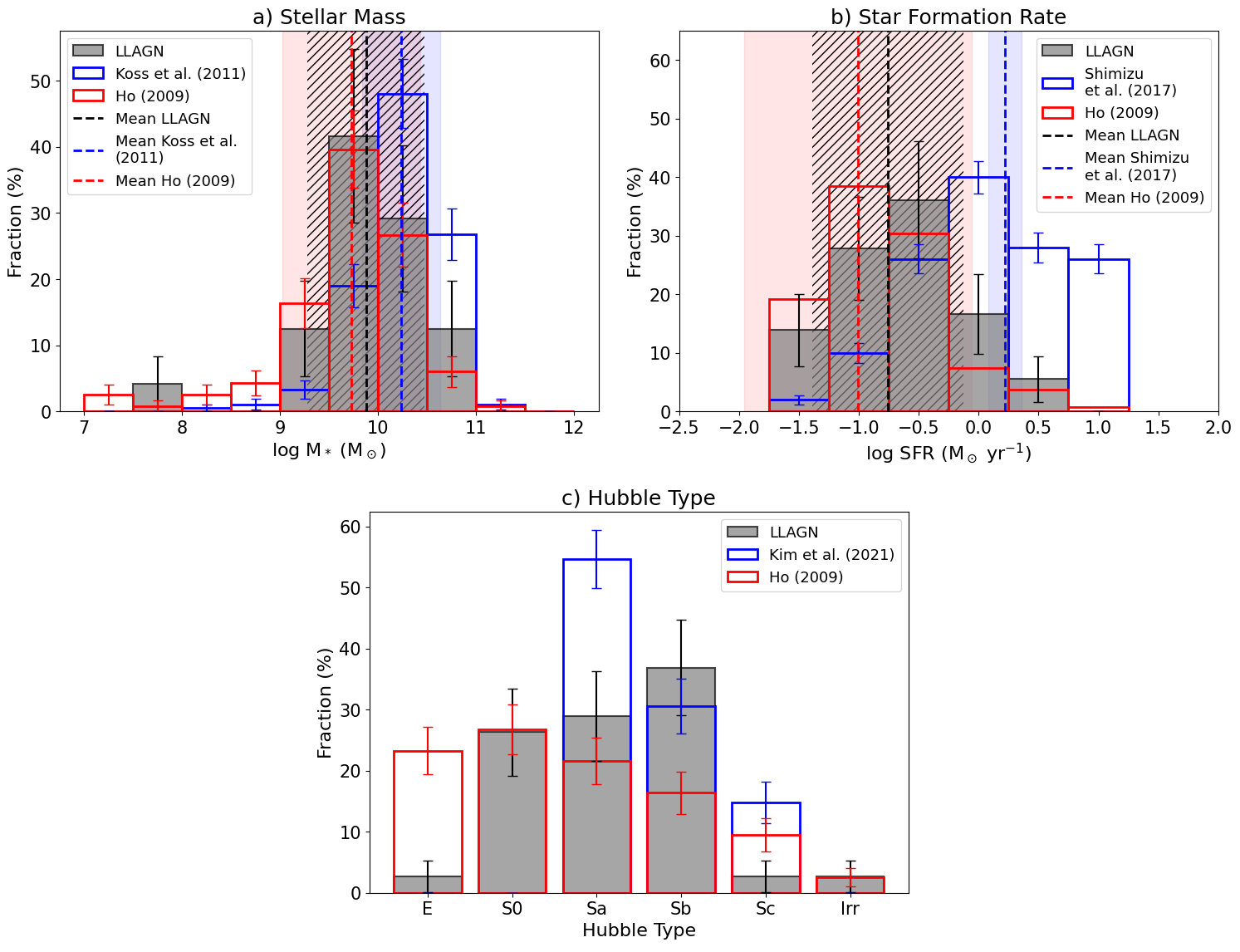}
\caption{Distributions of host galaxy properties for our LLAGN sample (grey) compared with the \textit{Swift}--BAT AGN sample (blue; \citealt{Shimizu2017, Koss2011, Kim2021}) and an additional LLAGN sample from \citealt{Ho2009} (red). Panels show stellar rate (a), star formation rate (b), and host galaxy Hubble type (c).}
\label{fig:hostgalaxyproperties}
\end{figure*}

\subsection{{Eddington Ratio}}
\label{sec:Eddington_ratio}

Eddington ratios were calculated for all LLAGN with available $L_{\mathrm{bol}}$ and $M_{\mathrm{BH}}$, as determined in Sections \ref{sec:Lbol} and \ref{subsec: black hole mass}, respectively. 
For each galaxy, the Eddington ratio was computed.
Uncertainties were propagated from both $L_{\mathrm{bol}}$ and $M_{\mathrm{BH}}$ estimates. 
The resulting distribution of $\lambda_{\mathrm{Edd}}$ was then compared to that of the \textit{Swift}--BAT AGN sample \citep{Koss2017} to assess differences in accretion efficiency. 
A statistical comparison between the two samples was performed using a two-sample KS test.

Our LLAGN sample exhibits Eddington ratios spanning $\log \lambda_{\mathrm{Edd}} \approx$ -6.5 to    $-1.3$, with a mean value of $\langle \log \lambda_{\mathrm{Edd}} \rangle  $  = -4.9 $\pm$ 0.5. 
Most nuclei accrete below $\log \lambda_{\mathrm{Edd}} < -2$, consistent with RIAF-type accretion flows. 
For comparison, the LLAGN population from \citet{Ho2009} exhibits a mean Eddington ratio of $\langle \log \lambda_{\mathrm{Edd}} \rangle = -5.6 \pm 1.1$, which is consistent with our sample within uncertainties, indicating that both populations occupy the same low-accretion regime. However, a two-sample KS test between our LLAGN and the \citet{Ho2009} sample yields $D = 0.37$ and $p = 3.6 \times 10^{-4}$, demonstrating that the underlying distributions are statistically distinct.  
Nevertheless, the mean values remain consistent between the two samples.

In contrast, the \textit{Swift}--BAT AGN cluster around $\log \lambda_{\mathrm{Edd}} \sim -0.7$ with $\langle \log \lambda_{\mathrm{Edd}} \rangle = -0.7 \pm 1.0 $, corresponding to radiatively efficient thin-disk accretion.  
A KS test confirms that the two distributions are statistically distinct, yielding $D = 0.97$ and $p = 6.6 \times 10^{-40}$. 
These findings confirm that our LLAGN occupy the low-accretion regime, with the exception of NGC 4395, which has an Eddington ratio of $\lambda_{\mathrm{Edd}} = 4.54 \times 10^{-2}$, consistent with previous studies suggesting that it is still growing (Brum et al. 2019).

\section{{Host galaxy properties}}

The properties of a host galaxy play a fundamental role in shaping the growth and visibility of its central black hole.  
In LLAGN, the interplay between morphology, stellar mass, and star formation rate (SFR) offers valuable clues about how nuclear activity persists at low accretion rates.  
In this section, we examine these host galaxy parameters for our LLAGN sample and compare their distributions with those of the \textit{Swift}--BAT AGN.  
The goal is to determine whether LLAGN occupy a different parameter regime in terms of host environment, structural type, and stellar content compared to more luminous AGN.

\subsection{{Stellar Mass}} 

To estimate stellar mass, we utilized both optical and infrared magnitudes. Optical magnitudes were compiled at three specific wavelengths: 445 nm (B-band), 551 nm (V-band), and 658 nm (R-band), while the near-infrared magnitude was taken from K-band, which centered at 2.16~$\mu$m. These photometric values were obtained via the Set of Identifications, Measurements, and Bibliography for Astronomical Data (SIMBAD)\footnote{SIMBAD data are publicly available via the CDS database at \url{https://simbad.cds.unistra.fr/simbad/}} database. The stellar mass was then computed using the empirical relations provided by \citet{Bell2003}, which relate optical and near-infrared luminosities to mass-to-light ratios. We adopt a systematic uncertainty of 0.1 dex due to stellar population modeling.

The stellar masses of our LLAGN hosts range from $\log(M_{\ast}/M_{\odot}) = 7.7$ to 10.8, with a mean value of $\langle \log(M_{\ast}/M_{\odot}) \rangle = 9.9 \pm 0.6$. 
In contrast, the \textit{Swift}--BAT AGN hosts exhibit higher average stellar masses, with $\langle \log(M_{\ast}/M_{\odot}) \rangle = 10.2$  $\pm 0.4$ \citep{Koss2011}. 
A two-sample KS test confirms that the two distributions differ significantly ($D = 0.343$, $p = 0.01$). 
While the two samples partially overlap, our LLAGN extend toward lower host masses ($\log M_{\ast} < 9.5$), whereas the BAT AGN populate the upper end of the distribution (up to $\log M_{\ast} \approx 11.1$).
For comparison with another LLAGN population from \citet{Ho2009} yields a mean of $\langle \log(M_{\ast}/M_{\odot}) \rangle = 9.8 \pm 0.7$, fully consistent with our value. A KS test between our LLAGN and the \citet{Ho2009} sample gives $D = 0.1006$ and $p = 0.9735$, indicating no statistically significant difference between the two distributions.  This agreement suggests that the lower stellar masses observed in our sample are a genuine property of LLAGN hosts rather than a selection effect specific to our dataset.

However, the stellar masses derived via the \citet{Bell2003} mass-to-light relations are subject to systematic uncertainties that depend on the underlying stellar population. Galaxies with younger stellar populations and ongoing star formation (typically late-type systems) may have lower mass-to-light ratios, leading to the values calculated being underestimated,  while older, bulge-dominated systems may exhibit the opposite effect. In our sample (Table~\ref{tab:Physical And host LLAGN properties}), 16 out of 38 galaxies ($42.1^{+21.8}_{-15.7}\%$) are classified as late-type (Sb/Sc/Irr), while the remaining 22 out of 38 ($57.9^{+24.8}_{-18.7}\%$) are earlier-type systems. These fractions are broadly comparable, suggesting that the sample is approximately morphologically balanced. As a result, no significant difference is expected.

\subsection{{Star Formation Rate}} 
\label{Sec: SFR}

Star formation rates for our LLAGN sample were estimated using the mid-infrared WISE 12~$\mu$m ($W3$) luminosity, which traces warm dust heated primarily by young, massive stars. 
In this work, we adopt the empirical relation proposed by \citet{Lee2013} to compute the SFRs from W3 luminosities, which is based on the luminous flux and distance.  

Our LLAGN sample exhibits systematically low star formation activity, with $\log(\mathrm{SFR}/M_{\odot}\,\mathrm{yr}^{-1})$ values ranging from $-2.5$ to $0.5$, and a mean of $\langle \log(\mathrm{SFR}/M_{\odot}\,\mathrm{yr}^{-1}) \rangle = -0.73 \pm 0.63$. 
By contrast, the \textit{Swift}--BAT AGN hosts studied by \citet{Shimizu2017} show higher star formation levels, with $\langle \log(\mathrm{SFR}/M_{\odot}\,\mathrm{yr}^{-1}) \rangle = 0.22 \pm 0.14$.
A KS test between the two samples yields a statistic of $D = 0.61$ and a $p$-value of $2.3 \times 10^{-11}$, confirming that the difference is highly significant. This reflects a clear separation between our LLAGN hosts, which are preferentially quiescent or weakly star-forming systems, with the more actively star-forming AGN hosts in the BAT sample.
We further compare our results with LLAGN population from \citet{Ho2009}, which exhibits a mean  of $\langle \log(\mathrm{SFR}/M_{\odot}\,\mathrm{yr}^{-1}) \rangle = -1.01 \pm 0.95$, consistent with our measured value within the uncertainties. A KS test between our LLAGN sample and the \citet{Ho2009} population yields $D = 0.12$ and $p = 0.71$, indicating no statistically significant difference between the two distributions. This agreement reinforces the picture that LLAGN are generally hosted by galaxies with suppressed or low star formation activity.

\begin{figure}
\centering
\includegraphics[width=\columnwidth]{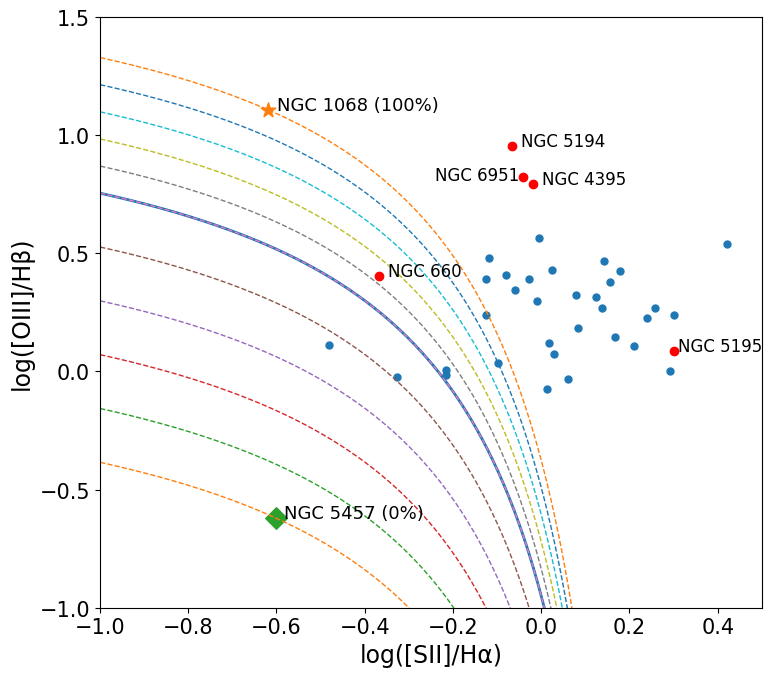}
\caption{BPT diagram of $\log([\ion{O}{III}]/\mathrm{H}\beta)$ vs. $\log([\ion{S}{II}]/\mathrm{H}\alpha)$. The solid curve (\citet{Kewley2006}) marks the AGN–star formation division (assumed 50\% AGN), while dashed curves trace mixing from pure star formation (0\%) to pure AGN (100\%). The AGN fraction is estimated from a galaxy’s position along this sequence.}
\label{fig:BPT_SII_mixing}
\end{figure}

However, the \citet{Lee2013} relation may overestimate SFRs in AGN-dominated systems due to AGN-heated dust and feedback (\citealt{Lopez2024, Yuk2025}) which also peaks at 12~$\mu$m ($W3$). To quantify this effect, we estimated the AGN contribution to the $W3$ flux. Since \textit{WISE} diagrams are optimised for luminous AGN, they are less suitable for this purpose. Instead, optical emission-line diagnostics may provide a more sensitive constraint. Therefore, we estimated the AGN contribution using BPT diagnostics ([O\,\textsc{iii}]/H$\beta$ and [S\,\textsc{ii}]/H$\alpha$) as shown in Fig.~\ref{fig:BPT_SII_mixing} via a mixing-sequence approach. The [S\,\textsc{ii}]-BPT diagram is adopted due to its sensitivity to AGN ionisation and less affected by metallicity and star formation contamination than [N\,\textsc{ii}], while the [O\,\textsc{i}] line is often weak in low-luminosity systems and therefore less reliable (\citealt{Kewley2006, Kauffmann2003}). We adopt NGC~1068 and NGC~5457 as reference points for 100\% AGN contribution  and 0\% AGN contribution, which correspond to archetypal AGN (\citealt{Leftley2024, Mutie2025}) and a star-forming galaxy (\citealt{Smirnova2023,Bresolin2025}). The \citet{Kewley2006} line is used to indicate 50\% AGN contribution as it is interpreted as a transition regime between star formation and AGN ionisation within a mixing-sequence framework. Based on Fig.~\ref{fig:BPT_SII_mixing}, we can see that most sources are strongly AGN-dominated ($\gtrsim$90\%). Among the five galaxies ($\sim$13\%) with mid-infrared AGN signatures, four are $\sim$100\% AGN-dominated, suggesting that their $W3$ emissions are strongly contaminated by AGN-heated dust. As a result, the SFRs derived for these sources are likely significantly overestimated and should therefore be treated as upper limits. The remaining source, NGC~660, shows a mixed contribution ($\sim$60\%), but is still AGN-dominated, implying a substantial AGN contribution to its mid-infrared emission. Consequently, its derived SFR should also be considered an upper limit. The remaining $\sim$87\% lack clear mid-infrared AGN signatures despite being AGN-dominated as shown in our diagnostics. This is consistent with LLAGN where optical emission traces AGN activity but mid-infrared emission is dominated by star formation. In these cases, AGN contamination to the 12~$\mu$m emission is likely small and their SFRs are more reliable.

The suppressed star formation in our LLAGN hosts is consistent with their low accretion rates (Section \ref{sec:Eddington_ratio}).
These results suggest that our sample probes a population of AGN that are not only radiatively inefficient but also reside in galaxies where star formation has already declined, in contrast to the more star-forming, luminous AGN typically identified in larger surveys.

\subsection{{Hubble Type}}

Host galaxy morphological types were compiled primarily from the NASA/IPAC Extragalactic Database (NED) using standard Hubble classifications (E, S0, Sa--Sc, Irr). The Hubble-type distribution of our LLAGN sample is shown in Figure~\ref{fig:hostgalaxyproperties}(c).

Our LLAGN sample spans a broad range of morphological types, from ellipticals to irregulars, but peaks at Sb, indicating a preference for intermediate disk galaxies. For comparison, the \textit{Swift}--BAT AGN sample of \cite{Kim2021} covers morphologies from Sa to Sc and peaks at Sa, consistent with more bulge-dominated spiral hosts. A KS test confirms that the two morphological distributions are significantly different ($D = 0.29$, $p = 0.01$).
In contrast, when compared to the LLAGN sample of \cite{Ho2009}, we find no statistically significant difference in both morphologies ($D = 0.21$, $p = 0.13$). This suggests that the morphological distribution of our LLAGN is consistent with previously identified LLAGN populations, reinforcing the view that LLAGN predominantly reside in a diverse range of host galaxies spanning both disk- and bulge-dominated systems.

Compared to \textit{Swift}--BAT AGN, our LLAGN shows a higher fraction of S0--Sb galaxies, including a notable number of bulge-dominated systems, and a relatively lower fraction of late-type (Sc) spirals.
To further quantify the structural properties of the host galaxies, we compiled Sérsic indices from the NASA--Sloan Atlas\footnote{NASA-Sloan Atlas can be reached via the https://www.nsatlas.org/getAtlas} for a subset of the sample. We interpreted these values following standard Sérsic index–based structural classifications in the literature, where low values ($n \leq 2$) are generally associated with disk-dominated systems and higher values ($n \geq 3$) with bulge-dominated components (e.g. \citealt{Graham2005, Fisher2008, Fisher2010}). For those with available measurements (29/38), we found that 16 ($55.2^{+28.6}_{-20.6}\%$) exhibited high Sérsic indices ($n \geq 3.0$), indicative of bulge-dominated systems, while 8 ($27.6^{+22.2}_{-13.9}\%$) showed low values ($n \leq 2.0$) consistent with disk-dominated morphologies. The remaining 5 sources ($17.2^{+19.0}_{-10.4}\%$) fell in the intermediate regime ($2.1 \leq n \leq 2.9$). This result confirmed that our LLAGN hosts span a wide range of structural types, with a substantial contribution from bulge-prominent systems.

\section{Discussion}
\label{Discussion}

In this section, we interpreted the main results of our study.  We first consider potential selection effects and how they may influence the representativeness of our LLAGN sample. We then discuss the implications of our findings for the demographics of LLAGN in the local Universe. Next, we compare our observational results with theoretical expectations, particularly in relation to RIAF models. Finally, we explore the broader implications of our results for AGN–galaxy co-evolution at low luminosities, including how weak nuclear activity relates to host galaxy properties and evolutionary pathways.

\subsection{Sample Representativeness \& Implications for LLAGN Demographics}

The parent sample from \citet{Saikia2018} is based on compact 15~GHz radio core detections, which preferentially select AGN with jet activity or relatively strong radio emission. As a result, our sample is biased toward radio-loud systems and may underrepresent the faintest and most weakly accreting LLAGN. This is in fact what we found when comparing our sample with the optically selected LLAGN by \citet{Ho2009}, as discussed in section \ref{sec:Lbol} and \ref{sec:Eddington_ratio}. While both samples probe the low-luminosity regime, our LLAGN systematically occupy higher bolometric luminosities and Eddington ratios, consistent with expectations for a radio-selected sample. This suggests that our sample preferentially traces the more active end of the LLAGN population rather than providing a complete census. Nevertheless, both samples remain broadly consistent within the observed scatter.

In addition, our results demonstrated that LLAGN demographics in the local Universe are strongly dependent on observational wavelength. We showed that AGN detection fractions vary from $\sim84\%$ in the optical to $\sim63\%$ in X-rays and only $\sim13\%$ in the infrared, indicating that a significant fraction of LLAGN would be missed in single-wavelength surveys, particularly in the mid-infrared. This reflects the incompleteness of current AGN censuses and their biases against weakly accreting nuclei when relying on a single waveband. Consistent with this, the nuclear properties indicate that LLAGN are not simply scaled-down versions of luminous AGN. Their low bolometric luminosities ($\log L_{\rm bol} \approx 39.6$--$41.9$~erg~s$^{-1}$) and extremely low Eddington ratios ($\langle \log \lambda_{\rm Edd} \rangle \approx -4.7$) point to a distinct accretion regime. They are also preferentially hosted by lower-mass galaxies with suppressed star formation compared to typical AGN. Bringing these results together, they indicate that LLAGN are likely common but frequently missed in single-wavelength studies. Since our sample is biased toward the more active LLAGN due to its radio-based selection, future multiwavelength surveys will be essential to fully characterize the underlying LLAGN population.

\subsection{Comparison with Theoretical Predictions for RIAF Physics}

Our analysis reveals that the observed properties of our LLAGN sample are consistent with theoretical expectations for RIAF. The low Eddington ratios, with $\langle \log \lambda_{\rm Edd} \rangle \approx -4.7$, lie well below the expected transition threshold of $\log \lambda_{\rm Edd} \sim -2$ to $-3$, indicating accretion in a radiatively inefficient regime \citep{Narayan1995, Yuan2014, Draper2010}.
Their low bolometric luminosities, approximately $4.1$ dex below those of more luminous AGN, further suggest that a large fraction of the accretion energy is not radiated efficiently but instead advected or lost through outflows \citep{Narayan1995, Blandford1999, Yuan2014}. In addition, the weak infrared detection fraction ($\sim 13\%$) supports reduced AGN-heated dust emission, consistent with a diminished or absent torus \citep{Ho2008, Elitzur2006, Elitzur2009}. The predominance of LINER-like spectra is also consistent with the harder ionizing continuum expected from optically thin accretion flows \citep{Ho2008, Eracleous2010, Yuan2014}.
However, some sources show detectable infrared emission, [Ne~V] lines, or signs of obscuration, indicating that not all LLAGN follow a simple RIAF picture. This suggests that LLAGN may span a range of accretion states or evolutionary stages, requiring models that allow for gradual transitions rather than a sharp dichotomy between thin-disk and RIAF regimes.

\subsection{Implications for AGN--Galaxy Co-evolution at Low Luminosities}

The observed nuclear and host galaxy properties of our LLAGN sample show that these objects occupy a different regime of AGN--galaxy co-evolution. Unlike luminous AGN, which are typically found in massive, star-forming galaxies \citep{Kauffmann2003, Heckman2014}, we found that LLAGN reside in lower-mass systems with suppressed star formation. Their hosts span a wide range of morphologies, including both disk- and bulge-dominated galaxies.
However, the presence of prominent bulges in a subset is consistent with a scenario in which some LLAGN represent galaxies that previously hosted stronger AGN activity and are now observed in a fading phase with weak accretion (e.g.\citealp{Couto2023}). The episodic nature of AGN activity, with variability on $\sim10^4$--$10^6$ yr timescales \citep{hickox2014, Schawinski2015, King2015}, further supports this interpretation.
The combination of low star formation, evolved host structures, and very low accretion rates suggests that LLAGN are preferentially found in relatively quiescent systems where both black hole growth and star formation proceed inefficiently \citep{Ho2009}. In this regime, accretion are likely dominated by radiatively inefficient flows, while compact radio emission may be associated with maintenance-mode feedback \citep{Fabian2012}.
These results suggest that the connection between black hole growth and host galaxy evolution persists at low luminosities, but both processes occur at reduced efficiency, with LLAGN representing a long-lived, low-activity phase of the AGN duty cycle.

\section{Summary and Conclusions}
\label{Summary and conclusion}

In this work, we presented a systematic census of LLAGN using a sample of 38 sources identified via compact core radio detections at 15~GHz. By combining X-ray, optical, and infrared diagnostics, we investigated the detectability of LLAGN across different wavebands, as well as their nuclear and host galaxy properties. Our main results can be summarised as follows:

\begin{enumerate}[label=(\roman*), leftmargin=*]
    
    \item We found that optical diagnostics recovered the largest fraction of the sample, identifying 32 out of 38 sources ($84.2^{+15.8}_{-22.9}\%$), while X-rays detected 24 sources ($63.2^{+25.7}_{-19.6}\%$) and infrared diagnostics identified only 5 sources ($13.2^{+14.5}_{-8.0}\%$). Compared to more luminous AGN, our LLAGN show a higher optical detection fraction than typically reported in previous studies ($\sim$10--42\%), likely due to the conservative classification criteria adopted in this work. In contrast, the infrared detection fraction is substantially lower than that of normal AGN ($\sim$60--80\%), reflecting the dominance of host galaxy emission in the low-luminosity regime. However, their X-ray detection rate is comparable to standard AGN ($\sim$20--60\%).

    \item The nuclear properties of our LLAGN differed markedly from those of standard AGN. Bolometric luminosities ranged from $10^{39.3}$ to $10^{41.9}$~erg~s$^{-1}$, with a mean value approximately 4.1~dex lower than that of the \textit{Swift}--BAT AGN sample. Black hole masses are also lower, with a mean of $\log(M_{\rm BH}/M_\odot) = 7.6 \pm 0.9$, corresponding to an offset of about 0.7~dex relative to \textit{Swift}--BAT AGN. Eddington ratios are extremely low (mean $\log \lambda_{\rm Edd}$ $= -4.9 \pm 0.5$), approximately 4.2~dex lower than \textit{Swift}--BAT AGN. These results confirm that the LLAGN in our sample host smaller black holes accreting at lower rates, more consistent with RIAFs model rather than the efficient thin-disk accretion typical of higher luminosity AGN. While our LLAGN exhibit slightly higher bolometric luminosities and Eddington ratios compared to the broader local LLAGN population, likely reflecting a bias toward radio-detected nuclei, their overall properties remain consistent within the observed scatter.

    \item LLAGN reside in less massive galaxies, with a mean of $\log(M_{*}/M_\odot) = 9.9 \pm 0.6$, roughly 0.3~dex lower than \textit{Swift}--BAT AGN. Star formation rates are also suppressed, with a mean of $\log(\mathrm{SFR}/M_{\odot}\,\mathrm{yr}^{-1}) = -0.73 \pm 0.63$, offset by $\sim$0.5~dex relative to \textit{Swift}--BAT AGN. Morphologically, the LLAGN span a range of types from E to Irr, with a modest peak at Sb, indicating a mix of bulge- and disk-dominated systems. This distribution differ from previous studies that primarily associated LLAGN with disk-dominated hosts. These results suggest that LLAGN reside in relatively quiescent, low-mass galaxies hosting small black holes with low accretion rates. These host galaxy properties are consistent with those of the other local LLAGN population, indicating that our sample is broadly representative of the nearby LLAGN population within the observed scatter.

\end{enumerate}

Based on our study, we show that LLAGN reside in systematically lower-mass and more quiescent galaxies while spanning a wide morphological range, including bulge-dominated systems. This suggests that at least a subset of LLAGN may represent a fading phase of nuclear activity rather than simply scaled-down AGN.
Overall, these findings indicate that our LLAGN systematically differ from higher luminosity AGN population, probing the low-accretion regime of black hole activity and its connection to galaxy evolution in the local Universe.
However, we note that our conclusions formally apply to compact radio-core LLAGN and may not necessarily represent the full LLAGN population.

\section*{Acknowledgements}

MNR acknowledges support from the MyBrainSc Scholarship programme under the Ministry of Higher Education Malaysia (MoHE). The authors thank N. S. A. Hamid for useful discussions and comments on the manuscript. This research has made use of data and/or software provided by the HEASARC, which is a service of the Astrophysics Science Division at NASA/GSFC. This research has made use of IRSA, which is funded by the National Aeronautics and Space Administration and operated by the California Institute of Technology. This research has made use of the SIMBAD database, operated at CDS, Strasbourg, France, \citet{Wenger2000}. This research has made use of the NED, which is operated by the Jet Propulsion Laboratory, California Institute of Technology, under contract with the National Aeronautics and Space Administration.

\section*{Data Availability}

The data utilized in this study is accessible for public download and access as follows: 

\begin{enumerate}[label=(\roman*), leftmargin=*]
    \item X-ray data (\textit{Chandra} and \textit{XMM-Newton})
        \begin{enumerate}
            \item From NASA’s HEASARC (\url{https://heasarc.gsfc.nasa.gov/cgi-bin/W3Browse/w3browse.pl}l).
        \end{enumerate}
    \item Mid-IR data (WISE)
        \begin{enumerate}
            \item From NASA/IPAC IRSA (\url{https://irsa.ipac.caltech.edu/Missions/wise.html}). Table \ref{tab:basic properties 38 LLAGN}. lists the data details (W1, W2 \& W3).
        \end{enumerate}
\end{enumerate}

\bibliographystyle{mnras}
\bibliography{example}

@ARTICLE{Antonucci1993,
       author = {{Antonucci}, Robert},
        title = "{Unified models for active galactic nuclei and quasars.}",
      journal = {\araa},
         year = 1993,
        month = jan,
       volume = {31},
        pages = {473-521},
          doi = {10.1146/annurev.aa.31.090193.002353},
       adsurl = {https://ui.adsabs.harvard.edu/abs/1993ARA&A..31..473A}
}

@ARTICLE{Ding2022,
       author = {{Ding}, Nan and {Gu}, Qiusheng and {Tang}, Yunyong and {Xiong}, Dingrong and {Guo}, Xiaotong and {Xu}, Xinpeng and {Geng}, Xiongfei and {Ge}, Xue and {Chen}, Yongyun},
        title = "{The variability and soft X-ray excess properties of narrow-line Seyfert 1 galaxies: The Swift view}",
      journal = {\aap},
         year = 2022,
        month = mar,
       volume = {659},
          eid = {A172},
        pages = {A172},
          doi = {10.1051/0004-6361/202142650},
       adsurl = {https://ui.adsabs.harvard.edu/abs/2022A&A...659A.172D}
}

@INCOLLECTION{Murase2022,
       author = {{Murase}, Kohta and {Stecker}, Floyd W.},
        title = "{High-Energy Neutrinos from Active Galactic Nuclei}",
    booktitle = {The Encyclopedia of Cosmology. Set 2: Frontiers in Cosmology. Volume 2: Neutrino Physics and Astrophysics},
    publisher = {World Scientific Publishing},
         year = 2023,
       editor = {{Stecker}, Floyd W.},
       pages = {483}
       }

@ARTICLE{McHardy2018,
       author = {{McHardy}, I.~M. and {Connolly}, S.~D. and {Horne}, K. and {Cackett}, E.~M. and {Gelbord}, J. and {Peterson}, B.~M. and {Pahari}, M. and {Gehrels}, N. and {Goad}, M. and {Lira}, P. and {Arevalo}, P. and {Baldi}, R.~D. and {Brandt}, N. and {Breedt}, E. and {Chand}, H. and {Dewangan}, G. and {Done}, C. and {Elvis}, M. and {Emmanoulopoulos}, D. and {Fausnaugh}, M.~M. and {Kaspi}, S. and {Kochanek}, C.~S. and {Korista}, K. and {Papadakis}, I.~E. and {Rao}, A.~R. and {Uttley}, P. and {Vestergaard}, M. and {Ward}, M.~J.},
        title = "{X-ray/UV/optical variability of NGC 4593 with Swift: reprocessing of X-rays by an extended reprocessor}",
      journal = {\mnras},
         year = 2018,
        month = nov,
       volume = {480},
       number = {3},
        pages = {2881-2897},
          doi = {10.1093/mnras/sty1983},
archivePrefix = {arXiv},
       eprint = {1712.04852},
 primaryClass = {astro-ph.HE},
       adsurl = {https://ui.adsabs.harvard.edu/abs/2018MNRAS.480.2881M}
}

@ARTICLE{Jin2012,
       author = {{Jin}, Chichuan and {Ward}, Martin and {Done}, Chris and {Gelbord}, Jonathan},
        title = "{A combined optical and X-ray study of unobscured type 1 active galactic nuclei - I. Optical spectra and spectral energy distribution modelling}",
      journal = {\mnras},
         year = 2012,
        month = mar,
       volume = {420},
       number = {3},
        pages = {1825-1847},
          doi = {10.1111/j.1365-2966.2011.19805.x},
archivePrefix = {arXiv},
       eprint = {1109.2069},
 primaryClass = {astro-ph.HE},
       adsurl = {https://ui.adsabs.harvard.edu/abs/2012MNRAS.420.1825J}
}

@ARTICLE{Kamraj2018,
       author = {{Kamraj}, N. and {Harrison}, F.~A. and {Balokovi{\'c}}, M. and {Lohfink}, A. and {Brightman}, M.},
        title = "{Coronal Properties of Swift/BAT-selected Seyfert 1 AGNs Observed with NuSTAR}",
      journal = {\apj},
         year = 2018,
        month = oct,
       volume = {866},
       number = {2},
          eid = {124},
        pages = {124},
          doi = {10.3847/1538-4357/aadd0d},
archivePrefix = {arXiv},
       eprint = {1809.01757},
 primaryClass = {astro-ph.HE},
       adsurl = {https://ui.adsabs.harvard.edu/abs/2018ApJ...866..124K}
}

@INPROCEEDINGS{Fukazawa2020,
       author = {{Fukazawa}, Yasushi},
        title = "{X-ray probing of NGC 1275 nuclear region with Hitomi, Swift, and Suzaku}",
    booktitle = {Perseus in Sicily: From Black Hole to Cluster Outskirts},
         year = 2020,
       editor = {{Asada}, Keiichi and {de Gouveia Dal Pino}, Elisabete and {Giroletti}, Marcello and {Nagai}, Hiroshi and {Nemmen}, Rodrigo},
       series = {IAU Symposium},
       volume = {342},
        month = jan,
        pages={118}
        }

@ARTICLE{Araudo2010,
       author = {{Araudo}, A.~T. and {Bosch-Ramon}, V. and {Romero}, G.~E.},
        title = "{Gamma rays from cloud penetration at the base of AGN jets}",
      journal = {\aap},
         year = 2010,
        month = nov,
       volume = {522},
          eid = {A97},
        pages = {A97},
          doi = {10.1051/0004-6361/201014660},
archivePrefix = {arXiv},
       eprint = {1007.2199},
 primaryClass = {astro-ph.HE},
       adsurl = {https://ui.adsabs.harvard.edu/abs/2010A&A...522A..97A}
}

@ARTICLE{Smith2016,
       author = {{Smith}, Krista Lynne and {Mushotzky}, Richard F. and {Vogel}, Stuart and {Shimizu}, Thomas T. and {Miller}, Neal},
        title = "{Radio Properties of the BAT AGNs: the FIR-radio Relation, the Fundamental Plane, and the Main Sequence of Star Formation}",
      journal = {\apj},
         year = 2016,
        month = dec,
       volume = {832},
       number = {2},
          eid = {163},
        pages = {163},
          doi = {10.3847/0004-637X/832/2/163},
archivePrefix = {arXiv},
       eprint = {1610.01167},
 primaryClass = {astro-ph.GA},
       adsurl = {https://ui.adsabs.harvard.edu/abs/2016ApJ...832..163S}
}

@ARTICLE{Saikia2018,
       author = {{Saikia}, Payaswini and {K{\"o}rding}, Elmar and {Coppejans}, Deanne L. and {Falcke}, Heino and {Williams}, David and {Baldi}, Ranieri D. and {Mchardy}, Ian and {Beswick}, Rob},
        title = "{15-GHz radio emission from nearby low-luminosity active galactic nuclei}",
      journal = {\aap},
         year = 2018,
        month = sep,
       volume = {616},
          eid = {A152},
        pages = {A152},
          doi = {10.1051/0004-6361/201833233},
archivePrefix = {arXiv},
       eprint = {1805.06696},
 primaryClass = {astro-ph.HE},
       adsurl = {https://ui.adsabs.harvard.edu/abs/2018A&A...616A.152S}
}

@article{Ptak2001,
       author = {{Ptak}, A.},
        title = "{Low-luminosity AGN and normal galaxies}",
    journal  = {X-ray Astronomy: Stellar Endpoints, AGN, and the Diffuse X-ray Background},
    booktitle = {X-ray Astronomy: Stellar Endpoints, AGN, and the Diffuse X-ray Background},
         year = 2001,
       editor = {{White}, Nicholas E. and {Malaguti}, Giuseppe and {Palumbo}, Giorgio G.~C.},
       series = {American Institute of Physics Conference Series},
       volume = {599},
        month = dec,
        pages = {326-335},
          doi = {10.1063/1.1434645},
archivePrefix = {arXiv},
       eprint = {astro-ph/0008459},
 primaryClass = {astro-ph},
       adsurl = {https://ui.adsabs.harvard.edu/abs/2001AIPC..599..326P}
}

@ARTICLE{Lambrides2020,
       author = {{Lambrides}, Erini L. and {Chiaberge}, Marco and {Heckman}, Timothy and {Gilli}, Roberto and {Vito}, Fabio and {Norman}, Colin},
        title = "{A Large Population of Obscured AGN in Disguise as Low-luminosity AGN in Chandra Deep Field South}",
      journal = {\apj},
         year = 2020,
        month = jul,
       volume = {897},
       number = {2},
          eid = {160},
        pages = {160},
          doi = {10.3847/1538-4357/ab919c},
archivePrefix = {arXiv},
       eprint = {2002.00955},
 primaryClass = {astro-ph.GA},
       adsurl = {https://ui.adsabs.harvard.edu/abs/2020ApJ...897..160L}
}

@ARTICLE{Tomar2021,
       author = {{Tomar}, Gunjan and {Gupta}, Nayantara and {Prince}, Raj},
        title = "{Broadband Modeling of Low-luminosity Active Galactic Nuclei Detected in Gamma Rays}",
      journal = {\apj},
         year = 2021,
        month = oct,
       volume = {919},
       number = {2},
          eid = {137},
        pages = {137},
          doi = {10.3847/1538-4357/ac1588},
archivePrefix = {arXiv},
       eprint = {2107.08256},
 primaryClass = {astro-ph.HE},
       adsurl = {https://ui.adsabs.harvard.edu/abs/2021ApJ...919..137T}
}

@ARTICLE{Terashima2002,
       author = {{Terashima}, Yuichi and {Iyomoto}, Naoko and {Ho}, Luis C. and {Ptak}, Andrew F.},
        title = "{X-Ray Properties of LINERs and Low-Luminosity Seyfert Galaxies Observed with ASCA. I. Observations and Results}",
      journal = {\apjs},
         year = 2002,
        month = mar,
       volume = {139},
       number = {1},
        pages = {1-36},
          doi = {10.1086/324373},
archivePrefix = {arXiv},
       eprint = {astro-ph/0203005},
 primaryClass = {astro-ph},
       adsurl = {https://ui.adsabs.harvard.edu/abs/2002ApJS..139....1T}
}

@ARTICLE{Elitzur2006,
       author = {{Elitzur}, Moshe and {Shlosman}, Isaac},
        title = "{The AGN-obscuring Torus: The End of the ``Doughnut'' Paradigm?}",
      journal = {\apjl},
         year = 2006,
        month = sep,
       volume = {648},
       number = {2},
        pages = {L101-L104},
          doi = {10.1086/508158},
archivePrefix = {arXiv},
       eprint = {astro-ph/0605686},
 primaryClass = {astro-ph},
       adsurl = {https://ui.adsabs.harvard.edu/abs/2006ApJ...648L.101E}
}

@ARTICLE{Gonzalez-Martin2017,
       author = {{Gonz{\'a}lez-Mart{\'\i}n}, Omaira and {Masegosa}, Josefa and {Hern{\'a}n-Caballero}, Antonio and {M{\'a}rquez}, Isabel and {Ramos Almeida}, Cristina and {Alonso-Herrero}, Almudena and {Aretxaga}, Itziar and {Rodr{\'\i}guez-Espinosa}, Jos{\'e} Miguel and {Acosta-Pulido}, Jose Antonio and {Hern{\'a}ndez-Garc{\'\i}a}, Lorena and {Esparza-Arredondo}, Donaji and {Mart{\'\i}nez-Paredes}, Mariela and {Bonfini}, Paolo and {Pasetto}, Alice and {Dultzin}, Deborah},
        title = "{Hints on the Gradual Resizing of the Torus in AGNs through Decomposition of Spitzer/IRS Spectra}",
      journal = {\apj},
         year = 2017,
        month = may,
       volume = {841},
       number = {1},
          eid = {37},
        pages = {37},
          doi = {10.3847/1538-4357/aa6f16},
archivePrefix = {arXiv},
       eprint = {1704.06739},
 primaryClass = {astro-ph.GA},
       adsurl = {https://ui.adsabs.harvard.edu/abs/2017ApJ...841...37G}
}

@ARTICLE{Baumgartner2013,
       author = {{Baumgartner}, W.~H. and {Tueller}, J. and {Markwardt}, C.~B. and {Skinner}, G.~K. and {Barthelmy}, S. and {Mushotzky}, R.~F. and {Evans}, P.~A. and {Gehrels}, N.},
        title = "{The 70 Month Swift-BAT All-sky Hard X-Ray Survey}",
      journal = {\apjs},
         year = 2013,
        month = aug,
       volume = {207},
       number = {2},
          eid = {19},
        pages = {19},
          doi = {10.1088/0067-0049/207/2/19},
archivePrefix = {arXiv},
       eprint = {1212.3336},
 primaryClass = {astro-ph.HE},
       adsurl = {https://ui.adsabs.harvard.edu/abs/2013ApJS..207...19B}
}

@ARTICLE{Assef2013,
       author = {{Assef}, R.~J. and {Stern}, D. and {Kochanek}, C.~S. and {Blain}, A.~W. and {Brodwin}, M. and {Brown}, M.~J.~I. and {Donoso}, E. and {Eisenhardt}, P.~R.~M. and {Jannuzi}, B.~T. and {Jarrett}, T.~H. and {Stanford}, S.~A. and {Tsai}, C. -W. and {Wu}, J. and {Yan}, L.},
        title = "{Mid-infrared Selection of Active Galactic Nuclei with the Wide-field Infrared Survey Explorer. II. Properties of WISE-selected Active Galactic Nuclei in the NDWFS Bo{\"o}tes Field}",
      journal = {\apj},
         year = 2013,
        month = jul,
       volume = {772},
       number = {1},
          eid = {26},
        pages = {26},
          doi = {10.1088/0004-637X/772/1/26},
archivePrefix = {arXiv},
       eprint = {1209.6055},
 primaryClass = {astro-ph.CO},
       adsurl = {https://ui.adsabs.harvard.edu/abs/2013ApJ...772...26A}
}

@ARTICLE{Goulding2009,
       author = {{Goulding}, A.~D. and {Alexander}, D.~M.},
        title = "{Towards a complete census of AGN in nearby Galaxies: a large population of optically unidentified AGN}",
      journal = {\mnras},
         year = 2009,
        month = sep,
       volume = {398},
       number = {3},
        pages = {1165-1193},
          doi = {10.1111/j.1365-2966.2009.15194.x},
archivePrefix = {arXiv},
       eprint = {0906.0772},
 primaryClass = {astro-ph.CO},
       adsurl = {https://ui.adsabs.harvard.edu/abs/2009MNRAS.398.1165G}
}

@ARTICLE{Beckmann2009,
       author = {{Beckmann}, V. and {Soldi}, S. and {Ricci}, C. and {Alfonso-Garz{\'o}n}, J. and {Courvoisier}, T.~J. -L. and {Domingo}, A. and {Gehrels}, N. and {Lubi{\'n}ski}, P. and {Mas-Hesse}, J.~M. and {Zdziarski}, A.~A.},
        title = "{The second INTEGRAL AGN catalogue}",
      journal = {\aap},
         year = 2009,
        month = oct,
       volume = {505},
       number = {1},
        pages = {417-439},
          doi = {10.1051/0004-6361/200912111},
archivePrefix = {arXiv},
       eprint = {0907.0654},
 primaryClass = {astro-ph.CO},
       adsurl = {https://ui.adsabs.harvard.edu/abs/2009A&A...505..417B}
}

@ARTICLE{Rigby2006,
       author = {{Rigby}, J.~R. and {Rieke}, G.~H. and {Donley}, J.~L. and {Alonso-Herrero}, A. and {P{\'e}rez-Gonz{\'a}lez}, P.~G.},
        title = "{Why X-Ray-selected Active Galactic Nuclei Appear Optically Dull}",
      journal = {\apj},
         year = 2006,
        month = jul,
       volume = {645},
       number = {1},
        pages = {115-133},
          doi = {10.1086/504067},
archivePrefix = {arXiv},
       eprint = {astro-ph/0603313},
 primaryClass = {astro-ph},
       adsurl = {https://ui.adsabs.harvard.edu/abs/2006ApJ...645..115R}
}

@ARTICLE{Shao2017,
       author = {{Shao}, Zhenzhen and {Jiang}, B.~W. and {Li}, Aigen},
        title = "{On the Optical-to-silicate Extinction Ratio as a Probe of the Dust Size in Active Galactic Nuclei}",
      journal = {\apj},
         year = 2017,
        month = may,
       volume = {840},
       number = {1},
          eid = {27},
        pages = {27},
          doi = {10.3847/1538-4357/aa6ba4},
archivePrefix = {arXiv},
       eprint = {1704.01103},
 primaryClass = {astro-ph.GA},
       adsurl = {https://ui.adsabs.harvard.edu/abs/2017ApJ...840...27S}
}

@ARTICLE{Mullaney2011,
       author = {{Mullaney}, J.~R. and {Alexander}, D.~M. and {Goulding}, A.~D. and {Hickox}, R.~C.},
        title = "{Defining the intrinsic AGN infrared spectral energy distribution and measuring its contribution to the infrared output of composite galaxies}",
      journal = {\mnras},
         year = 2011,
        month = jun,
       volume = {414},
       number = {2},
        pages = {1082-1110},
          doi = {10.1111/j.1365-2966.2011.18448.x},
archivePrefix = {arXiv},
       eprint = {1102.1425},
 primaryClass = {astro-ph.CO},
       adsurl = {https://ui.adsabs.harvard.edu/abs/2011MNRAS.414.1082M}
}

@ARTICLE{Satyapal2014,
       author = {{Satyapal}, S. and {Secrest}, N.~J. and {McAlpine}, W. and {Ellison}, S.~L. and {Fischer}, J. and {Rosenberg}, J.~L.},
        title = "{Discovery of a Population of Bulgeless Galaxies with Extremely Red Mid-IR Colors: Obscured AGN Activity in the Low-mass Regime?}",
      journal = {\apj},
         year = 2014,
        month = apr,
       volume = {784},
       number = {2},
          eid = {113},
        pages = {113},
          doi = {10.1088/0004-637X/784/2/113},
archivePrefix = {arXiv},
       eprint = {1401.5483},
 primaryClass = {astro-ph.GA},
       adsurl = {https://ui.adsabs.harvard.edu/abs/2014ApJ...784..113S}
}

@ARTICLE{Kaviraj2019,
       author = {{Kaviraj}, Sugata and {Martin}, Garreth and {Silk}, Joseph},
        title = "{AGN in dwarf galaxies: frequency, triggering processes and the plausibility of AGN feedback}",
      journal = {\mnras},
         year = 2019,
        month = oct,
       volume = {489},
       number = {1},
        pages = {L12-L16},
          doi = {10.1093/mnrasl/slz102},
archivePrefix = {arXiv},
       eprint = {1906.10697},
 primaryClass = {astro-ph.GA},
       adsurl = {https://ui.adsabs.harvard.edu/abs/2019MNRAS.489L..12K}
}

@ARTICLE{Brandt2005,
       author = {{Brandt}, W.~N. and {Hasinger}, G.},
        title = "{Deep Extragalactic X-Ray Surveys}",
      journal = {\araa},
         year = 2005,
        month = sep,
       volume = {43},
       number = {1},
        pages = {827-859},
          doi = {10.1146/annurev.astro.43.051804.102213},
archivePrefix = {arXiv},
       eprint = {astro-ph/0501058},
 primaryClass = {astro-ph},
       adsurl = {https://ui.adsabs.harvard.edu/abs/2005ARA&A..43..827B}
}

@ARTICLE{Ichikawa2019,
       author = {{Ichikawa}, Kohei and {Ricci}, Claudio and {Ueda}, Yoshihiro and {Bauer}, Franz E. and {Kawamuro}, Taiki and {Koss}, Michael J. and {Oh}, Kyuseok and {Rosario}, David J. and {Shimizu}, T. Taro and {Stalevski}, Marko and {Fuller}, Lindsay and {Packham}, Christopher and {Trakhtenbrot}, Benny},
        title = "{BAT AGN Spectroscopic Survey. XI. The Covering Factor of Dust and Gas in Swift/BAT Active Galactic Nuclei}",
      journal = {\apj},
         year = 2019,
        month = jan,
       volume = {870},
       number = {1},
          eid = {31},
        pages = {31},
          doi = {10.3847/1538-4357/aaef8f},
archivePrefix = {arXiv},
       eprint = {1811.02568},
 primaryClass = {astro-ph.GA},
       adsurl = {https://ui.adsabs.harvard.edu/abs/2019ApJ...870...31I}
}

@ARTICLE{Masoura2020,
       author = {{Masoura}, V.~A. and {Georgantopoulos}, I. and {Mountrichas}, G. and {Vignali}, C. and {Koulouridis}, E. and {Chiappetti}, L. and {Fotopoulou}, S. and {Paltani}, S. and {Pierre}, M.},
        title = "{The XXL Survey. XL. Obscuration properties of red AGNs in XXL-N}",
      journal = {\aap},
         year = 2020,
        month = jun,
       volume = {638},
          eid = {A45},
        pages = {A45},
          doi = {10.1051/0004-6361/201937161},
archivePrefix = {arXiv},
       eprint = {2003.04346},
 primaryClass = {astro-ph.GA},
       adsurl = {https://ui.adsabs.harvard.edu/abs/2020A&A...638A..45M}
}

@ARTICLE{Meulen2023,
       author = {{Vander Meulen}, Bert and {Camps}, Peter and {Stalevski}, Marko and {Baes}, Maarten},
        title = "{X-ray radiative transfer in full 3D with SKIRT}",
      journal = {\aap},
         year = 2023,
        month = jun,
       volume = {674},
          eid = {A123},
        pages = {A123},
          doi = {10.1051/0004-6361/202245783},
archivePrefix = {arXiv},
       eprint = {2304.10563},
 primaryClass = {astro-ph.HE},
       adsurl = {https://ui.adsabs.harvard.edu/abs/2023A&A...674A.123V}
}

@ARTICLE{Fruscione2007,
       author = {{Fruscione}, Antonella and {McDowell}, Jonathan C. and {Allen}, Glenn E. and {Brickhouse}, Nancy S. and {Burke}, Douglas J. and {Davis}, John E. and {Durham}, Nick and {Elvis}, Martin and {Galle}, Elizabeth C. and {Harris}, Daniel E. and {Huenemoerder}, David P. and {Houck}, John C. and {Ishibashi}, Bish and {Karovska}, Margarita and {Nicastro}, Fabrizio and {Nowak}, Michael S. Noble. Michael A. and {Primini}, Frank A. and {Siemiginowska}, Aneta and {Smith}, Randall K. and {Wise}, Michael},
        title = "{CIAO: Chandra's Data Analysis System}",
      journal = {Chandra News},
         year = 2007,
        month = mar,
       volume = {14},
        pages = {36},
       adsurl = {https://ui.adsabs.harvard.edu/abs/2007ChNew..14...36F}
}

@INPROCEEDINGS{Joye2003,
       author = {{Joye}, W.~A. and {Mandel}, E.},
        title = "{New Features of SAOImage DS9}",
    booktitle = {Astronomical Data Analysis Software and Systems XII.  Astron. Soc. Pac., San Francisco,},
         year = 2003,
       editor = {{Payne}, H.~E. and {Jedrzejewski}, R.~I. and {Hook}, R.~N.},
       series = {ASP Conf. Ser.},
       volume = {295},
        month = jan,
        pages = {489},
       adsurl = {https://ui.adsabs.harvard.edu/abs/2003ASPC..295..489J}
}

@ARTICLE{Mateos2012,
       author = {{Mateos}, S. and {Alonso-Herrero}, A. and {Carrera}, F.~J. and {Blain}, A. and {Watson}, M.~G. and {Barcons}, X. and {Braito}, V. and {Severgnini}, P. and {Donley}, J.~L. and {Stern}, D.},
        title = "{Using the Bright Ultrahard XMM-Newton survey to define an IR selection of luminous AGN based on WISE colours}",
      journal = {\mnras},
         year = 2012,
        month = nov,
       volume = {426},
       number = {4},
        pages = {3271-3281},
          doi = {10.1111/j.1365-2966.2012.21843.x},
archivePrefix = {arXiv},
       eprint = {1208.2530},
 primaryClass = {astro-ph.CO},
       adsurl = {https://ui.adsabs.harvard.edu/abs/2012MNRAS.426.3271M}
}

@ARTICLE{Iwasawa2011,
       author = {{Iwasawa}, K. and {Mazzarella}, J.~M. and {Surace}, J.~A. and {Sanders}, D.~B. and {Armus}, L. and {Evans}, A.~S. and {Howell}, J.~H. and {Komossa}, S. and {Petric}, A. and {Teng}, S.~H. and {U}, Vivian and {Veilleux}, S.},
        title = "{The location of an active nucleus and a shadow of a tidal tail in the ULIRG Mrk 273}",
      journal = {\aap},
         year = 2011,
        month = apr,
       volume = {528},
          eid = {A137},
        pages = {A137},
          doi = {10.1051/0004-6361/201015872},
archivePrefix = {arXiv},
       eprint = {1101.3659},
 primaryClass = {astro-ph.CO},
       adsurl = {https://ui.adsabs.harvard.edu/abs/2011A&A...528A.137I}
}

@ARTICLE{Izotov2012,
       author = {{Izotov}, Y.~I. and {Thuan}, T.~X. and {Privon}, G.},
        title = "{The detection of [Ne V] emission in five blue compact dwarf galaxies}",
      journal = {\mnras},
         year = 2012,
        month = dec,
       volume = {427},
       number = {2},
        pages = {1229-1237},
          doi = {10.1111/j.1365-2966.2012.22051.x},
archivePrefix = {arXiv},
       eprint = {1209.5265},
 primaryClass = {astro-ph.CO},
       adsurl = {https://ui.adsabs.harvard.edu/abs/2012MNRAS.427.1229I}
}

@ARTICLE{Spoon2022,
       author = {{Spoon}, H.~W.~W. and {Hern{\'a}n-Caballero}, A. and {Rupke}, D. and {Waters}, L.~B.~F.~M. and {Lebouteiller}, V. and {Tielens}, A.~G.~G.~M. and {Loredo}, T. and {Su}, Y. and {Viola}, V.},
        title = "{The Infrared Database of Extragalactic Observables from Spitzer. II. The Database and Diagnostic Power of Crystalline Silicate Features in Galaxy Spectra}",
      journal = {\apjs},
         year = 2022,
        month = apr,
       volume = {259},
       number = {2},
          eid = {37},
        pages = {37},
          doi = {10.3847/1538-4365/ac4989},
       adsurl = {https://ui.adsabs.harvard.edu/abs/2022ApJS..259...37S}
}

@ARTICLE{Baldwin1981,
       author = {{Baldwin}, J.~A. and {Phillips}, M.~M. and {Terlevich}, R.},
        title = "{Classification parameters for the emission-line spectra of extragalactic objects.}",
      journal = {\pasp},
         year = 1981,
        month = feb,
       volume = {93},
        pages = {5-19},
          doi = {10.1086/130766},
       adsurl = {https://ui.adsabs.harvard.edu/abs/1981PASP...93....5B}
}

@ARTICLE{Ho1997,
       author = {{Ho}, Luis C. and {Filippenko}, Alexei V. and {Sargent}, Wallace L.~W.},
        title = "{A Search for ``Dwarf'' Seyfert Nuclei. III. Spectroscopic Parameters and Properties of the Host Galaxies}",
      journal = {\apjs},
         year = 1997,
        month = oct,
       volume = {112},
       number = {2},
        pages = {315-390},
          doi = {10.1086/313041},
archivePrefix = {arXiv},
       eprint = {astro-ph/9704107},
 primaryClass = {astro-ph},
       adsurl = {https://ui.adsabs.harvard.edu/abs/1997ApJS..112..315H}
}

@ARTICLE{Veilleux1987,
       author = {{Veilleux}, Sylvain and {Osterbrock}, Donald E.},
        title = "{Spectral Classification of Emission-Line Galaxies}",
      journal = {\apjs},
         year = 1987,
        month = feb,
       volume = {63},
        pages = {295},
          doi = {10.1086/191166},
       adsurl = {https://ui.adsabs.harvard.edu/abs/1987ApJS...63..295V}
}

@ARTICLE{Kewley2006,
       author = {{Kewley}, Lisa J. and {Groves}, Brent and {Kauffmann}, Guinevere and {Heckman}, Tim},
        title = "{The host galaxies and classification of active galactic nuclei}",
      journal = {\mnras},
         year = 2006,
        month = nov,
       volume = {372},
       number = {3},
        pages = {961-976},
          doi = {10.1111/j.1365-2966.2006.10859.x},
archivePrefix = {arXiv},
       eprint = {astro-ph/0605681},
 primaryClass = {astro-ph},
       adsurl = {https://ui.adsabs.harvard.edu/abs/2006MNRAS.372..961K}
}

@ARTICLE{Kewley2001,
       author = {{Kewley}, L.~J. and {Dopita}, M.~A. and {Sutherland}, R.~S. and {Heisler}, C.~A. and {Trevena}, J.},
        title = "{Theoretical Modeling of Starburst Galaxies}",
      journal = {\apj},
         year = 2001,
        month = jul,
       volume = {556},
       number = {1},
        pages = {121-140},
          doi = {10.1086/321545},
archivePrefix = {arXiv},
       eprint = {astro-ph/0106324},
 primaryClass = {astro-ph},
       adsurl = {https://ui.adsabs.harvard.edu/abs/2001ApJ...556..121K}
}

@ARTICLE{Kauffmann2003,
       author = {{Kauffmann}, Guinevere and {Heckman}, Timothy M. and {Tremonti}, Christy and {Brinchmann}, Jarle and {Charlot}, St{\'e}phane and {White}, Simon D.~M. and {Ridgway}, Susan E. and {Brinkmann}, Jon and {Fukugita}, Masataka and {Hall}, Patrick B. and {Ivezi{\'c}}, {\v{Z}}eljko and {Richards}, Gordon T. and {Schneider}, Donald P.},
        title = "{The host galaxies of active galactic nuclei}",
      journal = {\mnras},
         year = 2003,
        month = dec,
       volume = {346},
       number = {4},
        pages = {1055-1077},
          doi = {10.1111/j.1365-2966.2003.07154.x},
archivePrefix = {arXiv},
       eprint = {astro-ph/0304239},
 primaryClass = {astro-ph},
       adsurl = {https://ui.adsabs.harvard.edu/abs/2003MNRAS.346.1055K}
}

@ARTICLE{Asmus2011,
       author = {{Asmus}, D. and {Gandhi}, P. and {Smette}, A. and {H{\"o}nig}, S.~F. and {Duschl}, W.~J.},
        title = "{Mid-infrared properties of nearby low-luminosity AGN at high angular resolution}",
      journal = {\aap},
         year = 2011,
        month = dec,
       volume = {536},
          eid = {A36},
        pages = {A36},
          doi = {10.1051/0004-6361/201116693},
archivePrefix = {arXiv},
       eprint = {1109.4873},
 primaryClass = {astro-ph.CO},
       adsurl = {https://ui.adsabs.harvard.edu/abs/2011A&A...536A..36A}
}

@ARTICLE{Zhang2009,
       author = {{Zhang}, Wei Ming and {Soria}, Roberto and {Zhang}, Shuang Nan and {Swartz}, Douglas A. and {Liu}, Ji Feng},
        title = "{A Census of X-ray Nuclear Activity in Nearby Galaxies}",
      journal = {\apj},
         year = 2009,
        month = jul,
       volume = {699},
       number = {1},
        pages = {281-297},
          doi = {10.1088/0004-637X/699/1/281},
archivePrefix = {arXiv},
       eprint = {0904.1091},
 primaryClass = {astro-ph.HE},
       adsurl = {https://ui.adsabs.harvard.edu/abs/2009ApJ...699..281Z}
}

@ARTICLE{Birchall2022,
       author = {{Birchall}, Keir L. and {Watson}, M.~G. and {Aird}, J. and {Starling}, R.~L.~C.},
        title = "{The incidence of X-ray selected AGN in nearby galaxies}",
      journal = {\mnras},
         year = 2022,
        month = mar,
       volume = {510},
       number = {3},
        pages = {4556-4572},
          doi = {10.1093/mnras/stab3573},
archivePrefix = {arXiv},
       eprint = {2112.03142},
 primaryClass = {astro-ph.GA},
       adsurl = {https://ui.adsabs.harvard.edu/abs/2022MNRAS.510.4556B}
}

@ARTICLE{Yuk2025,
       author = {{Yuk}, Heechan and {Dai}, Xinyu and {Mi{\'c}i{\'c}}, Marko},
        title = "{Multiwavelength and Environmental Properties of Variability Selected Low-luminosity Active Galactic Nuclei}",
      journal = {\apj},
         year = 2025,
        month = jul,
       volume = {987},
       number = {2},
          eid = {112},
        pages = {112},
          doi = {10.3847/1538-4357/addf42},
archivePrefix = {arXiv},
       eprint = {2501.17844},
 primaryClass = {astro-ph.GA},
       adsurl = {https://ui.adsabs.harvard.edu/abs/2025ApJ...987..112Y}
}

@ARTICLE{Koss2017,
       author = {{Koss}, Michael and {Trakhtenbrot}, Benny and {Ricci}, Claudio and {Lamperti}, Isabella and {Oh}, Kyuseok and {Berney}, Simon and {Schawinski}, Kevin and {Balokovi{\'c}}, Mislav and {Baronchelli}, Linda and {Crenshaw}, D. Michael and {Fischer}, Travis and {Gehrels}, Neil and {Harrison}, Fiona and {Hashimoto}, Yasuhiro and {Hogg}, Drew and {Ichikawa}, Kohei and {Masetti}, Nicola and {Mushotzky}, Richard and {Sartori}, Lia and {Stern}, Daniel and {Treister}, Ezequiel and {Ueda}, Yoshihiro and {Veilleux}, Sylvain and {Winter}, Lisa},
        title = "{BAT AGN Spectroscopic Survey. I. Spectral Measurements, Derived Quantities, and AGN Demographics}",
      journal = {\apj},
         year = 2017,
        month = nov,
       volume = {850},
       number = {1},
          eid = {74},
        pages = {74},
          doi = {10.3847/1538-4357/aa8ec9},
archivePrefix = {arXiv},
       eprint = {1707.08123},
 primaryClass = {astro-ph.HE},
       adsurl = {https://ui.adsabs.harvard.edu/abs/2017ApJ...850...74K}
}

@ARTICLE{Vasudevan2010,
       author = {{Vasudevan}, R.~V. and {Fabian}, A.~C. and {Gandhi}, P. and {Winter}, L.~M. and {Mushotzky}, R.~F.},
        title = "{The power output of local obscured and unobscured AGN: crossing the absorption barrier with Swift/BAT and IRAS}",
      journal = {\mnras},
         year = 2010,
        month = feb,
       volume = {402},
       number = {2},
        pages = {1081-1098},
          doi = {10.1111/j.1365-2966.2009.15936.x},
archivePrefix = {arXiv},
       eprint = {0910.5256},
 primaryClass = {astro-ph.GA},
       adsurl = {https://ui.adsabs.harvard.edu/abs/2010MNRAS.402.1081V}
}

@ARTICLE{Kormendy2013,
       author = {{Kormendy}, John and {Ho}, Luis C.},
        title = "{Coevolution (Or Not) of Supermassive Black Holes and Host Galaxies}",
      journal = {\araa},
         year = 2013,
        month = aug,
       volume = {51},
       number = {1},
        pages = {511-653},
          doi = {10.1146/annurev-astro-082708-101811},
archivePrefix = {arXiv},
       eprint = {1304.7762},
 primaryClass = {astro-ph.CO},
       adsurl = {https://ui.adsabs.harvard.edu/abs/2013ARA&A..51..511K}
}

@ARTICLE{Heckman2014,
       author = {{Heckman}, Timothy M. and {Best}, Philip N.},
        title = "{The Coevolution of Galaxies and Supermassive Black Holes: Insights from Surveys of the Contemporary Universe}",
      journal = {\araa},
         year = 2014,
        month = aug,
       volume = {52},
        pages = {589-660},
          doi = {10.1146/annurev-astro-081913-035722},
archivePrefix = {arXiv},
       eprint = {1403.4620},
 primaryClass = {astro-ph.GA},
       adsurl = {https://ui.adsabs.harvard.edu/abs/2014ARA&A..52..589H}
}

@inproceedings{Mushotzky2004,
       author = {{Mushotzky}, R.},
        title = "{How are AGN Found?}",
    booktitle = {Supermassive Black Holes in the Distant Universe},
         year = 2004,
       editor = {{Barger}, A.~J.},
       series = {Astrophysics and Space Science Library},
       volume = {308},
        month = aug,
        pages = {53}
        }

@ARTICLE{Tozzi2006,
       author = {{Tozzi}, P. and {Gilli}, R. and {Mainieri}, V. and {Norman}, C. and {Risaliti}, G. and {Rosati}, P. and {Bergeron}, J. and {Borgani}, S. and {Giacconi}, R. and {Hasinger}, G. and {Nonino}, M. and {Streblyanska}, A. and {Szokoly}, G. and {Wang}, J.~X. and {Zheng}, W.},
        title = "{X-ray spectral properties of active galactic nuclei in the Chandra Deep Field South}",
      journal = {\aap},
         year = 2006,
        month = may,
       volume = {451},
       number = {2},
        pages = {457-474},
          doi = {10.1051/0004-6361:20042592},
archivePrefix = {arXiv},
       eprint = {astro-ph/0602127},
 primaryClass = {astro-ph},
       adsurl = {https://ui.adsabs.harvard.edu/abs/2006A&A...451..457T}
}

@ARTICLE{Hickox2018,
       author = {{Hickox}, Ryan C. and {Alexander}, David M.},
        title = "{Obscured Active Galactic Nuclei}",
      journal = {\araa},
         year = 2018,
        month = sep,
       volume = {56},
        pages = {625-671},
          doi = {10.1146/annurev-astro-081817-051803},
archivePrefix = {arXiv},
       eprint = {1806.04680},
 primaryClass = {astro-ph.GA},
       adsurl = {https://ui.adsabs.harvard.edu/abs/2018ARA&A..56..625H}
}

@ARTICLE{Mohanadas2023,
       author = {{Mohanadas}, Pavithra and {Annuar}, Adlyka},
        title = "{NGC 4117: A New Compton-thick AGN Revealed by Broadband X-Ray Spectral Analysis}",
      journal = {RAA},
         year = 2023,
        month = may,
       volume = {23},
       number = {5},
          eid = {055002},
        pages = {055002},
          doi = {10.1088/1674-4527/acc151},
       adsurl = {https://ui.adsabs.harvard.edu/abs/2023RAA....23e5002M}
}

@ARTICLE{Annuar2020,
       author = {{Annuar}, A. and {Alexander}, D.~M. and {Gandhi}, P. and {Lansbury}, G.~B. and {Asmus}, D. and {Balokovi{\'c}}, M. and {Ballantyne}, D.~R. and {Bauer}, F.~E. and {Boorman}, P.~G. and {Brandt}, W.~N. and {Brightman}, M. and {Chen}, C.-T.~J. and {Del Moro}, A. and {Farrah}, D. and {Harrison}, F.~A. and {Koss}, M.~J. and {Lanz}, L. and {Marchesi}, S. and {Masini}, A. and {Nardini}, E. and {Ricci}, C. and {Stern}, D. and {Zappacosta}, L.},
        title = "{NuSTAR observations of four nearby X-ray faint AGNs: low luminosity or heavy obscuration?}",
      journal = {\mnras},
         year = 2020,
        month = sep,
       volume = {497},
       number = {1},
        pages = {229-245},
          doi = {10.1093/mnras/staa1820},
archivePrefix = {arXiv},
       eprint = {2006.13583},
 primaryClass = {astro-ph.HE},
       adsurl = {https://ui.adsabs.harvard.edu/abs/2020MNRAS.497..229A}
}

@ARTICLE{Ho2009,
       author = {{Ho}, Luis C.},
        title = "{Radiatively Inefficient Accretion in Nearby Galaxies}",
      journal = {\apj},
         year = 2009,
        month = jul,
       volume = {699},
       number = {1},
        pages = {626-637},
          doi = {10.1088/0004-637X/699/1/626},
archivePrefix = {arXiv},
       eprint = {0906.4104},
 primaryClass = {astro-ph.GA},
       adsurl = {https://ui.adsabs.harvard.edu/abs/2009ApJ...699..626H}
}

@ARTICLE{Jana2023,
       author = {{Jana}, Arghajit and {Chatterjee}, Arka and {Chang}, Hsiang-Kuang and {Nandi}, Prantik and {Rubinur}, K. and {Kumari}, Neeraj and {Naik}, Sachindra and {Safi-Harb}, Samar and {Ricci}, Claudio},
        title = "{Coronal properties of low-accreting AGNs using Swift, XMM-Newton, and NuSTAR observations}",
      journal = {\mnras},
         year = 2023,
        month = sep,
       volume = {524},
       number = {3},
        pages = {4670-4687},
          doi = {10.1093/mnras/stad2140},
archivePrefix = {arXiv},
       eprint = {2307.07966},
 primaryClass = {astro-ph.HE},
       adsurl = {https://ui.adsabs.harvard.edu/abs/2023MNRAS.524.4670J}
}

@ARTICLE{Gu2009,
       author = {{Gu}, Minfeng and {Cao}, Xinwu},
        title = "{The anticorrelation between the hard X-ray photon index and the Eddington ratio in low-luminosity active galactic nuclei}",
      journal = {\mnras},
         year = 2009,
        month = oct,
       volume = {399},
       number = {1},
        pages = {349-356},
          doi = {10.1111/j.1365-2966.2009.15277.x},
archivePrefix = {arXiv},
       eprint = {0906.3560},
 primaryClass = {astro-ph.GA},
       adsurl = {https://ui.adsabs.harvard.edu/abs/2009MNRAS.399..349G}
}

@ARTICLE{Ho1995,
       author = {{Ho}, L.~C. and {Filippenko}, A.~V. and {Sargent}, W.~L.},
        title = "{A Search for ``Dwarf'' Seyfert Nuclei. II. an Optical Spectral Atlas of the Nuclei of Nearby Galaxies}",
      journal = {\apjs},
         year = 1995,
        month = jun,
       volume = {98},
        pages = {477},
          doi = {10.1086/192170},
       adsurl = {https://ui.adsabs.harvard.edu/abs/1995ApJS...98..477H}
}

@ARTICLE{Flohic2006,
       author = {{Flohic}, H{\'e}l{\`e}ne M.~L.~G. and {Eracleous}, Michael and {Chartas}, George and {Shields}, Joseph C. and {Moran}, Edward C.},
        title = "{The Central Engines of 19 LINERs as Viewed by Chandra}",
      journal = {\apj},
         year = 2006,
        month = aug,
       volume = {647},
       number = {1},
        pages = {140-160},
          doi = {10.1086/505296},
archivePrefix = {arXiv},
       eprint = {astro-ph/0604487},
 primaryClass = {astro-ph},
       adsurl = {https://ui.adsabs.harvard.edu/abs/2006ApJ...647..140F}
}

@ARTICLE{Ho2008,
       author = {{Ho}, L.~C.},
        title = "{Nuclear activity in nearby galaxies.}",
      journal = {\araa},
         year = 2008,
        month = sep,
       volume = {46},
        pages = {475-539},
          doi = {10.1146/annurev.astro.45.051806.110546},
archivePrefix = {arXiv},
       eprint = {0803.2268},
 primaryClass = {astro-ph},
       adsurl = {https://ui.adsabs.harvard.edu/abs/2008ARA&A..46..475H}
}

@ARTICLE{Yu-wei2025,
       author = {{Yu}, Yu-Wei and {Zhang}, Jin},
        title = "{Hard X-ray view of two {\ensuremath{\gamma}}-ray-detected low-luminosity active galactic nuclei: NGC 315 and NGC 4261}",
      journal = {\aap},
         year = 2025,
        month = nov,
       volume = {703},
          eid = {A272},
        pages = {A272},
          doi = {10.1051/0004-6361/202556509},
       adsurl = {https://ui.adsabs.harvard.edu/abs/2025A&A...703A.272Y}
}

@ARTICLE{Narayan1994,
       author = {{Narayan}, Ramesh and {Yi}, Insu},
        title = "{Advection-dominated Accretion: A Self-similar Solution}",
      journal = {\apjl},
         year = 1994,
        month = jun,
       volume = {428},
        pages = {L13},
          doi = {10.1086/187381},
archivePrefix = {arXiv},
       eprint = {astro-ph/9403052},
 primaryClass = {astro-ph},
       adsurl = {https://ui.adsabs.harvard.edu/abs/1994ApJ...428L..13N}
}

@ARTICLE{Narayan1995,
       author = {{Narayan}, Ramesh and {Yi}, Insu},
        title = "{Advection-dominated Accretion: Underfed Black Holes and Neutron Stars}",
      journal = {\apj},
         year = 1995,
        month = oct,
       volume = {452},
        pages = {710},
          doi = {10.1086/176343},
archivePrefix = {arXiv},
       eprint = {astro-ph/9411059},
 primaryClass = {astro-ph},
       adsurl = {https://ui.adsabs.harvard.edu/abs/1995ApJ...452..710N}
}

@INPROCEEDINGS{Arnaud1996,
       author = {{Arnaud}, K.~A.},
        title = "{XSPEC: The First Ten Years}",
    booktitle = {Astronomical Data Analysis Software and Systems V. Astron. Soc. Pac., San Francisco},
         year = 1996,
       editor = {{Jacoby}, George H. and {Barnes}, Jeannette},
       series = {ASP Conf. Ser.},
       volume = {101},
        month = jan,
        pages = {17},
       adsurl = {https://ui.adsabs.harvard.edu/abs/1996ASPC..101...17A}
}

@ARTICLE{Kammoun2019,
       author = {{Kammoun}, E.~S. and {Nardini}, E. and {Zoghbi}, A. and {Miller}, J.~M. and {Cackett}, E.~M. and {Gallo}, E. and {Reynolds}, M.~T. and {Risaliti}, G. and {Barret}, D. and {Brandt}, W.~N. and {Brenneman}, L.~W. and {Kaastra}, J.~S. and {Koss}, M. and {Lohfink}, A.~M. and {Mushotzky}, R.~F. and {Raymond}, J. and {Stern}, D.},
        title = "{The Nature of the Broadband X-Ray Variability in the Dwarf Seyfert Galaxy NGC 4395}",
      journal = {\apj},
         year = 2019,
        month = dec,
       volume = {886},
       number = {2},
          eid = {145},
        pages = {145},
          doi = {10.3847/1538-4357/ab5110},
archivePrefix = {arXiv},
       eprint = {1910.11317},
 primaryClass = {astro-ph.HE},
       adsurl = {https://ui.adsabs.harvard.edu/abs/2019ApJ...886..145K}
}

@ARTICLE{Xu2016,
       author = {{Xu}, Weiwei and {Liu}, Zhu and {Gou}, Lijun and {Liu}, Jiren},
        title = "{X-ray fluorescent lines from the Compton-thick AGN in M51}",
      journal = {\mnras},
         year = 2016,
        month = jan,
       volume = {455},
       number = {1},
        pages = {L26-L30},
          doi = {10.1093/mnrasl/slv148},
archivePrefix = {arXiv},
       eprint = {1510.03123},
 primaryClass = {astro-ph.HE},
       adsurl = {https://ui.adsabs.harvard.edu/abs/2016MNRAS.455L..26X}
}

@ARTICLE{Negus2023,
       author = {{Negus}, James and {Comerford}, Julia M. and {S{\'a}nchez}, Francisco M{\"u}ller and {Revalski}, Mitchell and {Riffel}, Rogemar A. and {Bundy}, Kevin and {Nevin}, Rebecca and {Rembold}, Sandro B.},
        title = "{A Catalog of 71 Coronal Line Galaxies in MaNGA: [Ne V] Is an Effective AGN Tracer}",
      journal = {\apj},
         year = 2023,
        month = mar,
       volume = {945},
       number = {2},
          eid = {127},
        pages = {127},
          doi = {10.3847/1538-4357/acb772},
archivePrefix = {arXiv},
       eprint = {2301.13322},
 primaryClass = {astro-ph.GA},
       adsurl = {https://ui.adsabs.harvard.edu/abs/2023ApJ...945..127N}
}

@ARTICLE{Weedman2005,
       author = {{Weedman}, D.~W. and {Hao}, Lei and {Higdon}, S.~J.~U. and {Devost}, D. and {Wu}, Yanling and {Charmandaris}, V. and {Brandl}, B. and {Bass}, E. and {Houck}, J.~R.},
        title = "{Mid-Infrared Spectra of Classical AGNs Observed with the Spitzer Space Telescope}",
      journal = {\apj},
         year = 2005,
        month = nov,
       volume = {633},
       number = {2},
        pages = {706-716},
          doi = {10.1086/466520},
archivePrefix = {arXiv},
       eprint = {astro-ph/0507423},
 primaryClass = {astro-ph},
       adsurl = {https://ui.adsabs.harvard.edu/abs/2005ApJ...633..706W}
}

@ARTICLE{Bernard-Salas2009,
       author = {{Bernard-Salas}, J. and {Spoon}, H.~W.~W. and {Charmandaris}, V. and {Lebouteiller}, V. and {Farrah}, D. and {Devost}, D. and {Brandl}, B.~R. and {Wu}, Yanling and {Armus}, L. and {Hao}, L. and {Sloan}, G.~C. and {Weedman}, D. and {Houck}, J.~R.},
        title = "{A Spitzer High-resolution Mid-Infrared Spectral Atlas of Starburst Galaxies}",
      journal = {\apjs},
         year = 2009,
        month = oct,
       volume = {184},
       number = {2},
        pages = {230-247},
          doi = {10.1088/0067-0049/184/2/230},
archivePrefix = {arXiv},
       eprint = {0908.2812},
 primaryClass = {astro-ph.CO},
       adsurl = {https://ui.adsabs.harvard.edu/abs/2009ApJS..184..230B}
}

@ARTICLE{Awaki2001,
       author = {{Awaki}, Hisamitsu and {Terashima}, Yuichi and {Hayashida}, Kiyoshi and {Sakano}, Masaaki},
        title = "{Estimation of Central Black Hole Masses in Low-Luminosity Active Galactic Nuclei}",
      journal = {\pasj},
         year = 2001,
        month = aug,
       volume = {53},
       number = {4},
        pages = {647-652},
          doi = {10.1093/pasj/53.4.647},
       adsurl = {https://ui.adsabs.harvard.edu/abs/2001PASJ...53..647A}
}

@ARTICLE{Shen2011,
       author = {{Shen}, Yue and {Richards}, Gordon T. and {Strauss}, Michael A. and {Hall}, Patrick B. and {Schneider}, Donald P. and {Snedden}, Stephanie and {Bizyaev}, Dmitry and {Brewington}, Howard and {Malanushenko}, Viktor and {Malanushenko}, Elena and {Oravetz}, Dan and {Pan}, Kaike and {Simmons}, Audrey},
        title = "{A Catalog of Quasar Properties from Sloan Digital Sky Survey Data Release 7}",
      journal = {\apjs},
         year = 2011,
        month = jun,
       volume = {194},
       number = {2},
          eid = {45},
        pages = {45},
          doi = {10.1088/0067-0049/194/2/45},
archivePrefix = {arXiv},
       eprint = {1006.5178},
 primaryClass = {astro-ph.CO},
       adsurl = {https://ui.adsabs.harvard.edu/abs/2011ApJS..194...45S}
}

@ARTICLE{McConnell2011,
       author = {{McConnell}, Nicholas J. and {Ma}, Chung-Pei and {Gebhardt}, Karl and {Wright}, Shelley A. and {Murphy}, Jeremy D. and {Lauer}, Tod R. and {Graham}, James R. and {Richstone}, Douglas O.},
        title = "{Two ten-billion-solar-mass black holes at the centres of giant elliptical galaxies}",
      journal = {\nat},
         year = 2011,
        month = dec,
       volume = {480},
       number = {7376},
        pages = {215-218},
          doi = {10.1038/nature10636},
archivePrefix = {arXiv},
       eprint = {1112.1078},
 primaryClass = {astro-ph.CO},
       adsurl = {https://ui.adsabs.harvard.edu/abs/2011Natur.480..215M}
}

@ARTICLE{Reines2015,
       author = {{Reines}, Amy E. and {Volonteri}, Marta},
        title = "{Relations between Central Black Hole Mass and Total Galaxy Stellar Mass in the Local Universe}",
      journal = {\apj},
         year = 2015,
        month = nov,
       volume = {813},
       number = {2},
          eid = {82},
        pages = {82},
          doi = {10.1088/0004-637X/813/2/82},
archivePrefix = {arXiv},
       eprint = {1508.06274},
 primaryClass = {astro-ph.GA},
       adsurl = {https://ui.adsabs.harvard.edu/abs/2015ApJ...813...82R}
}

@ARTICLE{Masegosa2011,
       author = {{Masegosa}, J. and {M{\'a}rquez}, I. and {Ramirez}, A. and {Gonz{\'a}lez-Mart{\'\i}n}, O.},
        title = "{The nature of nuclear H$_{{\ensuremath{\alpha}}}$ emission in LINERs}",
      journal = {\aap},
         year = 2011,
        month = mar,
       volume = {527},
          eid = {A23},
        pages = {A23},
          doi = {10.1051/0004-6361/201015047},
archivePrefix = {arXiv},
       eprint = {1011.0865},
 primaryClass = {astro-ph.CO},
       adsurl = {https://ui.adsabs.harvard.edu/abs/2011A&A...527A..23M}
}

@ARTICLE{Hopkins2006,
       author = {{Hopkins}, Philip F. and {Hernquist}, Lars},
        title = "{Fueling Low-Level AGN Activity through Stochastic Accretion of Cold Gas}",
      journal = {\apjs},
         year = 2006,
        month = sep,
       volume = {166},
       number = {1},
        pages = {1-36},
          doi = {10.1086/505753},
archivePrefix = {arXiv},
       eprint = {astro-ph/0603180},
 primaryClass = {astro-ph},
       adsurl = {https://ui.adsabs.harvard.edu/abs/2006ApJS..166....1H}
}

@ARTICLE{Davis2011,
       author = {{Davis}, Shane W. and {Laor}, Ari},
        title = "{The Radiative Efficiency of Accretion Flows in Individual Active Galactic Nuclei}",
      journal = {\apj},
         year = 2011,
        month = feb,
       volume = {728},
       number = {2},
          eid = {98},
        pages = {98},
          doi = {10.1088/0004-637X/728/2/98},
archivePrefix = {arXiv},
       eprint = {1012.3213},
 primaryClass = {astro-ph.CO},
       adsurl = {https://ui.adsabs.harvard.edu/abs/2011ApJ...728...98D}
}

@ARTICLE{Yuan2014,
       author = {{Yuan}, Feng and {Narayan}, Ramesh},
        title = "{Hot Accretion Flows Around Black Holes}",
      journal = {\araa},
         year = 2014,
        month = aug,
       volume = {52},
        pages = {529-588},
          doi = {10.1146/annurev-astro-082812-141003},
archivePrefix = {arXiv},
       eprint = {1401.0586},
 primaryClass = {astro-ph.HE},
       adsurl = {https://ui.adsabs.harvard.edu/abs/2014ARA&A..52..529Y}
}

@ARTICLE{Draper2010,
       author = {{Draper}, A.~R. and {Ballantyne}, D.~R.},
        title = "{The Evolution and Eddington Ratio Distribution of Compton Thick Active Galactic Nuclei}",
      journal = {\apjl},
         year = 2010,
        month = jun,
       volume = {715},
       number = {2},
        pages = {L99-L103},
          doi = {10.1088/2041-8205/715/2/L99},
archivePrefix = {arXiv},
       eprint = {1004.0690},
 primaryClass = {astro-ph.CO},
       adsurl = {https://ui.adsabs.harvard.edu/abs/2010ApJ...715L..99D}
}

@ARTICLE{Gehrels1986,
       author = {{Gehrels}, N.},
        title = "{Confidence Limits for Small Numbers of Events in Astrophysical Data}",
      journal = {\apj},
         year = 1986,
        month = apr,
       volume = {303},
        pages = {336},
          doi = {10.1086/164079},
       adsurl = {https://ui.adsabs.harvard.edu/abs/1986ApJ...303..336G}
}

@ARTICLE{Bell2003,
       author = {{Bell}, Eric F. and {McIntosh}, Daniel H. and {Katz}, Neal and {Weinberg}, Martin D.},
        title = "{The Optical and Near-Infrared Properties of Galaxies. I. Luminosity and Stellar Mass Functions}",
      journal = {\apjs},
         year = 2003,
        month = dec,
       volume = {149},
       number = {2},
        pages = {289-312},
          doi = {10.1086/378847},
archivePrefix = {arXiv},
       eprint = {astro-ph/0302543},
 primaryClass = {astro-ph},
       adsurl = {https://ui.adsabs.harvard.edu/abs/2003ApJS..149..289B}
}

@ARTICLE{Trump2015,
       author = {{Trump}, Jonathan R. and {Sun}, Mouyuan and {Zeimann}, Gregory R. and {Luck}, Cuyler and {Bridge}, Joanna S. and {Grier}, Catherine J. and {Hagen}, Alex and {Juneau}, Stephanie and {Montero-Dorta}, Antonio and {Rosario}, David J. and {Brandt}, W. Niel and {Ciardullo}, Robin and {Schneider}, Donald P.},
        title = "{The Biases of Optical Line-Ratio Selection for Active Galactic Nuclei and the Intrinsic Relationship between Black Hole Accretion and Galaxy Star Formation}",
      journal = {\apj},
         year = 2015,
        month = sep,
       volume = {811},
       number = {1},
          eid = {26},
        pages = {26},
          doi = {10.1088/0004-637X/811/1/26},
archivePrefix = {arXiv},
       eprint = {1501.02801},
 primaryClass = {astro-ph.GA},
       adsurl = {https://ui.adsabs.harvard.edu/abs/2015ApJ...811...26T}
}

@ARTICLE{Eracleous2010,
       author = {{Eracleous}, Michael and {Hwang}, Jason A. and {Flohic}, H{\'e}l{\`e}ne M.~L.~G.},
        title = "{Spectral Energy Distributions of Weak Active Galactic Nuclei Associated with Low-Ionization Nuclear Emission Regions}",
      journal = {\apjs},
         year = 2010,
        month = mar,
       volume = {187},
       number = {1},
        pages = {135-148},
          doi = {10.1088/0067-0049/187/1/135},
archivePrefix = {arXiv},
       eprint = {1001.2924},
 primaryClass = {astro-ph.GA},
       adsurl = {https://ui.adsabs.harvard.edu/abs/2010ApJS..187..135E}
}

@ARTICLE{Shimizu2017,
       author = {{Shimizu}, T. Taro and {Mushotzky}, Richard F. and {Mel{\'e}ndez}, Marcio and {Koss}, Michael J. and {Barger}, Amy J. and {Cowie}, Lennox L.},
        title = "{Herschel far-infrared photometry of the Swift Burst Alert Telescope active galactic nuclei sample of the local universe - III. Global star-forming properties and the lack of a connection to nuclear activity}",
      journal = {\mnras},
         year = 2017,
        month = apr,
       volume = {466},
       number = {3},
        pages = {3161-3183},
          doi = {10.1093/mnras/stw3268},
archivePrefix = {arXiv},
       eprint = {1612.03941},
 primaryClass = {astro-ph.GA},
       adsurl = {https://ui.adsabs.harvard.edu/abs/2017MNRAS.466.3161S}
}

@ARTICLE{Oh2018,
       author = {{Oh}, Kyuseok and {Koss}, Michael and {Markwardt}, Craig B. and {Schawinski}, Kevin and {Baumgartner}, Wayne H. and {Barthelmy}, Scott D. and {Cenko}, S. Bradley and {Gehrels}, Neil and {Mushotzky}, Richard and {Petulante}, Abigail and {Ricci}, Claudio and {Lien}, Amy and {Trakhtenbrot}, Benny},
        title = "{The 105-Month Swift-BAT All-sky Hard X-Ray Survey}",
      journal = {\apjs},
         year = 2018,
        month = mar,
       volume = {235},
       number = {1},
          eid = {4},
        pages = {4},
          doi = {10.3847/1538-4365/aaa7fd},
archivePrefix = {arXiv},
       eprint = {1801.01882},
 primaryClass = {astro-ph.HE},
       adsurl = {https://ui.adsabs.harvard.edu/abs/2018ApJS..235....4O}
}

@ARTICLE{Koss2011,
       author = {{Koss}, Michael and {Mushotzky}, Richard and {Veilleux}, Sylvain and {Winter}, Lisa M. and {Baumgartner}, Wayne and {Tueller}, Jack and {Gehrels}, Neil and {Valencic}, Lynne},
        title = "{Host Galaxy Properties of the Swift Bat Ultra Hard X-Ray Selected Active Galactic Nucleus}",
      journal = {\apj},
         year = 2011,
        month = oct,
       volume = {739},
       number = {2},
          eid = {57},
        pages = {57},
          doi = {10.1088/0004-637X/739/2/57},
archivePrefix = {arXiv},
       eprint = {1107.1237},
 primaryClass = {astro-ph.CO},
       adsurl = {https://ui.adsabs.harvard.edu/abs/2011ApJ...739...57K}
}

@Article{Schawinski2010,
       author = {{Schawinski}, Kevin and {Urry}, C. Megan and {Virani}, Shanil and {Coppi}, Paolo and {Bamford}, Steven P. and {Treister}, Ezequiel and {Lintott}, Chris J. and {Sarzi}, Marc and {Keel}, William C. and {Kaviraj}, Sugata and {Cardamone}, Carolin N. and {Masters}, Karen L. and {Ross}, Nicholas P. and {Ross}},
        title = "{Black Hole Growth and Host Galaxy Morphology}",
    journal   = {Proc. Int. Astron. Union},
    booktitle = {Co-Evolution of Central Black Holes and Galaxies},
         year = 2010,
       editor = {{Peterson}, Bradley M. and {Somerville}, Rachel S. and {Storchi-Bergmann}, Thaisa},
       series = {IAU Symposium},
       volume = {267},
        month = may,
        pages = {438-441},
          doi = {10.1017/S1743921310006964},
archivePrefix = {arXiv},
       eprint = {1002.1488},
 primaryClass = {astro-ph.CO},
       adsurl = {https://ui.adsabs.harvard.edu/abs/2010IAUS..267..438S}
}

@ARTICLE{Elitzur2009,
       author = {{Elitzur}, Moshe and {Ho}, Luis C.},
        title = "{On the Disappearance of the Broad-Line Region in Low-Luminosity Active Galactic Nuclei}",
      journal = {\apjl},
         year = 2009,
        month = aug,
       volume = {701},
       number = {2},
        pages = {L91-L94},
          doi = {10.1088/0004-637X/701/2/L91},
archivePrefix = {arXiv},
       eprint = {0907.3752},
 primaryClass = {astro-ph.CO},
       adsurl = {https://ui.adsabs.harvard.edu/abs/2009ApJ...701L..91E}
}

@ARTICLE{Lee2013,
       author = {{Lee}, Jong Chul and {Hwang}, Ho Seong and {Ko}, Jongwan},
        title = "{The Calibration of Star Formation Rate Indicators for WISE 22 {\ensuremath{\mu}}m-Selected Galaxies in the Sloan Digital Sky Survey}",
      journal = {\apj},
         year = 2013,
        month = sep,
       volume = {774},
       number = {1},
          eid = {62},
        pages = {62},
          doi = {10.1088/0004-637X/774/1/62},
archivePrefix = {arXiv},
       eprint = {1307.4078},
 primaryClass = {astro-ph.CO},
       adsurl = {https://ui.adsabs.harvard.edu/abs/2013ApJ...774...62L}
}

@ARTICLE{Mason2013,
       author = {{Mason}, R.~E. and {Ramos Almeida}, C. and {Levenson}, N.~A. and {Nemmen}, R. and {Alonso-Herrero}, A.},
        title = "{The Role of the Accretion Disk, Dust, and Jets in the IR Emission of Low-luminosity Active Galactic Nuclei}",
      journal = {\apj},
         year = 2013,
        month = nov,
       volume = {777},
       number = {2},
          eid = {164},
        pages = {164},
          doi = {10.1088/0004-637X/777/2/164},
archivePrefix = {arXiv},
       eprint = {1310.1892},
 primaryClass = {astro-ph.CO},
       adsurl = {https://ui.adsabs.harvard.edu/abs/2013ApJ...777..164M}
}

@ARTICLE{Annuar2017,
       author = {{Annuar}, A. and {Alexander}, D.~M. and {Gandhi}, P. and {Lansbury}, G.~B. and {Asmus}, D. and {Ballantyne}, D.~R. and {Bauer}, F.~E. and {Boggs}, S.~E. and {Boorman}, P.~G. and {Brandt}, W.~N. and {Brightman}, M. and {Christensen}, F.~E. and {Craig}, W.~W. and {Farrah}, D. and {Goulding}, A.~D. and {Hailey}, C.~J. and {Harrison}, F.~A. and {Koss}, M.~J. and {LaMassa}, S.~M. and {Murray}, S.~S. and {Ricci}, C. and {Rosario}, D.~J. and {Stanley}, F. and {Stern}, D. and {Zhang}, W.},
        title = "{A New Compton-thick AGN in our Cosmic Backyard: Unveiling the Buried Nucleus in NGC 1448 with NuSTAR}",
      journal = {\apj},
         year = 2017,
        month = feb,
       volume = {836},
       number = {2},
          eid = {165},
        pages = {165},
          doi = {10.3847/1538-4357/836/2/165},
archivePrefix = {arXiv},
       eprint = {1701.00497},
 primaryClass = {astro-ph.HE},
       adsurl = {https://ui.adsabs.harvard.edu/abs/2017ApJ...836..165A}
}

@ARTICLE{Brightman2018,
       author = {{Brightman}, M. and {Balokovi{\'c}}, M. and {Koss}, M. and {Alexander}, D.~M. and {Annuar}, A. and {Earnshaw}, H. and {Gandhi}, P. and {Harrison}, F.~A. and {Hornschemeier}, A.~E. and {Lehmer}, B. and {Powell}, M.~C. and {Ptak}, A. and {Rangelov}, B. and {Roberts}, T.~P. and {Stern}, D. and {Walton}, D.~J. and {Zezas}, A.},
        title = "{A Long Hard-X-Ray Look at the Dual Active Galactic Nuclei of M51 with NuSTAR}",
      journal = {\apj},
         year = 2018,
        month = nov,
       volume = {867},
       number = {2},
          eid = {110},
        pages = {110},
          doi = {10.3847/1538-4357/aae1ae},
archivePrefix = {arXiv},
       eprint = {1805.12140},
 primaryClass = {astro-ph.HE},
       adsurl = {https://ui.adsabs.harvard.edu/abs/2018ApJ...867..110B}
}

@ARTICLE{Jansen2001,
       author = {{Jansen}, F. and {Lumb}, D. and {Altieri}, B. and {Clavel}, J. and {Ehle}, M. and {Erd}, C. and {Gabriel}, C. and {Guainazzi}, M. and {Gondoin}, P. and {Much}, R. and {Munoz}, R. and {Santos}, M. and {Schartel}, N. and {Texier}, D. and {Vacanti}, G.},
        title = "{XMM-Newton observatory. I. The spacecraft and operations}",
      journal = {\aap},
         year = 2001,
        month = jan,
       volume = {365},
        pages = {L1-L6},
          doi = {10.1051/0004-6361:20000036},
       adsurl = {https://ui.adsabs.harvard.edu/abs/2001A&A...365L...1J}
}

@ARTICLE{Weisskopf2002,
       author = {{Weisskopf}, M.~C. and {Brinkman}, B. and {Canizares}, C. and {Garmire}, G. and {Murray}, S. and {Van Speybroeck}, L.~P.},
        title = "{An Overview of the Performance and Scientific Results from the Chandra X-Ray Observatory}",
      journal = {\pasp},
         year = 2002,
        month = jan,
       volume = {114},
       number = {791},
        pages = {1-24},
          doi = {10.1086/338108},
archivePrefix = {arXiv},
       eprint = {astro-ph/0110308},
 primaryClass = {astro-ph},
       adsurl = {https://ui.adsabs.harvard.edu/abs/2002PASP..114....1W}
}

@ARTICLE{Wenger2000,
       author = {{Wenger}, M. and {Ochsenbein}, F. and {Egret}, D. and {Dubois}, P. and {Bonnarel}, F. and {Borde}, S. and {Genova}, F. and {Jasniewicz}, G. and {Lalo{\"e}}, S. and {Lesteven}, S. and {Monier}, R.},
        title = "{The SIMBAD astronomical database. The CDS reference database for astronomical objects}",
      journal = {\aaps},
         year = 2000,
        month = apr,
       volume = {143},
        pages = {9-22},
          doi = {10.1051/aas:2000332},
archivePrefix = {arXiv},
       eprint = {astro-ph/0002110},
 primaryClass = {astro-ph},
       adsurl = {https://ui.adsabs.harvard.edu/abs/2000A&AS..143....9W}
}

@ARTICLE{Kim2021,
       author = {{Kim}, Minjin and {Barth}, Aaron J. and {Ho}, Luis C. and {Son}, Suyeon},
        title = "{A Hubble Space Telescope Imaging Survey of Low-redshift Swift-BAT Active Galaxies}",
      journal = {\apjs},
         year = 2021,
        month = oct,
       volume = {256},
       number = {2},
          eid = {40},
        pages = {40},
          doi = {10.3847/1538-4365/ac133e},
archivePrefix = {arXiv},
       eprint = {2107.04213},
 primaryClass = {astro-ph.GA},
       adsurl = {https://ui.adsabs.harvard.edu/abs/2021ApJS..256...40K}
}

@ARTICLE{Nandi2023,
       author = {{Nandi}, Payel and {Stalin}, C.~S. and {Saikia}, D.~J. and {Muneer}, S. and {Mountrichas}, George and {Wylezalek}, Dominika and {Sagar}, R. and {Kissler-Patig}, Markus},
        title = "{Star Formation in the Dwarf Seyfert Galaxy NGC 4395: Evidence for Both AGN and SN Feedback?}",
      journal = {\apj},
         year = 2023,
        month = jun,
       volume = {950},
       number = {2},
          eid = {81},
        pages = {81},
          doi = {10.3847/1538-4357/accf1e},
archivePrefix = {arXiv},
       eprint = {2304.08986},
 primaryClass = {astro-ph.GA},
       adsurl = {https://ui.adsabs.harvard.edu/abs/2023ApJ...950...81N}
}

@ARTICLE{Yang2022,
       author = {{Yang}, Jun and {Yang}, Xiaolong and {Wrobel}, Joan M. and {Paragi}, Zsolt and {Gurvits}, Leonid I. and {Ho}, Luis C. and {Nyland}, Kristina and {Fan}, Lulu and {Tafoya}, Daniel},
        title = "{Is there a sub-parsec-scale jet base in the nearby dwarf galaxy NGC 4395?}",
      journal = {\mnras},
         year = 2022,
        month = aug,
       volume = {514},
       number = {4},
        pages = {6215-6224},
          doi = {10.1093/mnras/stac1753},
archivePrefix = {arXiv},
       eprint = {2206.10964},
 primaryClass = {astro-ph.HE},
       adsurl = {https://ui.adsabs.harvard.edu/abs/2022MNRAS.514.6215Y}
}

@ARTICLE{Boorman2025,
       author = {{Boorman}, Peter G. and {Gandhi}, Poshak and {Buchner}, Johannes and {Stern}, Daniel and {Ricci}, Claudio and {Balokovi{\'c}}, Mislav and {Asmus}, Daniel and {Harrison}, Fiona A. and {Svoboda}, Ji{\v{r}}{\'\i} and {Greenwell}, Claire and {Koss}, Michael J. and {Alexander}, David M. and {Annuar}, Adlyka and {Bauer}, Franz E. and {Brandt}, William N. and {Brightman}, Murray and {Civano}, Francesca and {Chen}, Chien-Ting J. and {Farrah}, Duncan and {Forster}, Karl and {Grefenstette}, Brian and {H{\"o}nig}, Sebastian F. and {Hill}, Adam B. and {Kammoun}, Elias and {Lansbury}, George and {Lanz}, Lauranne and {LaMassa}, Stephanie and {Madsen}, Kristin and {Marchesi}, Stefano and {Middleton}, Matthew and {Mingo}, Beatriz and {Parker}, Michael L. and {Treister}, Ezequiel and {Ueda}, Yoshihiro and {Urry}, C. Megan and {Zappacosta}, Luca},
        title = "{The NuSTAR Local AGN N $_{H}$ Distribution Survey (NuLANDS). I. Toward a Truly Representative Column Density Distribution in the Local Universe}",
      journal = {\apj},
         year = 2025,
        month = jan,
       volume = {978},
       number = {1},
          eid = {118},
        pages = {118},
          doi = {10.3847/1538-4357/ad8236},
archivePrefix = {arXiv},
       eprint = {2410.07339},
 primaryClass = {astro-ph.GA},
       adsurl = {https://ui.adsabs.harvard.edu/abs/2025ApJ...978..118B}
}

@ARTICLE{daSilva2021,
       author = {{da Silva}, Patr{\'\i}cia and {Menezes}, R.~B. and {D{\'\i}az}, Y. and {L{\'o}pez-Navas}, Elena and {Steiner}, J.~E.},
        title = "{The nuclear environment of NGC 2442: a Compton-thick low-luminosity AGN}",
      journal = {\mnras},
         year = 2021,
        month = jul,
       volume = {505},
       number = {1},
        pages = {223-235},
          doi = {10.1093/mnras/stab1249},
archivePrefix = {arXiv},
       eprint = {2105.09420},
 primaryClass = {astro-ph.GA},
       adsurl = {https://ui.adsabs.harvard.edu/abs/2021MNRAS.505..223D}
}

@ARTICLE{Balokovic2020,
       author = {{Balokovi{\'c}}, M. and {Harrison}, F.~A. and {Madejski}, G. and {Comastri}, A. and {Ricci}, C. and {Annuar}, A. and {Ballantyne}, D.~R. and {Boorman}, P. and {Brandt}, W.~N. and {Brightman}, M. and {Gandhi}, P. and {Kamraj}, N. and {Koss}, M.~J. and {Marchesi}, S. and {Marinucci}, A. and {Masini}, A. and {Matt}, G. and {Stern}, D. and {Urry}, C.~M.},
        title = "{NuSTAR Survey of Obscured Swift/BAT-selected Active Galactic Nuclei. II. Median High-energy Cutoff in Seyfert II Hard X-Ray Spectra}",
      journal = {\apj},
         year = 2020,
        month = dec,
       volume = {905},
       number = {1},
          eid = {41},
        pages = {41},
          doi = {10.3847/1538-4357/abc342},
archivePrefix = {arXiv},
       eprint = {2011.06583},
 primaryClass = {astro-ph.HE},
       adsurl = {https://ui.adsabs.harvard.edu/abs/2020ApJ...905...41B}
}

@ARTICLE{Annuar2025,
       author = {{Annuar}, A. and {Alexander}, D.~M. and {Gandhi}, P. and {Lansbury}, G.~B. and {Rosli}, M.~N. and {Stern}, D. and {Asmus}, D. and {Ballantyne}, D.~R. and {Balokovi{\'c}}, M. and {Bauer}, F.~E. and {Boorman}, P.~G. and {Brandt}, W.~N. and {Brightman}, M. and {Chen}, C.~T.~J. and {Del Moro}, A. and {Farrah}, D. and {Harrison}, F.~A. and {Koss}, M.~J. and {Lanz}, L. and {Marchesi}, S. and {Mohanadas}, P. and {Nardini}, E. and {Ricci}, C. and {Zappacosta}, L.},
        title = "{The Compton-thick AGN population and the N$_{H}$ distribution of low-mass AGN in our cosmic backyard}",
      journal = {\mnras},
         year = 2025,
        month = jul,
       volume = {540},
       number = {4},
        pages = {3827-3849},
          doi = {10.1093/mnras/staf956},
archivePrefix = {arXiv},
       eprint = {2506.08527},
 primaryClass = {astro-ph.GA},
       adsurl = {https://ui.adsabs.harvard.edu/abs/2025MNRAS.540.3827A}
}

@ARTICLE{She2017,
       author = {{She}, Rui and {Ho}, Luis C. and {Feng}, Hua},
        title = "{Chandra Survey of Nearby Galaxies: The Catalog}",
      journal = {\apj},
         year = 2017,
        month = feb,
       volume = {835},
       number = {2},
          eid = {223},
        pages = {223},
          doi = {10.3847/1538-4357/835/2/223},
archivePrefix = {arXiv},
       eprint = {1612.08507},
 primaryClass = {astro-ph.GA},
       adsurl = {https://ui.adsabs.harvard.edu/abs/2017ApJ...835..223S}
}

@ARTICLE{Desroches2009,
       author = {{Desroches}, Louis-Benoit and {Ho}, Luis C.},
        title = "{Candidate Active Nuclei in Late-Type Spiral Galaxies}",
      journal = {\apj},
         year = 2009,
        month = jan,
       volume = {690},
       number = {1},
        pages = {267-278},
          doi = {10.1088/0004-637X/690/1/267},
       adsurl = {https://ui.adsabs.harvard.edu/abs/2009ApJ...690..267D}
}

@ARTICLE{Miller2015,
       author = {{Miller}, Brendan P. and {Gallo}, Elena and {Greene}, Jenny E. and {Kelly}, Brandon C. and {Treu}, Tommaso and {Woo}, Jong-Hak and {Baldassare}, Vivienne},
        title = "{X-Ray Constraints on the Local Supermassive Black Hole Occupation Fraction}",
      journal = {\apj},
         year = 2015,
        month = jan,
       volume = {799},
       number = {1},
          eid = {98},
        pages = {98},
          doi = {10.1088/0004-637X/799/1/98},
archivePrefix = {arXiv},
       eprint = {1403.4246},
 primaryClass = {astro-ph.GA},
       adsurl = {https://ui.adsabs.harvard.edu/abs/2015ApJ...799...98M}
}

@ARTICLE{Mazzolari2025,
       author = {{Mazzolari}, Giovanni and {Scholtz}, Jan and {Maiolino}, Roberto and {Gilli}, Roberto and {Traina}, Alberto and {L{\'o}pez}, Ivan E. and {{\"U}bler}, Hannah and {Trefoloni}, Bartolomeo and {D'Eugenio}, Francesco and {Ji}, Xihan and {Mignoli}, Marco and {Vito}, Fabio and {Vignali}, Cristian and {Brusa}, Marcella},
        title = "{Narrow-line AGN selection in CEERS: Spectroscopic selection, physical properties, and X-ray and radio analysis}",
      journal = {\aap},
         year = 2025,
        month = aug,
       volume = {700},
          eid = {A12},
        pages = {A12},
          doi = {10.1051/0004-6361/202451860},
archivePrefix = {arXiv},
       eprint = {2408.15615},
 primaryClass = {astro-ph.GA},
       adsurl = {https://ui.adsabs.harvard.edu/abs/2025A&A...700A..12M}
}

@ARTICLE{Mickaelian2021,
       author = {{Mickaelian}, A.~M. and {Abrahamyan}, H.~V. and {Paronyan}, G.~M. and {Mikayelyan}, G.~A.},
        title = "{Fine classification of QSOs and Seyferts for activity types based on SDSS spectroscopy}",
      journal = {Frontiers in Astronomy and Space Sciences},
         year = 2021,
        month = mar,
       volume = {7},
          eid = {82},
        pages = {82},
          doi = {10.3389/fspas.2020.505043},
       adsurl = {https://ui.adsabs.harvard.edu/abs/2021FrASS...7...82M}
}

@ARTICLE{Mickaelian2024,
       author = {{Mickaelian}, A.~M. and {Abrahamyan}, H.~V. and {Paronyan}, G.~M. and {Mikayelyan}, G.~A. and {Sukiasyan}, A.~G. and {Mkrtchyan}, V.~K.},
        title = "{Classification of Flat Spectrum Radio Quasars by Optical Activity Types}",
      journal = {Astrophysics},
         year = 2024,
        month = mar,
       volume = {67},
       number = {1},
        pages = {1-8},
          doi = {10.1007/s10511-024-09811-8},
       adsurl = {https://ui.adsabs.harvard.edu/abs/2024Ap.....67....1M}
}

@ARTICLE{Mezcua2024,
       author = {{Mezcua}, M. and {Dom{\'\i}nguez S{\'a}nchez}, H.},
        title = "{MaNGA AGN dwarf galaxies (MAD) - I. A new sample of AGNs in dwarf galaxies with spatially-resolved spectroscopy}",
      journal = {\mnras},
         year = 2024,
        month = mar,
       volume = {528},
       number = {3},
        pages = {5252-5268},
          doi = {10.1093/mnras/stae292},
archivePrefix = {arXiv},
       eprint = {2401.15152},
 primaryClass = {astro-ph.GA},
       adsurl = {https://ui.adsabs.harvard.edu/abs/2024MNRAS.528.5252M}
}

@ARTICLE{Zaw2019,
       author = {{Zaw}, Ingyin and {Chen}, Yan-Ping and {Farrar}, Glennys R.},
        title = "{A Uniformly Selected, All-sky, Optical AGN Catalog}",
      journal = {\apj},
         year = 2019,
        month = feb,
       volume = {872},
       number = {2},
          eid = {134},
        pages = {134},
          doi = {10.3847/1538-4357/aaffaf},
archivePrefix = {arXiv},
       eprint = {1902.03799},
 primaryClass = {astro-ph.GA},
       adsurl = {https://ui.adsabs.harvard.edu/abs/2019ApJ...872..134Z}
}

@ARTICLE{Couto2023,
       author = {{Couto}, Guilherme S. and {Storchi-Bergmann}, Thaisa},
        title = "{The Interplay between Radio AGN Activity and Their Host Galaxies}",
      journal = {Galaxies},
         year = 2023,
        month = mar,
       volume = {11},
       number = {2},
          eid = {47},
        pages = {47},
          doi = {10.3390/galaxies11020047},
archivePrefix = {arXiv},
       eprint = {2303.12033},
 primaryClass = {astro-ph.GA},
       adsurl = {https://ui.adsabs.harvard.edu/abs/2023Galax..11...47C}
}

@ARTICLE{Ho2009a,
       author = {{Ho}, Luis C. and {Greene}, Jenny E. and {Filippenko}, Alexei V. and {Sargent}, Wallace L.~W.},
        title = "{A Search for ``Dwarf'' Seyfert Nuclei. VII. A Catalog of Central Stellar Velocity Dispersions of Nearby Galaxies}",
      journal = {\apjs},
         year = 2009,
        month = jul,
       volume = {183},
       number = {1},
        pages = {1-16},
          doi = {10.1088/0067-0049/183/1/1},
archivePrefix = {arXiv},
       eprint = {0906.4105},
 primaryClass = {astro-ph.GA},
       adsurl = {https://ui.adsabs.harvard.edu/abs/2009ApJS..183....1H}
}

@ARTICLE{Lopez2024,
       author = {{L{\'o}pez}, I.~E. and {Yang}, G. and {Mountrichas}, G. and {Brusa}, M. and {Alexander}, D.~M. and {Baldi}, R.~D. and {Bertola}, E. and {Bonoli}, S. and {Comastri}, A. and {Shankar}, F. and {Acharya}, N. and {Alonso Tetilla}, A.~V. and {Lapi}, A. and {Laloux}, B. and {L{\'o}pez L{\'o}pez}, X. and {Mu{\~n}oz Rodr{\'\i}guez}, I. and {Musiimenta}, B. and {Osorio Clavijo}, N. and {Sala}, L. and {Sengupta}, D.},
        title = "{A CIGALE module tailored (not only) for low-luminosity active galactic nuclei}",
      journal = {\aap},
         year = 2024,
        month = dec,
       volume = {692},
          eid = {A209},
        pages = {A209},
          doi = {10.1051/0004-6361/202450510},
archivePrefix = {arXiv},
       eprint = {2404.16938},
 primaryClass = {astro-ph.GA},
       adsurl = {https://ui.adsabs.harvard.edu/abs/2024A&A...692A.209L}
}

@ARTICLE{Hviding2022,
       author = {{Hviding}, Raphael E. and {Hainline}, Kevin N. and {Rieke}, Marcia and {Juneau}, St{\'e}phanie and {Lyu}, Jianwei and {Pucha}, Ragadeepika},
        title = "{A New Infrared Criterion for Selecting Active Galactic Nuclei to Lower Luminosities}",
      journal = {\aj},
         year = 2022,
        month = may,
       volume = {163},
       number = {5},
          eid = {224},
        pages = {224},
          doi = {10.3847/1538-3881/ac5e33},
archivePrefix = {arXiv},
       eprint = {2203.11217},
 primaryClass = {astro-ph.GA},
       adsurl = {https://ui.adsabs.harvard.edu/abs/2022AJ....163..224H}
}

@ARTICLE{King2015,
       author = {{King}, Andrew and {Nixon}, Chris},
        title = "{AGN flickering and chaotic accretion}",
      journal = {\mnras},
         year = 2015,
        month = oct,
       volume = {453},
       number = {1},
        pages = {L46-L47},
          doi = {10.1093/mnrasl/slv098},
archivePrefix = {arXiv},
       eprint = {1507.05960},
 primaryClass = {astro-ph.HE},
       adsurl = {https://ui.adsabs.harvard.edu/abs/2015MNRAS.453L..46K}
}

@ARTICLE{Schawinski2015,
       author = {{Schawinski}, Kevin and {Koss}, Michael and {Berney}, Simon and {Sartori}, Lia F.},
        title = "{Active galactic nuclei flicker: an observational estimate of the duration of black hole growth phases of {\ensuremath{\sim}}{}10$^{5}$ yr}",
      journal = {\mnras},
         year = 2015,
        month = aug,
       volume = {451},
       number = {3},
        pages = {2517-2523},
          doi = {10.1093/mnras/stv1136},
archivePrefix = {arXiv},
       eprint = {1505.06733},
 primaryClass = {astro-ph.GA},
       adsurl = {https://ui.adsabs.harvard.edu/abs/2015MNRAS.451.2517S}
}

@ARTICLE{Hickox2014,
       author = {{Hickox}, Ryan C. and {Mullaney}, James R. and {Alexander}, David M. and {Chen}, Chien-Ting J. and {Civano}, Francesca M. and {Goulding}, Andy D. and {Hainline}, Kevin N.},
        title = "{Black Hole Variability and the Star Formation-Active Galactic Nucleus Connection: Do All Star-forming Galaxies Host an Active Galactic Nucleus?}",
      journal = {\apj},
         year = 2014,
        month = feb,
       volume = {782},
       number = {1},
          eid = {9},
        pages = {9},
          doi = {10.1088/0004-637X/782/1/9},
archivePrefix = {arXiv},
       eprint = {1306.3218},
 primaryClass = {astro-ph.CO},
       adsurl = {https://ui.adsabs.harvard.edu/abs/2014ApJ...782....9H}
}

@ARTICLE{Fabian2012,
       author = {{Fabian}, A.~C.},
        title = "{Observational Evidence of Active Galactic Nuclei Feedback}",
      journal = {\araa},
         year = 2012,
        month = sep,
       volume = {50},
        pages = {455-489},
          doi = {10.1146/annurev-astro-081811-125521},
archivePrefix = {arXiv},
       eprint = {1204.4114},
 primaryClass = {astro-ph.CO},
       adsurl = {https://ui.adsabs.harvard.edu/abs/2012ARA&A..50..455F}
}

@ARTICLE{Young2012,
       author = {{Young}, M. and {Brandt}, W.~N. and {Xue}, Y.~Q. and {Paolillo}, M. and {Alexander}, D.~M. and {Bauer}, F.~E. and {Lehmer}, B.~D. and {Luo}, B. and {Shemmer}, O. and {Schneider}, D.~P. and {Vignali}, C.},
        title = "{Variability-selected Low-luminosity Active Galactic Nuclei in the 4 Ms Chandra Deep Field-South}",
      journal = {\apj},
         year = 2012,
        month = apr,
       volume = {748},
       number = {2},
          eid = {124},
        pages = {124},
          doi = {10.1088/0004-637X/748/2/124},
archivePrefix = {arXiv},
       eprint = {1201.4391},
 primaryClass = {astro-ph.CO},
       adsurl = {https://ui.adsabs.harvard.edu/abs/2012ApJ...748..124Y}
}

@ARTICLE{Nemer2025,
       author = {{Nemer}, Ahmad and {Katkov}, Ivan Yu. and {Gelfand}, Joseph D. and {Cho}, Changhyun},
        title = "{A Closer Look at the Origin of LINER Emission in Later-type Galaxies and Its Connection to Evolved Stars with a Machine Learning Classification Scheme}",
      journal = {\apj},
         year = 2025,
        month = may,
       volume = {984},
       number = {2},
          eid = {106},
        pages = {106},
          doi = {10.3847/1538-4357/adc57e},
archivePrefix = {arXiv},
       eprint = {2410.20156},
 primaryClass = {astro-ph.GA},
       adsurl = {https://ui.adsabs.harvard.edu/abs/2025ApJ...984..106N}
}

@ARTICLE{Berney2015,
       author = {{Berney}, Simon and {Koss}, Michael and {Trakhtenbrot}, Benny and {Ricci}, Claudio and {Lamperti}, Isabella and {Schawinski}, Kevin and {Balokovi{\'c}}, Mislav and {Crenshaw}, D. Michael and {Fischer}, Travis and {Gehrels}, Neil and {Harrison}, Fiona and {Hashimoto}, Yasuhiro and {Ichikawa}, Kohei and {Mushotzky}, Richard and {Oh}, Kyuseok and {Stern}, Daniel and {Treister}, Ezequiel and {Ueda}, Yoshihiro and {Veilleux}, Sylvain and {Winter}, Lisa},
        title = "{BAT AGN spectroscopic survey-II. X-ray emission and high-ionization optical emission lines}",
      journal = {\mnras},
         year = 2015,
        month = dec,
       volume = {454},
       number = {4},
        pages = {3622-3634},
          doi = {10.1093/mnras/stv2181},
archivePrefix = {arXiv},
       eprint = {1509.05425},
 primaryClass = {astro-ph.GA},
       adsurl = {https://ui.adsabs.harvard.edu/abs/2015MNRAS.454.3622B}
}

@ARTICLE{Bresolin2025,
       author = {{Bresolin}, Fabio and {Kudritzki}, Rolf-Peter and {Urbaneja}, Miguel A. and {Sextl}, Eva and {Riess}, Adam G.},
        title = "{Blue Supergiants in the Pinwheel Galaxy M101: Comparison with H II Region Chemical Abundances, Spectroscopic Distance, and an Independent Determination of the Hubble Constant}",
      journal = {\apj},
         year = 2025,
        month = oct,
       volume = {991},
       number = {2},
          eid = {151},
        pages = {151},
          doi = {10.3847/1538-4357/adfc4c},
archivePrefix = {arXiv},
       eprint = {2508.11837},
 primaryClass = {astro-ph.GA},
       adsurl = {https://ui.adsabs.harvard.edu/abs/2025ApJ...991..151B}
}

@ARTICLE{Smirnova2023,
       author = {{Smirnova}, Ksenia Ildarovna and {Wiebe}, Dmitri Siegfriedovich},
        title = "{Dust and gas in star-forming complexes in NGC 3351, NGC 5055, and NGC 5457}",
      journal = {Open Astronomy},
         year = 2023,
        month = apr,
       volume = {32},
       number = {1},
          eid = {219},
        pages = {219},
          doi = {10.1515/astro-2022-0219},
       adsurl = {https://ui.adsabs.harvard.edu/abs/2023OAst...32..219S}
}

@ARTICLE{Leftley2024,
       author = {{Leftley}, J.~H. and {Petrov}, R. and {Moszczynski}, N. and {Vermot}, P. and {H{\"o}nig}, S.~F. and {Gamez Rosas}, V. and {Isbell}, J.~W. and {Jaffe}, W. and {Cl{\'e}net}, Y. and {Augereau}, J.-C. and {Berio}, P. and {Davies}, R.~I. and {Henning}, T. and {Lagarde}, S. and {Lopez}, B. and {Matter}, A. and {Meilland}, A. and {Millour}, F. and {Nesvadba}, N. and {Shimizu}, T.~T. and {Sturm}, E. and {Weigelt}, G.},
        title = "{Chromatically modeling the parsec-scale dusty structure in the center of NGC 1068}",
      journal = {\aap},
         year = 2024,
        month = jun,
       volume = {686},
          eid = {A204},
        pages = {A204},
          doi = {10.1051/0004-6361/202348977},
archivePrefix = {arXiv},
       eprint = {2312.12125},
 primaryClass = {astro-ph.GA},
       adsurl = {https://ui.adsabs.harvard.edu/abs/2024A&A...686A.204L}
}

@ARTICLE{Mutie2025,
       author = {{Mutie}, Isaac M. and {del Palacio}, Santiago and {Beswick}, Robert J. and {Williams-Baldwin}, David and {Gallimore}, Jack F. and {Gallagher}, John S. and {Aalto}, Susanne E. and {Baki}, Paul O.},
        title = "{A consistent radio to sub-mm pc-scale study of the nucleus of NGC 1068}",
      journal = {\mnras},
         year = 2025,
        month = may,
       volume = {539},
       number = {2},
        pages = {808-819},
          doi = {10.1093/mnras/staf524},
archivePrefix = {arXiv},
       eprint = {2503.20303},
 primaryClass = {astro-ph.GA},
       adsurl = {https://ui.adsabs.harvard.edu/abs/2025MNRAS.539..808M}
}

@ARTICLE{Pons2016,
       author = {{Pons}, E. and {Watson}, M.~G.},
        title = "{A new sample of X-ray selected narrow emission-line galaxies. II. Looking for True Seyfert 2}",
      journal = {\aap},
         year = 2016,
        month = oct,
       volume = {594},
          eid = {A72},
        pages = {A72},
          doi = {10.1051/0004-6361/201629194},
archivePrefix = {arXiv},
       eprint = {1608.01134},
 primaryClass = {astro-ph.HE},
       adsurl = {https://ui.adsabs.harvard.edu/abs/2016A&A...594A..72P}
}

@ARTICLE{Nandra1997,
       author = {{Nandra}, K. and {George}, I.~M. and {Mushotzky}, R.~F. and {Turner}, T.~J. and {Yaqoob}, T.},
        title = "{ASCA Observations of Seyfert 1 Galaxies. I. Data Analysis, Imaging, and Timing}",
      journal = {\apj},
         year = 1997,
        month = feb,
       volume = {476},
       number = {1},
        pages = {70-82},
          doi = {10.1086/303600},
       adsurl = {https://ui.adsabs.harvard.edu/abs/1997ApJ...476...70N}
}

@ARTICLE{Ford2025,
       author = {{Ford}, Nicole M. and {Nowak}, Michael and {Ramakrishnan}, Venkatessh and {Haggard}, Daryl and {Dage}, Kristen and {Nair}, Dhanya G. and {Chan}, Chi-kwan},
        title = "{Tracking X-Ray Variability in Next-generation EHT Low-luminosity Active Galactic Nucleus Targets}",
      journal = {\apj},
         year = 2025,
        month = mar,
       volume = {981},
       number = {2},
          eid = {126},
        pages = {126},
          doi = {10.3847/1538-4357/adae0f},
archivePrefix = {arXiv},
       eprint = {2501.14871},
 primaryClass = {astro-ph.HE},
       adsurl = {https://ui.adsabs.harvard.edu/abs/2025ApJ...981..126F}
}

@ARTICLE{Emmanoulopoulos2012,
       author = {{Emmanoulopoulos}, D. and {Papadakis}, I.~E. and {McHardy}, I.~M. and {Ar{\'e}valo}, P. and {Calvelo}, D.~E. and {Uttley}, P.},
        title = "{The 'harder when brighter' X-ray behaviour of the low-luminosity active galactic nucleus NGC 7213}",
      journal = {\mnras},
         year = 2012,
        month = aug,
       volume = {424},
       number = {2},
        pages = {1327-1334},
          doi = {10.1111/j.1365-2966.2012.21316.x},
archivePrefix = {arXiv},
       eprint = {1205.3524},
 primaryClass = {astro-ph.CO},
       adsurl = {https://ui.adsabs.harvard.edu/abs/2012MNRAS.424.1327E}
}

@ARTICLE{Graham2005,
       author = {{Graham}, Alister W. and {Driver}, Simon P.},
        title = "{A Concise Reference to (Projected) S{\'e}rsic R$^{1/n}$ Quantities, Including Concentration, Profile Slopes, Petrosian Indices, and Kron Magnitudes}",
      journal = {\pasa},
         year = 2005,
        month = jan,
       volume = {22},
       number = {2},
        pages = {118-127},
          doi = {10.1071/AS05001},
archivePrefix = {arXiv},
       eprint = {astro-ph/0503176},
 primaryClass = {astro-ph},
       adsurl = {https://ui.adsabs.harvard.edu/abs/2005PASA...22..118G}
}

@ARTICLE{Fisher2008,
       author = {{Fisher}, David B. and {Drory}, Niv},
        title = "{The Structure of Classical Bulges and Pseudobulges: the Link Between Pseudobulges and S{\'E}RSIC Index}",
      journal = {\aj},
         year = 2008,
        month = aug,
       volume = {136},
       number = {2},
        pages = {773-839},
          doi = {10.1088/0004-6256/136/2/773},
archivePrefix = {arXiv},
       eprint = {0805.4206},
 primaryClass = {astro-ph},
       adsurl = {https://ui.adsabs.harvard.edu/abs/2008AJ....136..773F}
}

@ARTICLE{Fisher2010,
       author = {{Fisher}, David B. and {Drory}, Niv},
        title = "{Bulges of Nearby Galaxies with Spitzer: Scaling Relations in Pseudobulges and Classical Bulges}",
      journal = {\apj},
         year = 2010,
        month = jun,
       volume = {716},
       number = {2},
        pages = {942-969},
          doi = {10.1088/0004-637X/716/2/942},
archivePrefix = {arXiv},
       eprint = {1004.5393},
 primaryClass = {astro-ph.CO},
       adsurl = {https://ui.adsabs.harvard.edu/abs/2010ApJ...716..942F}
}

@ARTICLE{Blandford1999,
       author = {{Blandford}, Roger D. and {Begelman}, Mitchell C.},
        title = "{On the fate of gas accreting at a low rate on to a black hole}",
      journal = {\mnras},
         year = 1999,
        month = feb,
       volume = {303},
       number = {1},
        pages = {L1-L5},
          doi = {10.1046/j.1365-8711.1999.02358.x},
archivePrefix = {arXiv},
       eprint = {astro-ph/9809083},
 primaryClass = {astro-ph},
       adsurl = {https://ui.adsabs.harvard.edu/abs/1999MNRAS.303L...1B}
}

\appendix

\section{Sources Excluded from the Analysis}
\label{Appendix A}

\begin{table*}
\centering
\caption{Seven AGN in the original sample from \citet{Saikia2018} that have been excluded from our sample due having exhibit $\log L_{\rm bol} > 42$~erg~s$^{-1}$, higher that our defined luminosity threshold for LLAGN.}

\vspace{0.6cm}

\begin{tabular}{l c c c}
\hline
Galaxy & 
$\log L_{\mathrm{bol}}$ (X-ray; \citealt{Lopez2024}) & 
$\log L_{\mathrm{bol}}$ (X-ray; \citealt{Vasudevan2010}) & 
$\log L_{\mathrm{bol}}$ (Optical; \citealt{Heckman2014}) \\
 & (erg s$^{-1}$) & (erg s$^{-1}$) & (erg s$^{-1}$) \\
\hline
NGC410  & 41.77$^{+0.12}_{-0.18}$ & 42.09$^{+0.12}_{-0.18}$ & 42.18$^{+0.11}_{-0.16}$ \\
NGC2832 & 41.75$^{+0.16}_{-0.26}$ & 42.06$^{+0.16}_{-0.26}$ & 41.93$^{+0.11}_{-0.16}$ \\
NGC3735 & 41.57$^{+0.02}_{-0.02}$ & 41.88$^{+0.02}_{-0.02}$ & 42.63$^{+0.11}_{-0.16}$ \\
NGC3982 & 39.99$^{+0.11}_{-0.15}$ & 40.31$^{+0.11}_{-0.15}$ & 42.59$^{+0.11}_{-0.16}$ \\
NGC4051 & 42.01$^{+0.01}_{-0.01}$ & 42.33$^{+0.01}_{-0.01}$ & 42.53$^{+0.11}_{-0.16}$ \\
NGC5273 & 41.49$^{+0.05}_{-0.05}$ & 41.81$^{+0.05}_{-0.05}$ & 42.36$^{+0.11}_{-0.16}$ \\
NGC6482 & 42.48$^{+0.01}_{-0.01}$ & 42.80$^{+0.01}_{-0.01}$ & – \\
\hline
\end{tabular}
\label{Tab: Luminosity for Excluded 7 LLAGN}
\end{table*}

\bsp	
\label{lastpage}
\end{document}